\documentclass[a4paper, 12pt]{article}

\usepackage{graphicx}
\usepackage[utf8]{inputenc}
\usepackage{lineno}
\usepackage{blindtext}
\usepackage{url}
\usepackage{amsmath}
\usepackage{amsthm}
\usepackage{amssymb} 
\usepackage{float, placeins}
\usepackage{caption, subcaption}
\usepackage{dblfloatfix}
\usepackage[font={it}]{caption}
\usepackage[margin=2.0cm]{geometry}
\usepackage{cancel}
\usepackage{xfrac}
\usepackage{hyperref}
\usepackage{booktabs}
\usepackage{siunitx} 
\usepackage{accents}
\usepackage{titling}
\usepackage[super]{nth}
\usepackage[toc,page]{appendix}
\usepackage{algorithm}
\usepackage{algpseudocode}
\usepackage{bm} 
\usepackage{arydshln}
\usepackage{tikz}
\usepackage{tikz-3dplot}
\usepackage{authblk}
\usepackage{blindtext}
\usepackage{makecell}

\newcommand{\whatAmI}{paper\xspace}

\newcommand{\pytorch}{\textsc{PyTorch}\xspace}

\newcommand{\tomopt}{\textsc{TomOpt}\xspace}

\newcommand{\geant}{\textsc{Geant}~4\xspace}
\newcommand{\xo}{\ensuremath{X_0}\xspace}
\newcommand{\mev}{\mega\electronvolt}
\newcommand{\gev}{\giga\electronvolt}
\newcommand{\tev}{\tera\electronvolt}
\newcommand{\wrt}{with respect to\xspace}
\newcommand{\poca}{PoCA\xspace}
\newcommand{\pocas}{PoCAs\xspace}

\newcommand{\sfSmall}{0.49}

\begin{document}

\date{2023-09-26}
\title{\tomopt: Differential optimisation for task- and constraint-aware design of particle detectors in the context of muon tomography}

\author[1,2]{Giles C. Strong\footnote{giles.c.strong@gmail.com}}
\author[1,3]{Maxime Lagrange \footnote{maxime.lagrange@uclouvain.be}}
\author[4,5]{Aitor Orio\footnote{aitor.orio@muon.systems}}
\author[6]{Anna Bordignon}
\author[7]{Florian Bury}
\author[1,2,8]{Tommaso Dorigo}
\author[1,3]{Andrea Giammanco}
\author[9]{Mariam Heikal}
\author[1,10]{Jan Kieseler}
\author[1,11]{Max Lamparth}
\author[1,12]{Pablo Mart\'inez Ru\'iz del \'Arbol}
\author[1,13,14]{Federico Nardi}
\author[1,15]{Pietro Vischia}
\author[1,16,17]{Haitham Zaraket}
\affil[1]{MODE Collaboration, \url{https://mode-collaboration.github.io/}}
\affil[2]{Istituto Nazionale di Fisica Nucleare, Sezione di Padova, Italy}
\affil[3]{Centre for Cosmology, Particle Physics and Phenomenology (CP3), Universit\'e catholique de Louvain, Louvain la Neuve, Belgium}
\affil[4]{Muon Tomography Systems S.L., Bilbao, España}
\affil[5]{Universidad de Cantabria (UC), Santander, España}
\affil[6]{Department of Statistical Sciences, University of Padova, Italy}
\affil[7]{University of Bristol, United Kingdom}
\affil[8]{Lulea Tekniska Universitet, Laboratorievägen 14 Lulea, Sweden}
\affil[9]{American University of Beirut, Beirut, Lebanon}
\affil[10]{Karlsruhe Institute of Technology, Karlsruhe, Germany}
\affil[11]{Physik-Department, Technische Universität München, Germany}
\affil[12]{Instituto de F\'isica de Cantabria, España}
\affil[13]{Dipartimento di Fisica ed Astronomia, University of Padova, Italy}
\affil[14]{Université Clermont-Auvergne, Clermont-Ferrand, France}
\affil[15]{Universidad de Oviedo and ICTEA, Oviedo, España}
\affil[16]{Multi-Disciplinary Physics Laboratory, Optics and Fiber Optics Group, Faculty of
Sciences, Lebanese University, Lebanon}
\affil[17]{Laboratoire de Physique Subatomique et de Cosmologie, Université Grenoble-Alpes, CNRS/IN2P3,
53 Avenue des Martyrs, 38026 Grenoble, France}
\maketitle

\abstract{We describe a software package, TomOpt, developed to optimise the geometrical layout and specifications of detectors designed for tomography by scattering of cosmic-ray muons. The software exploits differentiable programming for the modeling of muon interactions with detectors and scanned volumes, the inference of volume properties, and the optimisation cycle performing the loss minimisation. In doing so, we provide the first demonstration of end-to-end-differentiable and inference-aware optimisation of particle physics instruments. We study the performance of the software on a relevant benchmark scenario and discuss its potential applications. Our code is available on Github~\cite{tomopt_git}}

\clearpage

\begin{center}
	\hrule
\end{center}
\tableofcontents
\begin{center}
	\hrule
\end{center}


\section{Introduction}\label{sec:intro}

\subsection{Differentiable programming for detector design optimisation}

Over the past two decades, the availability of high-performance computing and the development of neural networks of larger capacity have conspired to fuel a revolution in the way we think at the optimisation of complex systems. When the dimensionality of the space of relevant design parameters exceeds a few units, and brute-force scans cease to be a viable option for its exploration. We nowadays, have the option of letting automated systems find their way to configurations that correspond to advantageous extrema of carefully specified objective functions. The engine under the hood of these optimisation searches is automatic differentiation, which allows computer programs to keep track of the gradient of the objective function, through the chain rule of differential calculus, as computer code performs arbitrarily complex successions of operations to model the behaviour of the system. 

Crucial to a successful optimisation of the system is the inclusion in the model of all relevant effects that have an impact on the precision of the inference that the data generated by the system may produce. An incomplete description of the inference itself, or a mock up of the reconstruction techniques performing the dimensionality reduction step which translates raw data into high-level features informing the objective function, are likely to prevent the identification of designs that maximise the true objective, as they introduce a misalignment. Instead, it is possible to introduce a substantial margin of flexibility for the components of the model responsible for explaining factors that influence the absolute magnitude of the objective function but do not affect its gradient direction within the design space.The model, in other words, can be approximate as long as it correctly captures the interdependence of the various components, as the latter is what matters in determining the gradients of the objective function.

Particle detectors are among the most complex interconnected systems humans have ever designed and built: the large experiments at the CERN LHC, for example, include tens of millions of detection elements reading out highly interdependent physical phenomena produced when protons collide in the core of the detectors, at rates of 40 million times a second. For most of present-day experimental situations, we do not yet possess the capability of producing a fully differentiable, high-fidelity end-to-end model of the whole procedure that translates raw detector readouts into final measurements of physical quantities; this prevents the application of gradient descent methods to optimise the detector design. Because of the high complexity of that goal, we approach the problem by turning, for the time being, to easier modelling tasks that nonetheless share the same basic ingredients and structure with it. The successful optimisation of particle detectors of low complexity may allow us to teach ourselves how to tackle progressively harder problems, while accumulating working solutions to sub-tasks which are liable to be reassembled into larger modeling problems as needed.

Particle detectors used in muon tomography applications (see next subsection) appear especially fit to an end-to-end modelling. They involve physical phenomena of low complexity as the data-generating process usually involves the reconstruction of only one particle at a time, and they are liable to a successful parametric modeling of all the relevant effects. They thus constitute an ideal entry point in a long-term study of the end-to-end optimisation of full-fledged experiments. In this document we describe the studies we undertook to develop a software package for the optimisation of muon tomography apparatuses. In its present implementation, the code only considers a subset of the possible three-dimensional geometries of tomography systems, and is meant to provide only a proof of principle of the accessibility of a full, end-to-end optimisation of the problems these instruments are tasked to solve --most notably, imaging of inaccessible volumes, and hypothesis testing on their composition. It is our intention to develop a future version of the software which will provide a more complete solution of the concrete problems in tomography applications.

\subsection{Muon scattering tomography}
    \label{sec:scattering}

Muons, elementary particles related to the electrons but about 200 times heavier, are constantly and naturally produced by cosmic-ray interactions in the atmosphere. 
Their flux at sea level is of the order of \SI{100}{\hertz\per\metre\squared}, and their energy spectrum is very broad, peaking at a few \si{\gev} and extending up to the \si{\tev} scale. 
Muons are not subject to the strong nuclear interaction, and in the energy range \SIrange{1}{100}{\gev} their electromagnetic interactions with matter are relatively mild: they mostly loose energy by ionisation, at a rate of about \SI{200}{\mev} per meter of water. This makes them the most penetrating charged elementary particles.

The first known use of these cosmogenic muons for a practical application was the measurement of the overburden of a tunnel in an Australian mine, reported in 1955~\cite{George1955}. 
A decade later, muons were famously used for searching for hidden chambers in Chephren’s pyramid in Egypt~\cite{Alvarez1970}. While that pioneering attempt did not find any new chambers, it convincingly established their absence from the bulk of the volume of that pyramid. Half a century later, the ScanPyramids collaboration was more lucky and was able to report the actual identification of unexpected features in Khufu's Great Pyramid~\cite{Morishima2017, procureur2023precise}.
These results have been based on the absorption of muons, i.e. on measuring the reduction of the muon flux due to their passage through matter. 
The same method has been applied successfully to several other use cases in archaeology and geophysics (see \cite{MuographyBook} for reports on several recent examples). 
However, since 2003 another process of muon-matter interaction has been used for a variety of applications: multiple scattering, where the muon undergoes several elastic electromagnetic interactions with the nuclei of the traversed material. As the strength of each collision depends on the charge of the nucleus, the deflection of a muon trajectory has a known dependence on the atomic number, Z~\cite{Rutherford1911,LynchDahl1991}

The first proposal to use muon scattering for elemental classification~\cite{Borozdin2003} focused on the search for high-Z materials hidden in containers. 
Searches of illicit materials at border controls are still one of the main drivers for research in muon tomography, see Ref.~\cite{barnes2023cosmic}
 for a recent review. However, in the last two decades this method has been explored for a large variety of other applications, including material identification in nuclear waste drums~\cite{weekes2021material},
detection of nuclear warheads~\cite{morris2014horizontal}, 
and various industrial applications to process monitoring and preventive maintenance~\cite{IAEA2022}. 
While most applications of the method are focusing on the identification of high-Z elements, some teams are attacking the much more challenging task of identifying relatively light materials~\cite{Mrdja_2016,Bikit_2016,TUMUTY_yifan_2018,xuan_tao_ji_2022}, which demands excellent angular resolution given the aforementioned dependence on Z of the scattering angle.

Muon imaging, considering both absorption- and scattering-based methods, is a booming research direction with tens of new publications per year at a steadily growing rate~\cite{Holma_Joutsenvaara_Kuusiniemi_2022}. 
The interested reader can find more information in some recent reviews~\cite{Bonechi:2019ckl, IAEA2022}.

Inferring the atomic number of a passive volume irradiated by a flux of muons from cosmic rays requires modelling the muon flux at the Earth surface, and the multiple scattering muons undergo in matter due to multiple interactions with the Coulomb field of medium nuclei and electrons.

A formula for the flux of cosmic muons at sea level was first given by Gaisser in 1990~\cite{gaisser}, neglecting Earth's curvature and assuming muons do not decay while they traverse the atmosphere; the formula is therefore approximately valid only for muons arriving at sea level with a zenith angle $\theta<70^{\circ}$ and with an energy $E_\mu>100/\cos\theta$ GeV. An improved formula was given by Guan et al.~\cite{guan}, accounting for the curvature of the Earth and providing a modified term that improves the agreement of the formula with experimental data at lower muon energies. 
Other alternative formulas have been proposed in the literature, e.g. the model by Shukla et al.~\cite{shukla} 
provides a closed form for the zenith angle distribution which gives an improved description of the data at all angles, including at $\theta>70^{\circ}$.
\tomopt, the software package we developed and describe in \autoref{sec:tomopt}, enables the user to choose between the model by Guan et al. and that of Shukla et al.

For what concerns the scattering of muons in a material, a simple model is recommended by the Particle Data Group (PDG)~\cite{pdg_model} and used in TomOpt as the default scattering model (which we refer to as the \textit{PDG scattering model}). It assumes a small-angle Gaussian scattering, and does not take into account non-Gaussian tails from rare large-angle scatterings. The model consists of picking two random scattering angles independently in orthogonal directions, as well as correlated transverse displacements. The initial model was proposed to predict a homogeneous material propagation, but then used by several groups in a step-by-step algorithm to model propagation in a volume discretised in voxels (\textit{voxelised}) according to the radiation length \xo of the voxels the muons traverse.
            
The PDG scattering model begins by computing the width of the angular distribution $\theta_0$. For high energy muons with momentum $p$ travelling a distance $d$ the PDG model assumes:
\begin{equation}
   \theta_0 = \frac{\SI{0.0136}{\gev}}{p}\sqrt{n_{\xo}}\times\left(1 + 0.038\ln{n_{\xo}}\right)\,,
\label{eq:pdg_model}
\end{equation}
where $n_{\xo}=d/\xo$, $d$ is the distance traversed by the muon in the material in meters and $X_0$ is the material radiation length in meters. The logarithmic term is added in the model to account for possible low $Z$ materials and to be able to reproduce Moli\`ere's theory~\cite{moliere1948theorie,bethe1953moliere}.  
            The angular scattering is sampled from a zero-centred Gaussian distribution of this width with random numbers $z_1$ and $z_2$ taken from a normal distribution of mean 0 and variance 1:
            \begin{equation}
                \delta\theta = \theta_0\times z_2.
                \label{eq:sample_scattering}
            \end{equation}
            The lateral displacement due to scattering is computed as:
            \begin{equation}
                \delta x = \theta_0\left(z_1\frac{d}{\sqrt{12}} + z_2\frac{d}{2}\right).
                \label{eq:sample_displacement}
            \end{equation}

    \subsection{Overview of this document}
    The present document is organised as follows: in \autoref{sec:tomopt} we describe the software package we developed, which was used to obtain all results we discuss in the following sections; Section~\ref{sec:ladle} describes our benchmark of choice, involving the estimation of the fill level at a metal refinery; Finally, in \autoref{sec:conclusions} we summarise the results and discuss the future perspectives of this work.

\section{TomOpt}
\label{sec:tomopt}
    The simulation and optimisation routines used for this study are structured into a software package referred to as \tomopt. The code for \tomopt is available open-source on Github~\cite{tomopt_git}. In this section we will attempt to provide the reader with a sufficient understanding of the program.
        
    \tomopt is a highly modular Python-based package that provides the full suite of tools and resources required for the investigation of the general problem of optimization of a scattering tomography detector. Its modular design allows each stage of the process to be performed by inheriting classes written to be most suitable for the particular task being studied. The key feature of \tomopt is the fully differentiable nature of the detector modelling and inference pipeline, implemented using the automatic differentiation framework \pytorch~\cite{pytorch}. In this Section we describe the capabilities and workflow of \tomopt.

    The \tomopt project began back in April 2021 with the aim of providing a practical demonstration of the differentiable pipeline proposed by the MODE Collaboration in Ref.~\cite{mode_npni} (and later published in Ref.~\cite{mode_whitepaper}). Initially worked at by two of us, the project quickly gained contributors, helped by a proof-of-concept demonstration at the \nth{1} MODE workshop on differentiable programming~\cite{giles_1st_mode_ws} in September that year. Over 2022, progress focussed on improving the capabilities, stability, and physical accuracy of \tomopt, as well as explorations into advanced inference using maximum likelihood methods, and deep-learning approaches. The primary contributors met twice for extended in-person development and knowledge sharing (both times kindly hosted by CP3, Belgium). Additionally, \tomopt was used as the basis for a data-challenge preceding the \nth{2} MODE workshop on differentiable programming~\cite{2nd_mode_challenge}, where further progress on the project was presented~\cite{giles_2nd_mode_ws}. Progress in 2023 concerned the development of benchmarks for this publication. It was slowed slightly by one of the primary contributors leaving academia for industry and only able to offer evenings and weekends; but anyway, here we are at last!

     \subsection{Cosmic muon generation}
    \label{sec:muon_generation}
    In the context of detector optimisation, it is essential to have a realistic and flexible cosmic muon source that can adapt to the various muon scattering tomography applications and geographical locations. With that in mind, \tomopt samples muons from a horizontal generation surface $S_{\text{gen}}$ with customisable energy and angular distributions. The user is free to provide their own parameterisation of the differential cosmic muon flux, expressed as:

    \begin{equation}
        \Phi_z \left( E, \theta\right)= \frac{dN}{dE \cdot d\theta \cdot d\phi \cdot dS_z}\:\:\text{m}^{-2}\cdot \text{sr}^{-1}\cdot \text{GeV}^{-1},  
    \end{equation}

     where $dN$ is number of muon crossing a horizontal surface element $dS_z$, with energy between $E$ and $E + dE$, zenith angle between $\theta$ and $\theta + d\theta$, and azimuthal angle between $\phi$ and $\phi + d\phi$. Most of cosmic muon flux parametrisation are given as:

     \begin{equation}
        \Phi\left(E, \theta\right) = \frac{dN}{dE \cdot dS_n \cdot d\Omega}\:\:\text{m}^{-2}\cdot \text{sr}^{-1}\cdot \text{GeV}^{-1}, 
    \end{equation}

      where $dN$ is number of muon crossing a surface element $dS_n$ perpendicular to the muon direction, with energy between $E$ and $E + dE$, and solid angle between $\Omega$ and $\Omega + d\Omega$. Because $d\Omega = \text{sin}\theta\:d\theta\:d\phi$ and $dS_z = dS_n\:\text{cos}\theta$, $\Phi_z$ can be obtained from $\Phi$ as:

      \begin{equation}
          \Phi_z = \Phi \cdot \text{sin}\theta \cdot \text{cos}\theta\:.
      \end{equation}

    The present version of \tomopt embeds two default differential muon flux parameterisations: Guan et al.~\cite{guan} and Shukla et al.~\cite{shukla}.
    After providing the desired generation surface span $dx, dy$, energy range $E_{\text{range}}$, zenith angle range $\theta_{\text{range}}$, and number of muons $N_{\text{muon}}$, their position is uniformly distributed on $S_{\text{gen}}$. 
    
    \begin{figure}[ht]
        \begin{center}
            \includegraphics[width=.35\textwidth]{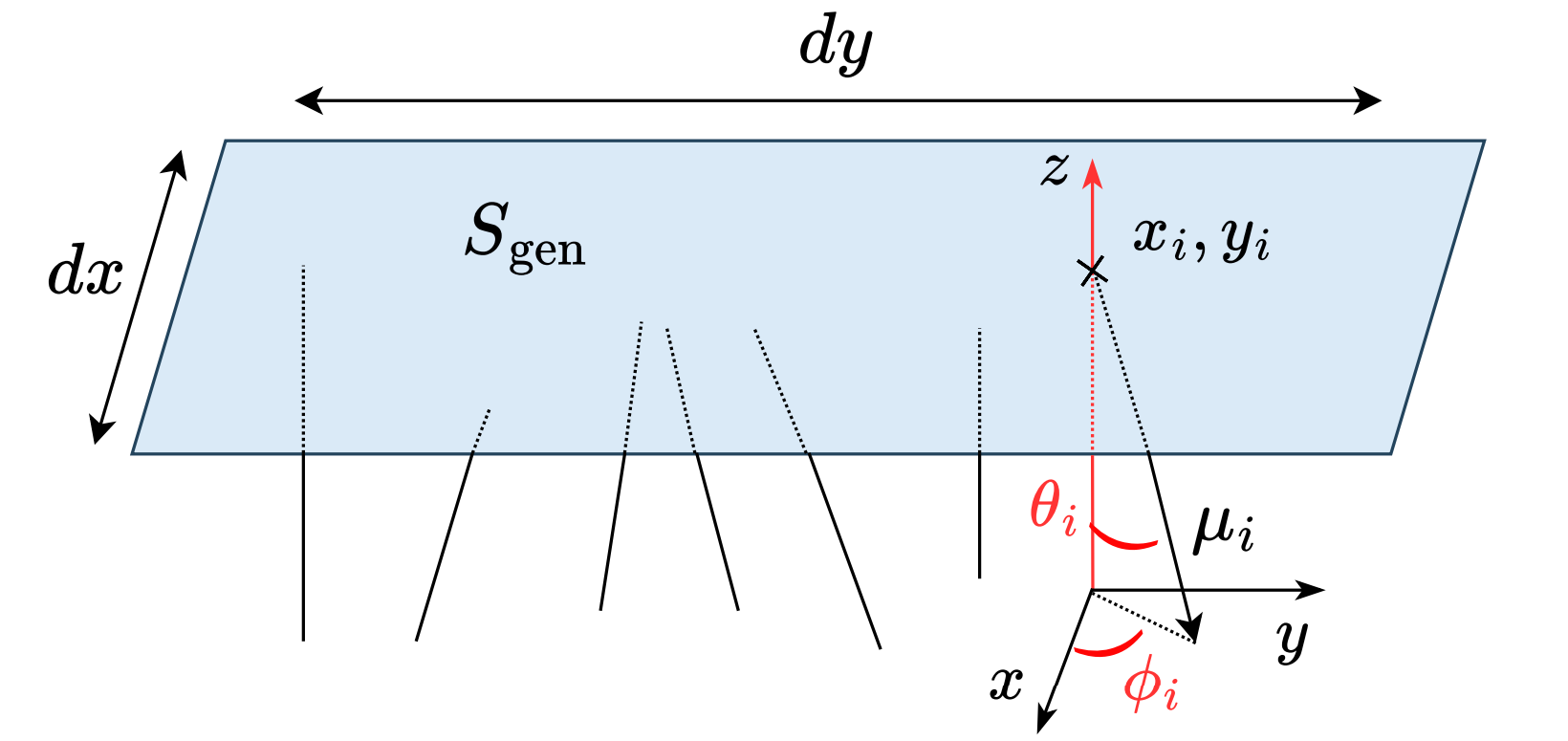}
            \includegraphics[width=.3\textwidth]{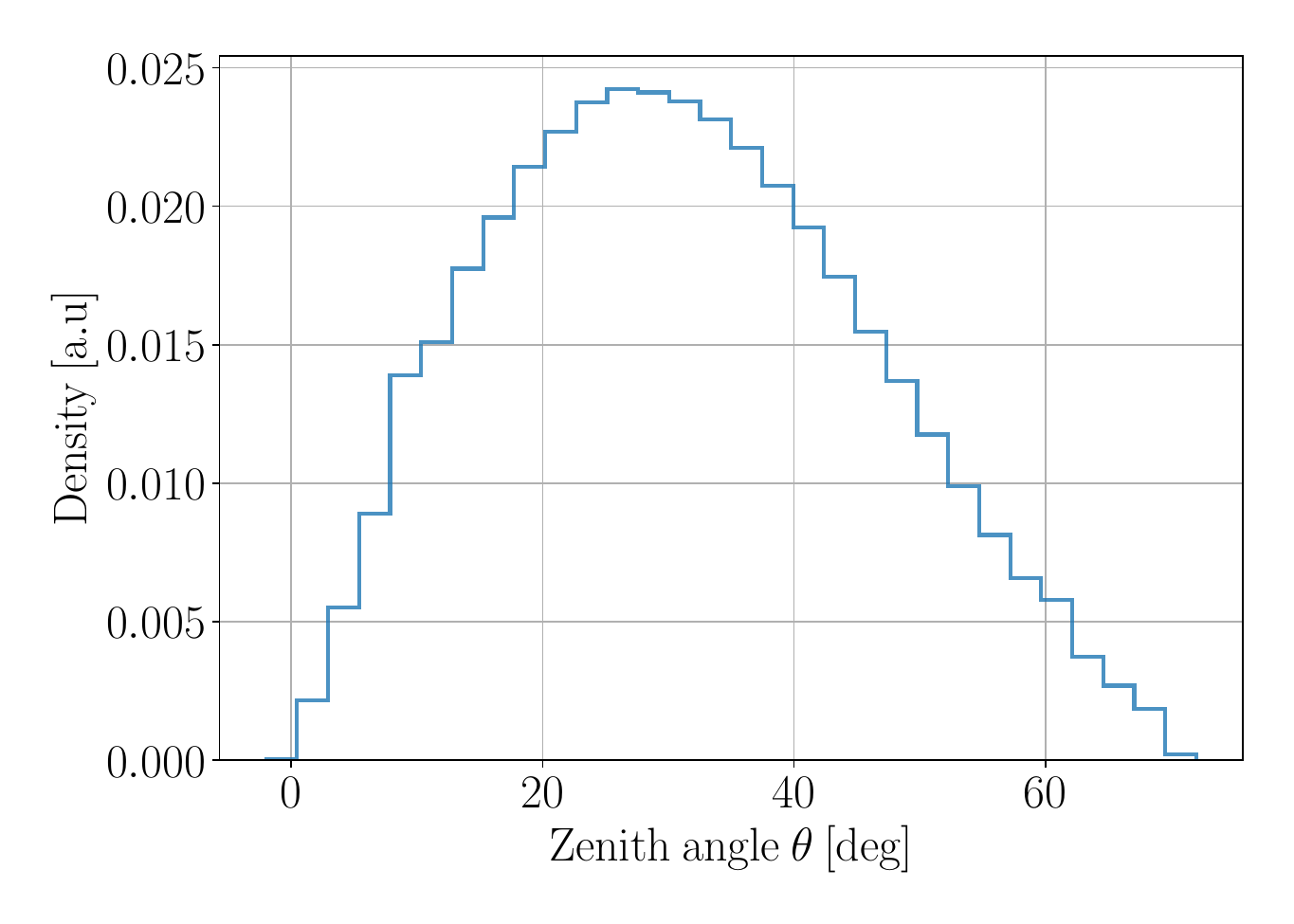}
            \includegraphics[width=.3\textwidth]{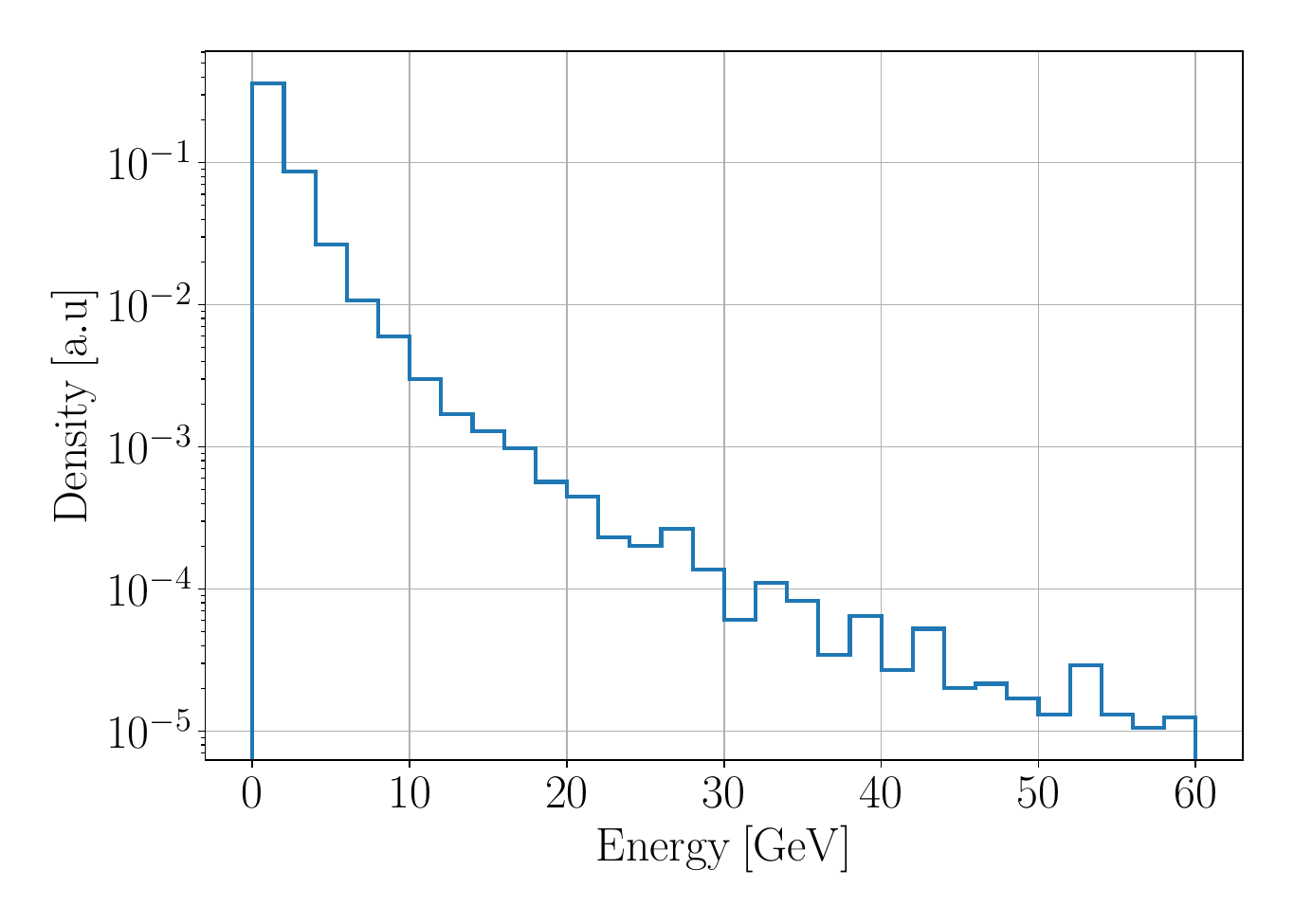}
            
            \caption{Representation of the generation surface in \tomopt (left), muon zenith angle (middle) and energy (right) distribution sampled from the Guan et al.~\cite{guan} parameterisation.}
            \label{fig:generation_surface}
        \end{center}
        \end{figure}

    For computing efficiency purposes, it might be relevant to choose $\theta_{\text{range}}$ upper bound $<70^\circ$ since horizontal muons are unlikely to reach both the passive volume and the detector layers. Once muons are generated, they will be propagated along negative $z$ direction. The simulation of their propagation through passive and active materials is described in \ref{sec:scatter_implementation} and \ref{sec:detector_modelling} respectively.

    \subsection{Passive volumes and multiple scattering}
    \label{sec:scatter_implementation}

        Once muons are generated, \tomopt models their passage the volume of interest (\textit{passive volume}) and their interaction with matter. 
        The passive volumes of interest are modelled in 3D using layers of voxels, and one can assign to each voxel a material (specified by its radiation length \xo). \tomopt has it own database of \xo values for various pure elements computed assuming nominal material density at \SI{20}{\celsius}. Alternatively, users can easily provide their custom materials radiation lengths and densities, which are then plugged into \autoref{eq:pdg_model}. They may also manually specify the layout of materials inside the volumes of interest, or instead provide a customised generator class to provide randomly generated volumes that are representative of the current task.

        \begin{figure}[ht]
            \begin{center}
                \includegraphics[width=.4\textwidth]{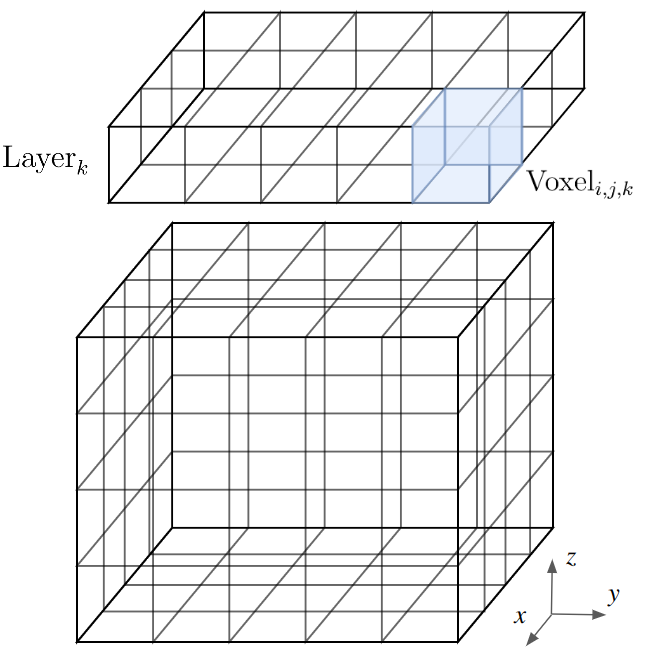}
                \caption{Example of voxelised volume of interest.}
                \label{fig:vox_volume}
            \end{center}
        \end{figure}
        
        For computational efficiency via vectorisation, batches of many muons traverse the volume simultaneously. Each muon is defined using its momentum, $\left(x,y,z\right)$ position, and $\left(\theta,\phi\right)$ trajectory, where $0\leq\theta<\pi/2$\footnote{Muons travelling upwards are removed, since all muons in the batch must share the same z position.} is the deviation from the $z$ axis, and $0\leq\phi<2\pi$ is the azimuthal angle in the $xy$ plane.

        The PDG scattering model described in \autoref{eq:pdg_model} is the default scattering model implemented in \tomopt. We show in \autoref{sec:limitations} that, although embedding such model in a step-by-step algorithm affects the precision of the Coulomb scattering simulation compared to, for instance, highly accurate but slow models such as those provided in \geant~\cite{AGOSTINELLI2003250,1610988,ALLISON2016186}, the optimisation-focussed nature of the procedure we present in this paper requires prioritising computational speed over precision.

        As described in \autoref{sec:scattering}, we sample the angular scattering according to \autoref{eq:sample_scattering} and subsequently compute the displacement due to the scattering according to \autoref{eq:sample_displacement}: this procedure is performed twice per scattering call, to compute scatterings in perpendicular directions ($x_\mu$ and $y_\mu$) in the muon reference frame. The muon-frame scatterings must then be converted into the volume frame using 3D rotations matrices. As mentioned earlier in this section, muons are propagated by steps of distance $dr$ along their flight path. The default value of $dr$ is set to \SI{1}{\centi\metre}, but can be changed to accommodate for the voxel size, the volume of interest size, and the remaining distance to the bottom of the voxel. The optimal choice of $dr$ is discussed in \autoref{sec:limitations}. The algorithm ~\ref{alg:cap} summarises \tomopt muon propagation implementation.\\
        \begin{algorithm}
        \caption{Muon propagation in \tomopt}\label{alg:cap}
        \begin{algorithmic}
        \For{each voxel layer}
        \For{each muon}
        \While{muon inside layer}
            
            \State Compute $\theta_0(dr,X_0)$ \autoref{eq:pdg_model}, $\delta\theta_{\mu}$ \autoref{eq:sample_scattering}, $\left(\delta x_{\mu}, \delta y_{\mu}\right)$ \autoref{eq:sample_displacement}
                \Comment{$\mu$ denotes variables computed in the muon reference frame}
                \State Compute $d\theta_{vol}$, $d\phi_{vol}$, $\delta x_{vol}$, $\delta y_{vol}$, $\delta z_{vol}$\Comment{$vol$ denotes variables computed in the volume reference frame}
            
                \State $x \gets x+\delta x_{vol}$ \Comment Update muon position
                \State $y \gets y+\delta y_{vol}$
                \State $z \gets z+\delta z_{vol}$

                \State $\theta \gets x + \delta \theta_{vol}$ \Comment Update muon direction
                \State $\phi \gets y + \delta \phi_{vol}$
                \If{$z_{\mu} < z_{bottom}$} 
                    \State $z_{\mu} = z_{bottom}$ \Comment Muons that get propagated outside of the layer are moved (up) to the bottom of the layer, $x$, $y$ and $z$ positions are updated according to $\theta$ and $\phi$
                \EndIf
            \EndWhile
            \EndFor
        \EndFor
        \end{algorithmic}
        \end{algorithm}

    \subsection{Detector modelling and hit recording}\label{sec:detector_modelling}
        The detectors modelled in \tomopt must have a flexible enough parameterisation to adapt to the use case, whilst not being constrained by a priori assumptions about the optimal configuration. The basic detector unit used is referred to as a \textit{panel}. In $xy$-space, each panel has a spatial resolution $\sigma_{\text{max}}$ and detection efficiency  $\epsilon_{\text{max}}$, along with a fixed cost per unit area. In this way, the user may simulate detectors that are either constructable in-house, or otherwise commercially available. Panels are free to be positioned in $xyz$ space within pre-specified regions, referred to as \textit{detector layers}. In optimisation, the aim is to learn the optimal position in $xyz$ and width in $xy$ ($xy$-span) per panel; five learnable parameters per panel. Currently, the number of panels per detector layer is fixed throughout the optimisation process, due the difficulties in computing gradients for integer variables.
            
        Muon positions (referred to as \textit{hits}) are recorded in $xyz$ by the panels. The effect of spatial resolution on the hits will depend on the detector technology used. For simplicity, the panels implemented in \tomopt assume that the recorded position of a hit can be sampled from a Gaussian distribution $\mathcal{G}\left(\mu=xy_{\text{true}},\sigma=\sigma_{xy, \text{eff}}\right)$ centered over the true position of the muon $xy_{\text{true}}$, the scale of which depends on the effective spatial resolution $\sigma_{xy, \text{eff}}$, as dictated by $\sigma_{\text{max}}$ of the detection panel, and the position of the muon relative to the panel centre.
            
        In reality, a muon passing through a panel would either record a hit, or not, depending on the efficiency of the panel. For differentiability reasons, \tomopt instead takes a probabilistic interpretation: hits are always recorded, but with an efficiency factor $\epsilon_{\text{hit}}$. The total efficiency of the muon $\epsilon_\mu$ is then computed given its individual hits as the probability of at least two hits before and after the passive volume\footnote{At least two hits are required to reconstruct a trajectory, but more hits will better constrain the uncertainty on the trajectory parameters}. Both the hit and overall efficiency enter various parts of the inference and optimisation process, and will be discussed in later sections.
            
        \paragraph{Optimisation mode}\label{sec:tomopt:opt_mode}
            Optimisation of the panel parameters requires that the muon trajectories be differentiable \wrt them. Trajectory reconstruction (discussed next) will naturally be differentiable \wrt the $z$ position of the hits, however derivatives for the $xy$-position and $xy$-span are non-trivial: these affect only whether a muon hits a panel or not, and hits are then recorded with a constant resolution and efficiency, i.e. the optimisation process would be insensitive to muons which could have been recorded but were not due to the position and size of the panel.
            
\begin{figure}[ht]
    			\begin{center}
    				\includegraphics[width=.85\textwidth]{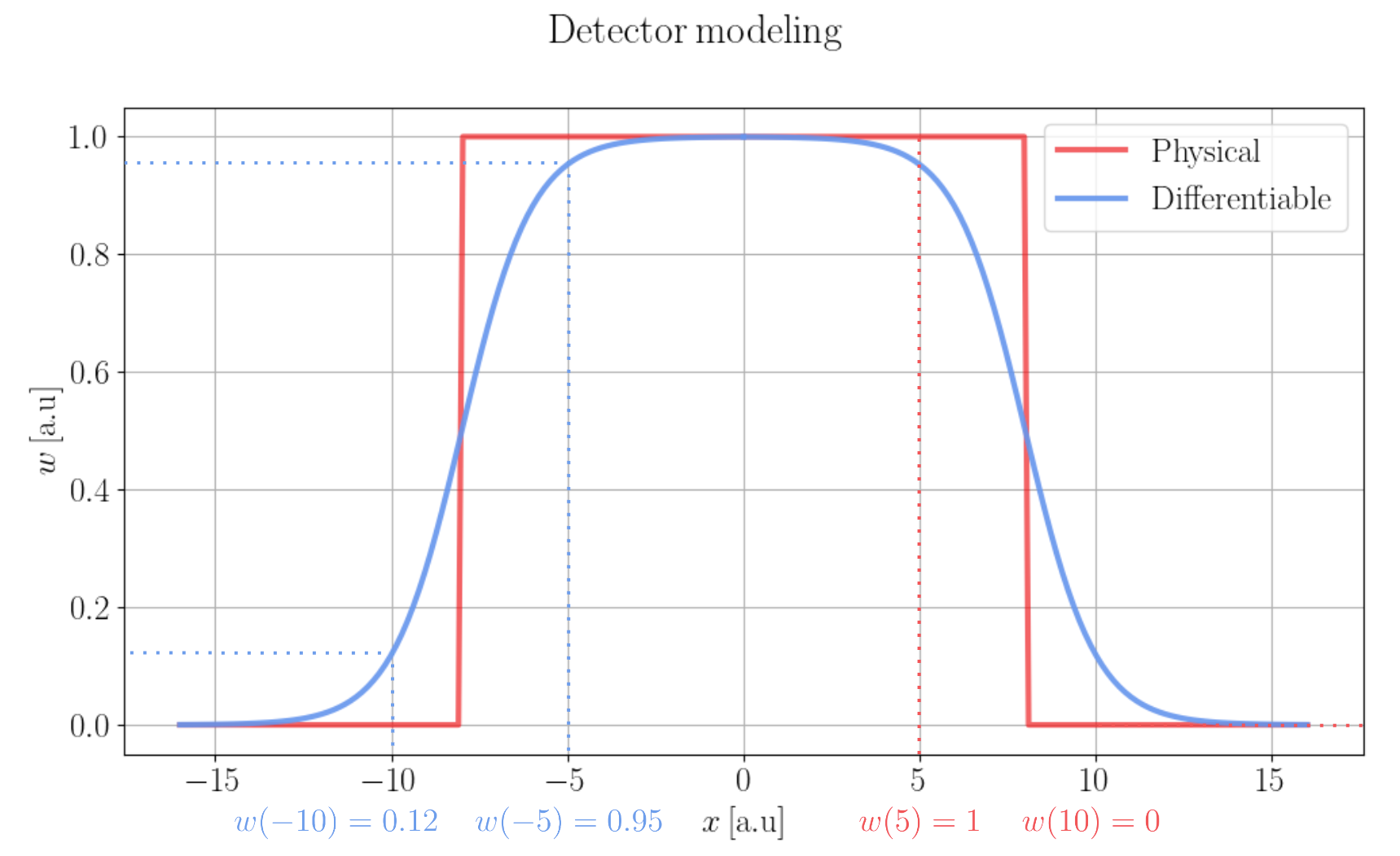}
    				\caption{Example of detector panel modeling with sigmoid function used during optimisation (blue) and with rectangle function used during validation (red).}
    				\label{fig:det_sigmoid}
    			\end{center}
    	    \end{figure}

            Instead, during optimisation mode, \tomopt takes an unphysical approach in which ``panels" will always record a hit for every muon at its current $z$ position, even if these hits are located outside the panel. The effective resolutions $\sigma_{xy, \text{eff}}$ and efficiencies $\epsilon_{\text{eff}}$ of hits, however decreases the further away the hits are from the centre of the panel, at a rate that scales with the span of the panel. They are computed by scaling the maximum panel resolution $\sigma_{\text{max}}$ and efficiency $\epsilon_{\text{max}}$, according to where the muon passes the panel's $z$-position:\:
            \begin{equation}
                \sigma_{xy, \text{eff}} = \sigma_{\text{max}} / w(xy_{\text{true}}), \:\: \epsilon_{\text{eff}} = \epsilon_{\text{max}}\cdot w(xy_{\text{true}}),  
            \end{equation}
            where $w(xy_{\text{true}})$ is the panel model evaluated at the true muon-position, as presented in \autoref{fig:det_sigmoid}. This scales between zero and one, reaching a maximum when the muon hits at the centre of the panel. The current model used in \tomopt is a pair of double-sigmoid functions in $x$ and $y$. The values of this pair are used to provide resolutions in $xy$, and their product provides the hit efficiency. These are then scaled such that a hit at the centre receives the full resolution and efficiency of the panel. Effectively, this can provide detection performance close to the actual detector for hits within the panel, and a smooth transition to zero efficiency and resolution outside the panel. In this way, each hit and muon trajectory will be differentiable \wrt all parameters of the panel. When derivatives are not required, e.g. for validating the detector, the panels can of course be reverted to a more physical interpretation. The algorithm \ref{alg:hit_rec} presented below summarises \tomopt hit recording implementation.

            The smoothness of the sigmoid model (how slowly the hit performance transitions at the panel edge) can be adjusted, and even evolved during optimisation. Inference with smoother panels will be more sensitive to muons that are far from the panel, and so are more suitable for optimising the $xy$ position and span of the panels: since the inference will account for the efficiency on the muons, muons far from the panel are able to affect the inference. Panels with a sharper transition have close to full hit performance within the majority of the area of the panel, and so inference is more sensitive to the $z$ position of the panels: muons far from the panel receive diminished weights, and muons inside the panel receive a more uniform efficiency and resolution therefore the only remaining parameter to have non-negligible gradient is the $z$-position. Again, the user is free to implemented their own panel model $w$ to reach the desired detector configuration. Examples could be different scalings in $x$ and $y$, or a periodic function to represent detection wires with a given density.

        \begin{algorithm}
        \caption{Hit recording in \tomopt}\label{alg:hit_rec}.
        \begin{algorithmic}
        \For{each muon}
            \For{each detector panel}
            \State Propagate muon to the next detector panel at $z_j$
            \State Record true hit: $\left(x_{\text{true}},y_{\text{true}},z_j\right)$
            \State Compute effective hit resolutions: $\sigma_{x, \text{eff}} = \sigma_{\text{max}} / w(x_{\text{true}}),\:\sigma_{y, \text{eff}} = \sigma_{\text{max}} / w(y_{\text{true}})$

            \State Compute reconstructed hit: $\left(x_{\text{true}}+ \delta_x,y_{\text{true}}+\delta_y,z_j\right)$ \Comment $\delta_{x,y}$ sampled from $\mathcal{G}\left(0, \sigma_{x,y,\text{eff}}\right)$
            \State Compute hit efficiency $\epsilon_{\text{eff}} = \epsilon_{\text{max}} \cdot w\!\left(x_{\text{true}}\right) \cdot w\!\left(y_{\text{true}}\right)$
            \EndFor
        \EndFor
        \end{algorithmic}
        \end{algorithm}        

    \subsection{Trajectory fitting and Point-of-Closest-Approach finding}
        The aim of trajectory fitting is to determine the paths of each muon as they enter and exit the passive volume using the hits recorded by the two sets of detector panels. Each trajectory is computed via analytic minimisation of a likelihood function that takes the $xyz$ hits as an input. When computing the fit, a weight is associated to each of the hit positions, that depends on the effective $xy$ resolution of the hits $\sigma_{xy, \text{eff}}$, and their efficiencies $\epsilon_{\text{eff}}$, i.e. low-resolution, or low efficiency, hits have less of an impact on the trajectory fit than higher resolution, or more probable hits. No uncertainty in $z$ is considered. From an optimisation perspective, the true trajectory of the muon is most likely to be recovered when the detector panels are positioned such that all hits for the muon are recorded with high efficiency and resolution. Potentially this would come at the cost of a poor reconstruction of other muon tracks, and so a global compromise must be found, as determined by the loss function of the muon tomography task being performed.
            
        Once the incoming and outgoing trajectories of the muons are computed, scattering variables such as the incoming and outgoing $\theta$ and $\phi$ angles, and the total amount of angular scattering may be differentiably computed. By exploiting the differentiable computation of these variables, their associated uncertainties can also be easily computed using the product of their partial derivatives \wrt the recorded hits and the $xy$ uncertainties on the hits.
        
\begin{figure}[ht]
            \begin{center}
                \includegraphics[width=.55\textwidth]{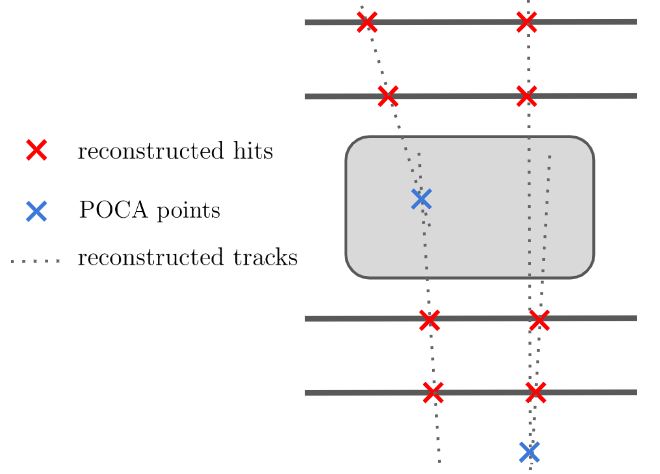}
                \caption{Example of \poca point finding. When the scattering angle is large, the \poca point is well defined and is inside the passive volume (left). For almost parallel incoming and outgoing tracks, \poca point might be reconstructed outside of the volume of interest.}
                \label{fig:poca_finding}
            \end{center}
        \end{figure}

        Once muon tracking is performed, \tomopt proceeds to a first inference phase in order to associate a region of space with the measured muon deflection. It is done using an extended version of the Point-of-Closest-Approach algorithm, which is a standard in muon scattering tomography. In this approach, the entirety of the muon scattering is assumed to have occurred at a single point in the passive volume, located by extrapolating the inferred trajectories, as shown in \autoref{fig:poca_finding}. Because the \poca points are computed from the incoming and outgoing reconstructed tracks, they also bear an uncertainty in $x,y$ and $z$. The \poca points may then be used for a variety of purposes, such as: directly producing a 3D image for human interpretation; or computing hit densities in voxels, perhaps scaling their weights by their associated angular scattering. Exactly which variables are required, and the way in which they are used depends on the task-specific inference approach used. We will show an example of this in \autoref{sec:ladel_inf}.

        \begin{algorithm}
        \caption{Track fitting and \poca finding in \tomopt}\label{alg:track_poca}.
        \begin{algorithmic}
        \For{each muon}
            \For{incoming and outgoing tracks}
                \State Fit track vector
            \EndFor
            \State Compute \poca point from the track vectors
            \State Compute \poca point uncertainty via automatic differentiation
        \EndFor
        \end{algorithmic}
        \end{algorithm} 
        
    \subsection{Volume inference}\label{sec:tomopt_inference}
        The purpose of ``volume inference" is to convert the muon-trajectory variables into a (set of) prediction(s) for the volume. The form of these predictions and the inference method used will depend on the muon tomography task being performed. A variety of methods may be used here, provided they allow the predictions to be differentiable \wrt the muon-trajectory variables, and a range of these will be explored in a second publication.
            
        Our implementation of \poca in \tomopt aims to compute estimates of the voxel X0 values by inverting the PDG scattering model. The basic approach for a given batch of muons with their \poca located in a given voxel of height $\delta z$ is:
        
        \begin{align}
            \theta_0^{\mathrm{rms}} = \frac{\theta^{\mathrm{rms}}_{\mathrm{tot.}}}{\sqrt{2}},\label{eq:scat_rms}\\
            \theta_0^{\mathrm{rms}} = \frac{\SI{0.0136}{\gev}}{p^{\mathrm{rms}}}\sqrt{n_{\xo}}^{\mathrm{rms}},\label{eq:scatter_inversion}\\
            n_{\xo}^{\mathrm{rms}} = \frac{\delta z}{\xo\cos\left(\bar{\theta}^{\mathrm{rms}}\right)},\notag\\
            \bar{\theta}^{\mathrm{rms}} = \frac{\theta_{\mathrm{in}}^{\mathrm{rms}}+\theta_{\mathrm{out}}^{\mathrm{rms}}}{2},\notag\\
            \xo = \left(\frac{\SI{0.0136}{\gev}}{p^{\mathrm{rms}}}\right)^2\frac{\delta z}{\cos\left(\bar{\theta}^{\mathrm{rms}}\right)}\frac{2}{{\theta^{\mathrm{rms}}_{\mathrm{tot.}}}^2}.
            \label{eq:scatter_inversion_final}
        \end{align}

        Where $\theta^{\mathrm{rms}}_{\mathrm{tot.}}$ is the root-mean-square (RMS) of the measured scattering angles. In \autoref{eq:scatter_inversion}, $\theta_{\mathrm{in}}^{\mathrm{rms}}$ and $\theta_{\mathrm{in}}^{\mathrm{rms}}$ respectively refer to the incoming and outgoing muon zenith angle. Note that the natural log term present in \autoref{eq:pdg_model} has been ignored to simplify the inversion. The effects of efficiency and resolution may be further included in \autoref{eq:scatter_inversion_final} by instead aggregating over the muons with a weighted RMS, where muon-wise weights are computed as the efficiency divided by the variance of the squared value of the variable being aggregated.
            
\begin{figure}[ht]
            \begin{center}
                \includegraphics[width=.6\textwidth]{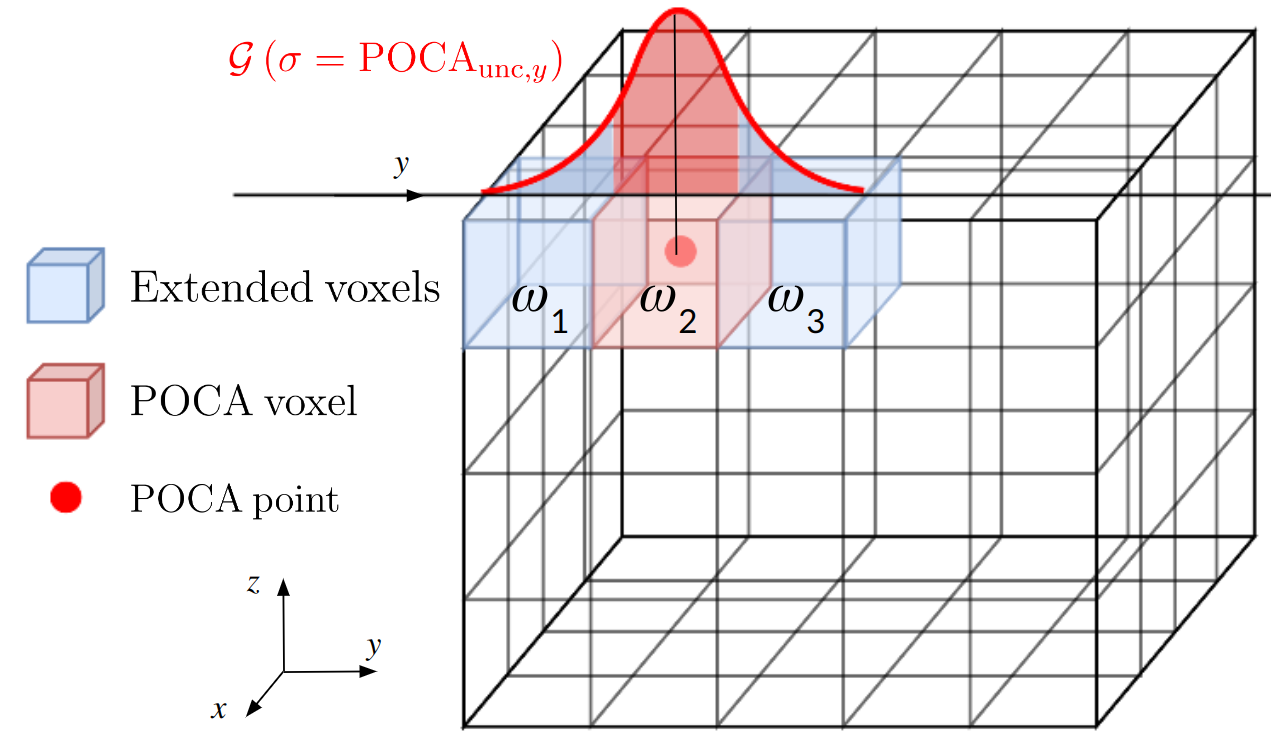}
                \caption{Illustration of \poca point being extended to other neighbouring voxels with a Gaussian model. For the sake of readability, only the extension along the $y$ dimension is shown.}
                \label{fig:poca_vox}
            \end{center}
        \end{figure}

        In the above inference, we only considered muons whose \poca was located inside the voxel in question, however we have an uncertainty on the \poca position, and some muons may have a \poca located outside the passive volume due to mis-reconstruction. An extension is then to consider that muons can contribute to inference in multiple voxels by considering that the \poca could have occurred in voxels surrounding its nominal location. The \poca uncertainty $\text{PoCA}_{\text{unc}}$ in $x$, $y$ and $z$ is computed via automatic differentiation. Then the probability density function (PDF) for a \poca to be located within a voxel is modelled using 3 independent Gaussian distributions centered over the nominal \poca location and scaled in the three directions according to $\text{PoCA}_{unc}$. A ``scattering probability" is then computed per voxel for each muon, by integrating the \poca PDFs over ever voxel in the passive volume.  An illustration of this procedure is presented on \autoref{fig:poca_vox}. This probability then enters into the weighted RMS as a multiplicative coefficient.

        \begin{algorithm}
        \caption{Volume inference in \tomopt}\label{alg:vol_inference}.
        \begin{algorithmic}
        \For{each \poca point}
        \State Compute scattering probability per voxel
        \State Compute muon weights per voxel as muon efficiency times scattering probability in each voxel
        \EndFor
        \For{each voxel}
            \State Compute weighted $\theta_{0}^{\text{rms}}$ from \autoref{eq:scat_rms}
            \State Compute \xo from \autoref{eq:scatter_inversion_final} 
        \EndFor
        \end{algorithmic}
        \end{algorithm} 
            
        This method of using \poca points, however, is biased to underestimate \xo, since the entirety of the scattering is assumed to happen in one scattering event. The muons of course undergo many scattering events as they traverse through the volume. Additionally, it  converts muon-wise data into voxel-wise predictions, however the muon tomography task may not require voxel-wise predictions, but instead perhaps predictions concerning the volume as a whole. Nonetheless, it can serve as a useful and generic first-stage of inference to serve as input to a second, more task-specific, stage of inference; indeed both of the benchmarks presented in this \whatAmI take this approach.
        \tomopt implements more advanced methods such as an Expectation-Maximisation algorithm, where the strong (and often unrealistic) assumption of a single scattering centre is dropped and all most likely paths of the muon are taken into account with a proper weight~\cite{4271541}.
        
        Alternatively, deep-learning based methods can be applied to muon-level data. Here, latent-space representations of the relevant muons in each voxel can be learnt using graph-neural-networks that are able to learn their own graphs in a latent-coordinate-space, such as those proposed in Ref.~\cite{Qasim:2019otl}. These representations can then be refined per voxel, based on the surround voxels through a learnable higher-level graph. An initial investigation of this is presented in Ref.~\cite{giles_5th_iml}: the representation in each voxel is then refined based on the surrounding voxels. More simply, deep-learning methods can be used as a second stage to \xo inference, to conveniently convert the dimensionality of voxel-level data to volume-wise predictions. Convolutional neural networks are appropriate for this task, as demonstrated during the Second MODE Workshop data-challenge~\cite{2nd_mode_challenge}.

    \subsection{Loss functions and optimisation}\label{sec:tomopt:optimisation}
        Similar to the volume inference, the most suitable loss function for the optimisation will depend on the muon tomography task being performed, e.g. cross-entropy for classification tasks. The total loss function, however, should not only consider the performance of the detector, but also its budget, for example the cost in currency units that the detector would take to build given the surface area of its detectors. The cost could enter directly into the loss via a scaling coefficient, however it is more likely that the user already has in mind a particular target budget, and wants to design the best detector that gets close to this budget.

        \tomopt implements two approaches: fixed-budget, where the user specifies the maximum budget allowed and the optimisation is constrained to not surpassing that value; and budget-penalisation, where the user specifies a loose target budget.
        
\begin{figure}[ht]
            \begin{center}
                \includegraphics[width=\sfSmall\textwidth]{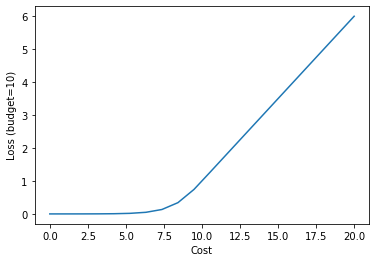}
                \caption{Example of the cost component of the loss as a function of detector cost, for a target budget of 10 A.U.. A sigmoid component below the budget provides a slowly increasing gradient as the cost approaches the target budget, beyond which the loss increases linearly. The function is fully differentiable.}
                \label{fig:budget_switch}
            \end{center}
        \end{figure}

        In the budget-penalisation case, a cost component of the loss added according to a function that is low up to the target budget, and then begins to rapidly increase if the detector exceeds the target budget. In this approach, the detector design can go over budget, but only if the increase in performance vastly surpasses the budget excess, as exemplified in \autoref{fig:budget_switch}. This mode introduces hyper-parameters for the weight of the new component in the total loss, and the functional form of the penalisation given the cost and budget.
            
        Alternatively, in ``fixed-budget optimisation mode", the user still specifies a total and fixed cost for the detector, and a fractional assignment of this cost to each panel is learnt, starting from a uniform assignment. The panels still learn a separate $xy$ span, however the the area of the panel is then rescaled such that the cost of the panel is equal to its assigned fraction of the cost of the detector. This mode of optimisation has two main advantages: the user is able to ensure that they meet their budget exactly; there is no balancing coefficients required in the loss function between performance and cost, the loss is purely based on performance.
            
        Once the value of the loss function has been computed, the analytic effects of each detector parameter can be computed through back-propagation~\cite{Linnainmaa_70, Werbos_81, backprop} of the the derivative of the loss through the inference and hit-recording steps. The parameters may then be updated via stepping in the direction of steepest descent of the loss surface. For this process, \tomopt leverages the standard optimisers built into \pytorch (SGD~\cite{gradient_descent}, Adam~\cite{adam}, etc.).
            
        Training is supported by a callback system, which allows aspects of the modules and the training procedure to be adjusted without requiring new modules to be written.
            
    \subsection{Update cycle}
        When computing a value for the loss function, it is important to ensure that it is computed under conditions that are generally representative of those that can be encountered during actual operation, in terms of e.g. number of muons and passive volumes. In effect, this means that optimisation should be performed using a range of passive volumes (whether generated on demand, or supplied by the user as a dataset), and that these should each be scanned using a number of muons that would be reasonable to expect given the position and exposure time of the proposed detector. Additionally, rather than computing the loss using just a single passive volume, it should instead be averaged over many passive volumes in order to provide a more generalising update to the detector parameters.
            
\begin{figure}[ht]
            \begin{center}
                \begin{subfigure}[t]{\sfSmall\textwidth}
                    \begin{center}
                        \includegraphics[width=\textwidth]{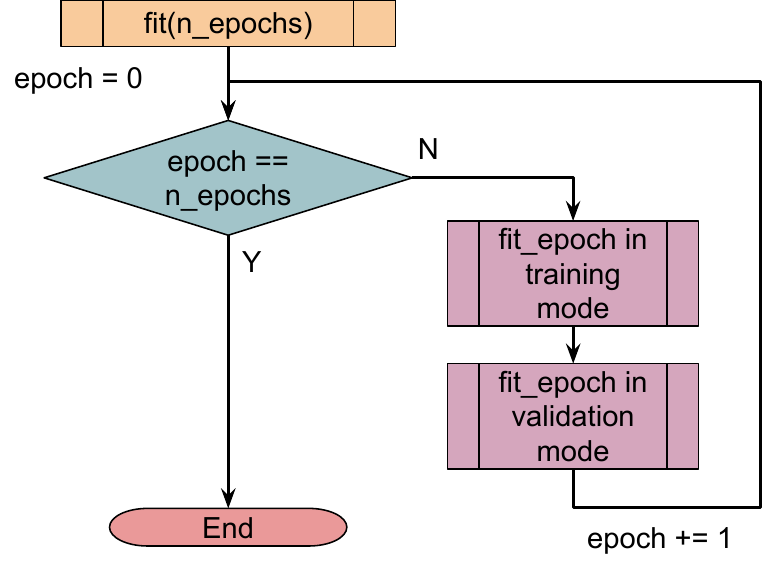}
                        \caption{Complete fit loop.}
                        \label{fig:tomopt_fit_loop:fit}
                    \end{center}
                \end{subfigure}
                \begin{subfigure}[t]{\sfSmall\textwidth}
                    \begin{center}
                        \includegraphics[width=\textwidth]{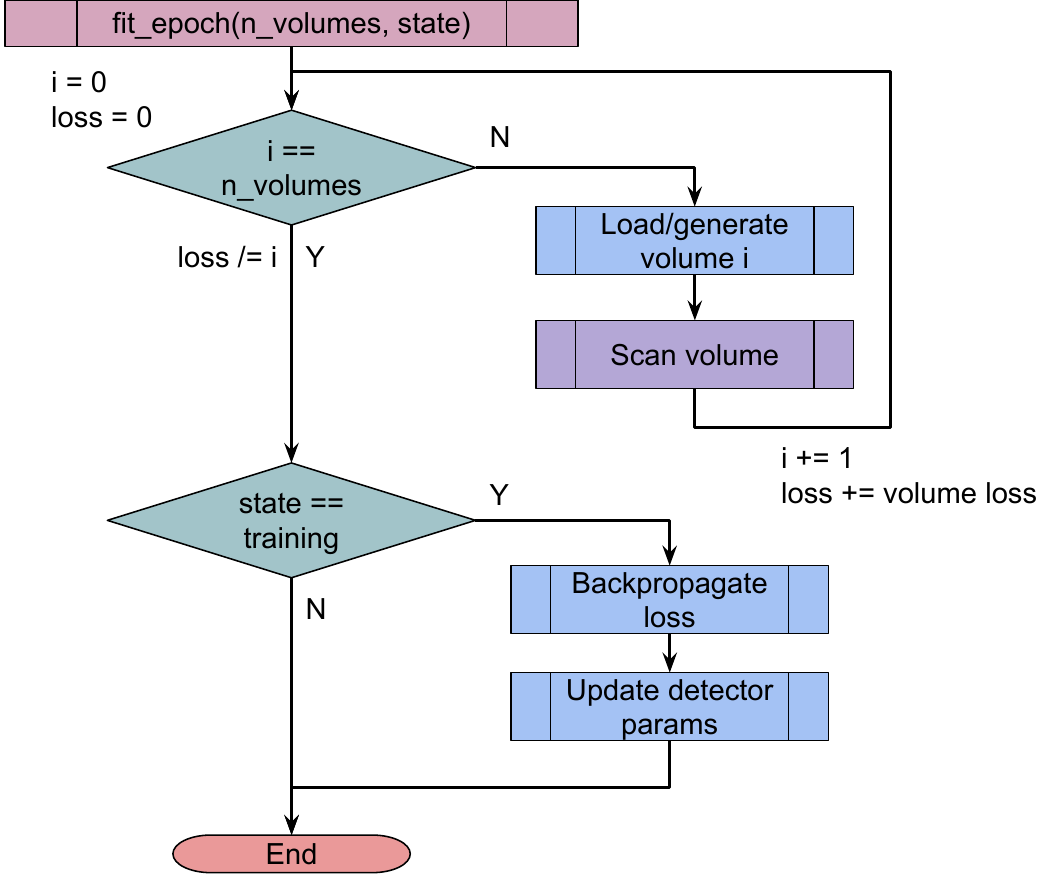}
                        \caption{Scan loop for a batch of passive volumes.}
                    \label{fig:tomopt_fit_loop:volumes}
                    \end{center}
                \end{subfigure}
                \begin{subfigure}[t]{\sfSmall\textwidth}
                    \begin{center}
                        \includegraphics[width=\textwidth]{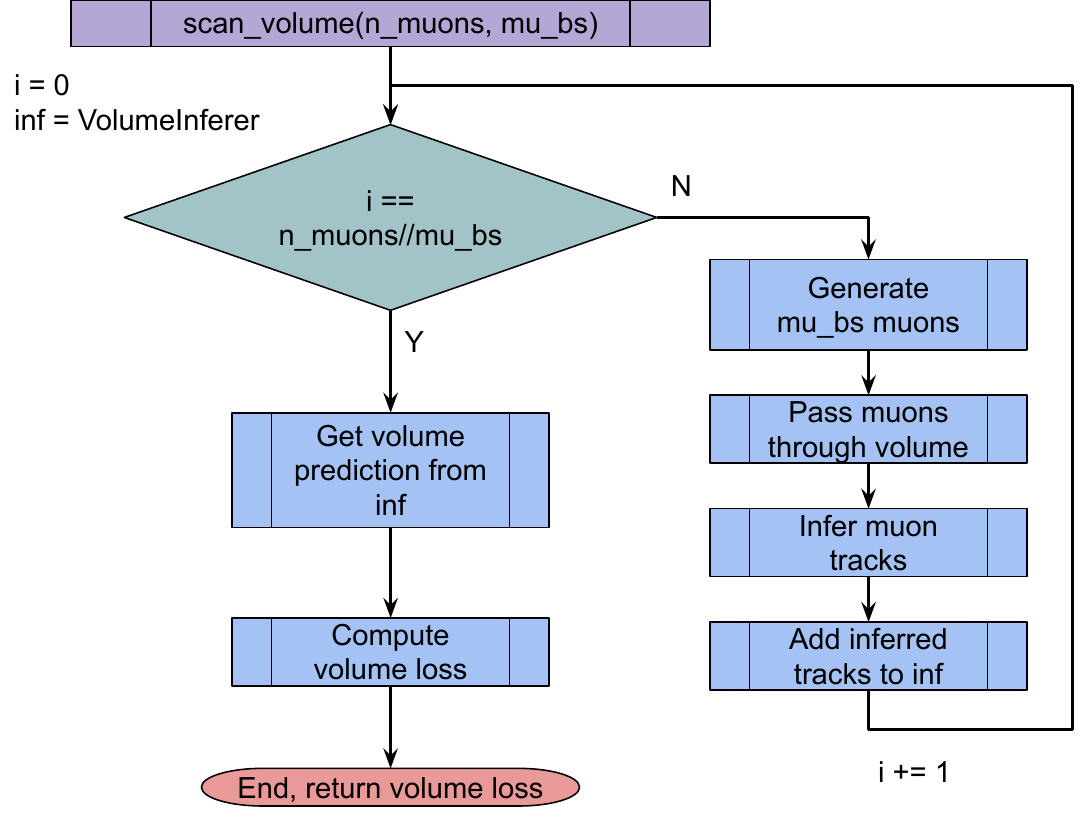}
                        \caption{Scan loop for muons over a single passive volume.}
                        \label{fig:tomopt_fit_loop:volume}
                    \end{center}
                \end{subfigure}
                \caption{Breakdown of the fitting procedure of detectors in \tomopt}
                \label{fig:tomotp_tomopt_fit_loopfit_loop}
            \end{center}
        \end{figure}

        The update cycle in \tomopt is performed thus in the following way:
        \begin{enumerate}
            \item \autoref{fig:tomopt_fit_loop:volume}: Scanning of a given passive volume involves passing a batch of muons through the volume and inferring their tracks. For computational reasons, the total number of muons may be split into smaller batches. Once all muons have been processed, the volume predictions can be made and the loss computed;
            \item \autoref{fig:tomopt_fit_loop:volumes}: This process is repeated for a batch of volumes, and the sum (or average) of their losses can be used to update the detector parameters. This is referred to as an \textit{epoch}, in keeping with Machine Learning terminology\footnote{This assumes that volumes are being generated. When working with a dataset of volumes, instead, this process can be repeated several times using a volume batch size that is less than the total number of volumes in the dataset.};
            \item \autoref{fig:tomopt_fit_loop:fit}: The fit runs for a specified number of epochs before finishing. It is also possible to run epochs in validation mode, where the detector is not updated: the detector panels can be therefore be considered, more realistically, as discrete detector panels rather than with an associated distributed resolution and efficiency, as discussed in \autoref{sec:detector_modelling}.
        \end{enumerate}

        The appendix in \autoref{sec:appendix} shows an overview of each of the steps presented above. 
            
    \subsection{Current limitations}\label{sec:limitations}

        \subsubsection{Muon propagation through matter}
\begin{figure}[ht]
            \begin{center}
                \includegraphics[width=.9\textwidth]{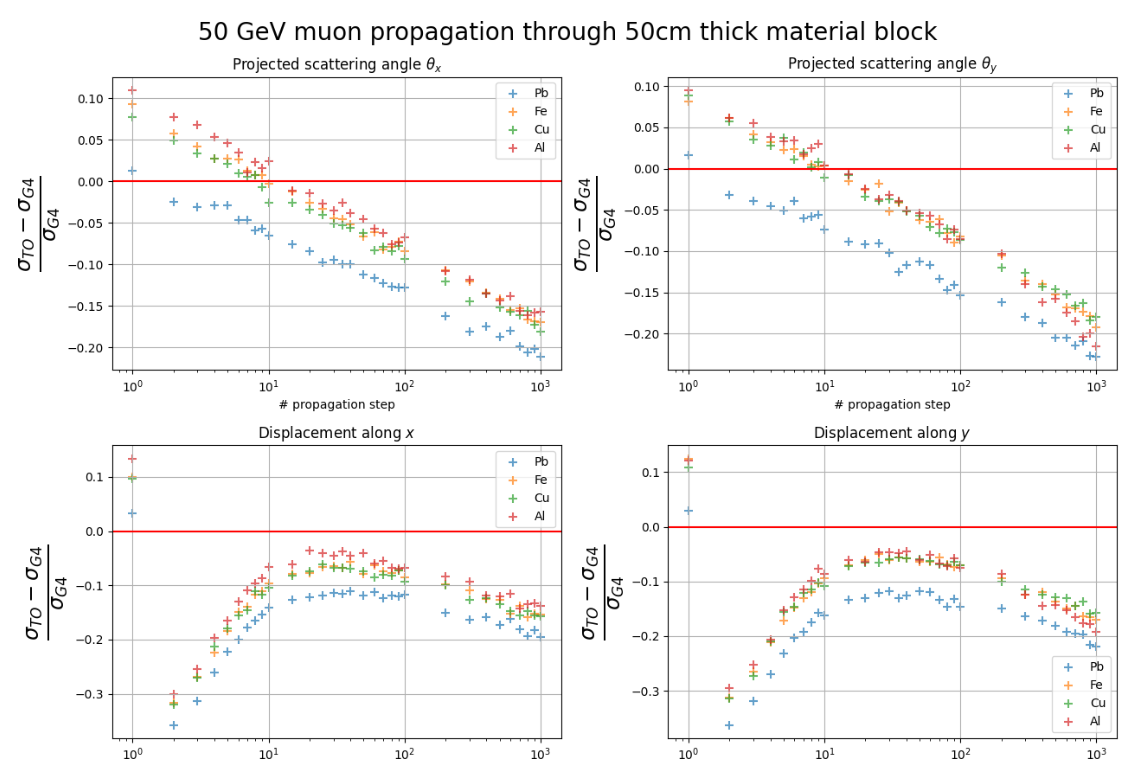}
                \caption{Comparison of standard deviations of projected scattering angles  and displacement distributions between \tomopt and \geant and for various number of propagation steps.}
                \label{fig:msc_TO_vs_G4}
            \end{center}
        \end{figure}

        As described in \autoref{sec:scatter_implementation}, the implementation of multiple Coulomb scattering is rather straight-forward since the computation efficiency is prioritised over precision. Scattering angle and displacement distributions are assumed to be a \SI{100}{$\%$} Gaussian, where in reality only \SI{98}{$\%$} of the bulk is Gaussian and the tails follow a $1/\theta^4$ law corresponding to large scattering events. In addition, The PDG model used is not designed for a step-by-step algorithm: when it is applied to such an algorithm, it results in underestimating the scattering and displacement amplitudes more and more with each step, as shown in \autoref{fig:msc_TO_vs_G4}.

        It becomes clear that the choice of step length $dr$ affects the precision of the multiple coulomb scattering, and thus has to be chosen carefully. In this context, it is advised to choose a step length $dr$ which represent about $1\%$ of the passive volume size.\\
        In addition to multiple Coulomb scattering, muons lose energy through ionisation thus sub-$\text{GeV}$ muons might decay in the passive volume. Energy loss has not been implemented in \tomopt yet, which results in a overestimation of the detected muon flux. Both of these limitations could be overcome by building a parametric model of \geant. Creating a data-base of muon energy loss, scattering and displacement for various material and at several energy scales from \geant simulations, \tomopt could simulate muon transport more accurately and effortlessly.
        
        \subsubsection{Detector and cost modeling}

        \tomopt's ambition is to guide the user through the choice of detector technology and not to provide accurately simulation of particle detection, signal generation or light transport. Thus, the difference between various detector technologies will be implemented at the level of cost modelling, meaning how the cost scales with the active detection surface and number of readout channel. Such features have not been implemented yet, but are planned to be included in future updates.\\
        Regarding detector modelling, \tomopt is still limited to horizontal panels with continuous spatial resolution and uniform granularity.This limits the phase space of possible detector geometry.\\
        
        \subsubsection{Volume inference}

        For now, \tomopt only provides a basic implementation of radiation length inference based on the inversion of the PDG scattering model utilising the Point of Closest Approach. The inversion of the scattering model \autoref{eq:scatter_inversion_final} assumes muon momentum to be known with a $100\%$ precision, which is quite unrealistic in the context of a muon scattering tomography experiment. If a momentum estimation is available, the user can still modify the inversion model and replace the nominal muon momentum by a the estimated momentum. If not, it can be replaced by the RMS computed over the whole cosmic muon momentum spectrum.
\section{Benchmark: Fill-level estimation at metal refinery}

The following section describes a benchmark involving the estimation of the fill level at a metal refinery, and aims at demonstrating a typical inference and optimisation chain using \tomopt. After evaluating the performance of an initial sub-optimal detector configuration, the latter will be optimised, and its performance will be compared to two human baseline detector configurations.
\label{sec:ladle}
    \subsection{Description}

\begin{figure}[ht]
            \begin{center}
                \includegraphics[width=.6\textwidth]{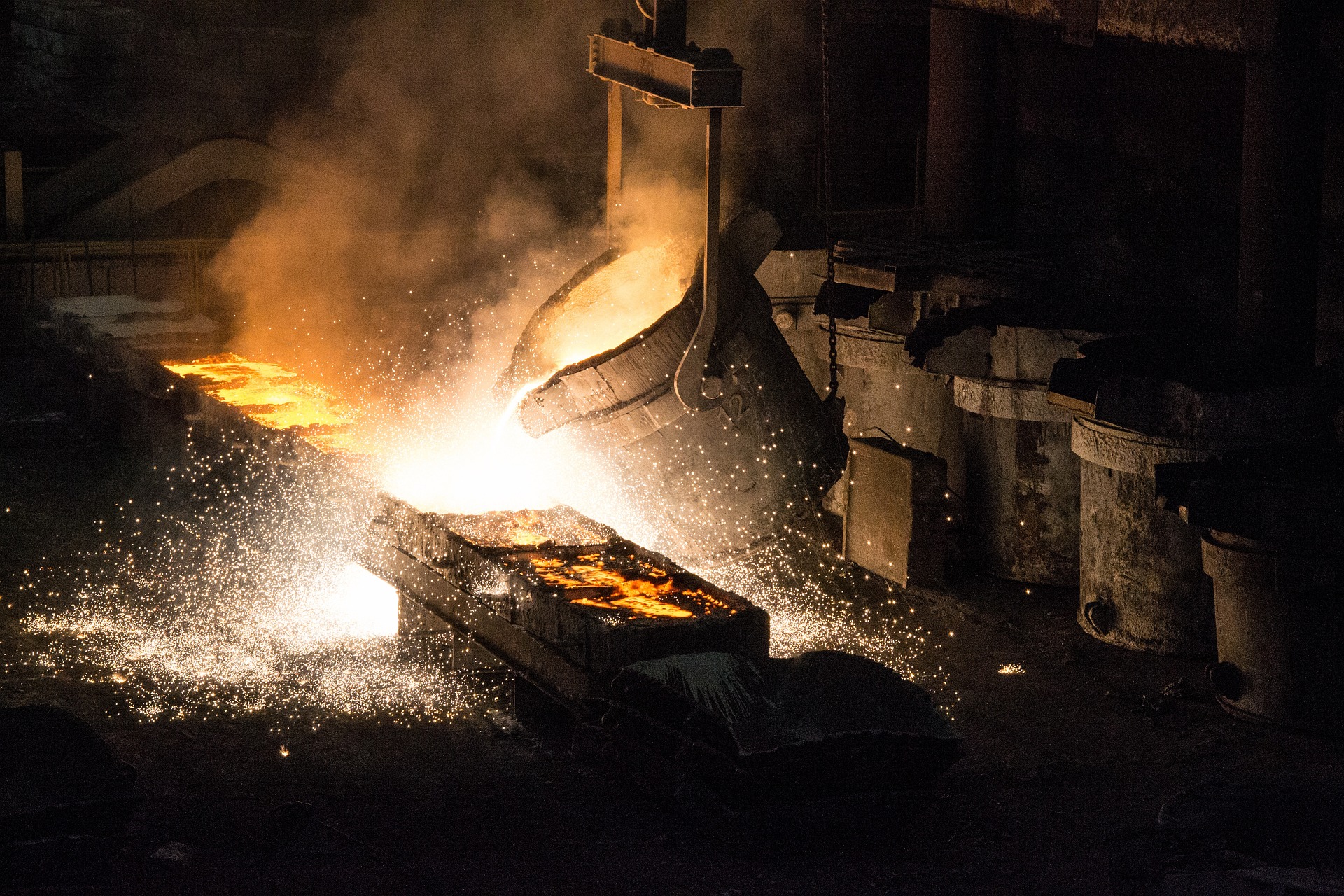}
                \caption{Example of a furnace ladle used in foundry industry. Image by Boris Bukovský, via Pixabay.}.
                \label{fig:ladle_real}
            \end{center}
        \end{figure}

        Furnace ladles are structures used to transport melted metal produced in an industrial plant, as shown in \autoref{fig:ladle_real}. A description of a generic industrial furnace ladle model used to simulate muography measurements can be found in \cite{Martinez2012}. A typical example of the part of the manufacturing process which involves the use of the ladle could be described in a general way as follows: firstly the metal is melted in a furnace, then it is transported with a furnace ladle, and finally it is poured into moulds to create metal parts with a defined shape. This parts can later be post-processed, in order to get the final product. 

        In this process, the amount of melted metal inside the furnace is a key parameter. The lack of metal in the ladle can result in moulds that are not completely filled. Therefore, they may not be used to produce the final product correctly. On the other hand, an excessive charge of metal in the ladle leaves remnants and scraps, causing problems and losses in the manufacturing process. It has to be mentioned that usually, in the top of the melted metal inside the ladle, there is a lighter layer of slag produced in the melting process itself. This layer, prevents the use of optical methods to determine the liquid metal level. The resolution needed by the industry to optimise the manufacturing process is of the order of a few centimetres, although it depends on the distinctive features of each case.

    \subsection{Detector and volume}

        Active layers, i.e., the detector panels, are located above and below the measurement target, in an usual scattering muography configuration. At least two panels at different $z$ locations providing $x-y$ positions of muons are required per tracking detector (upper, and lower detectors). As specified in \autoref{sec:detector_modelling}, they are modelled considering several parameters, such as position resolution $\sigma_{\text{max}}$, detection efficiency $\epsilon_{\text{max}}$, and cost per unit area. 

\begin{figure}[!ht]
            \centering
            \includegraphics[width=0.8\linewidth]{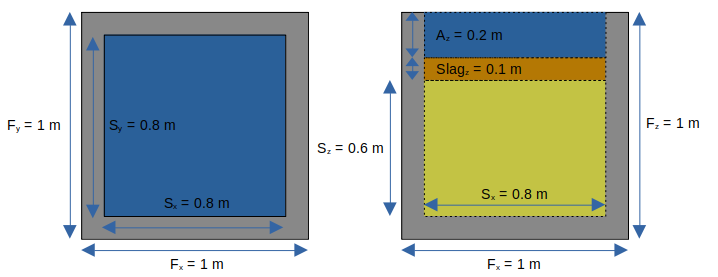}
            \caption{Top view (left), and front view (right) of the rectangular furnace ladle simulated. It is composed of walls (grey), liquid steel (yellow), slag (orange) and the air volume (blue) over the liquid steel and slag.}
            \label{fig:TomOpt_FurnaceSketch}
        \end{figure}

        Within the volume of interest between the detectors, a rectangular furnace ladle filled with liquid steel has been simulated (passive volume). In the \autoref{fig:TomOpt_FurnaceSketch} a sketch of the furnace ladle is shown. The grey parts correspond to ladle walls, made of solid steel, the molten steel is represented in yellow, the slag layer on top of it in orange, and the remaining void volume over them in blue. The definition of materials is detailed in \autoref{tab:Materials}.
        
        \begin{table}[!ht]
        \centering
        \begin{tabular}{lccccc}
        \toprule
        Material & Density, $d \ [\si{\gram\per\centi\metre\squared}]$ & Radiation length, $X_{0} \ [\si{\centi\metre}]$ & Composition \\
        \midrule
        Solid steel   & 7.818 & 1.782 & $Fe$ (99\%), $C$ (1\%)  \\
        Liquid steel  & 7.000 & 1.991 & $Fe$ (99\%), $C$ (1\%) \\
        Slag          & 2.700 & 8.211 & $CaO$ (53.5\%), $Al_{2}O_{3}$ (33.5\%) \\
        Air           & 0.001205 & 30390 & $N_{2}$ (78\%), $O_{2}$ (20\%) \\
        \bottomrule
        \end{tabular}
            \caption{Density, radiation length and composition of the simulated furnace ladle materials.}
            \label{tab:Materials}
        \end{table}

    \subsection{Inference method and loss function}
        The aim of this task is to design a detector that can provide an accurate measurement for the fill-heights of furnaces. This is essentially a single-variable regression task, and a natural choice for the loss function could be the mean squared-error (MSE) between the predicted and true fill-heights. However, as will be discussed below, MSE is not necessarily the most suitable loss for optimisation of this task.
        
        \subsubsection{Inference}\label{sec:ladel_inf}
\begin{figure}[h!]
                 \centering
                 \begin{subfigure}[b]{0.5\textwidth}
                     \centering
                     \includegraphics[width=\textwidth, height=7.5cm, keepaspectratio]{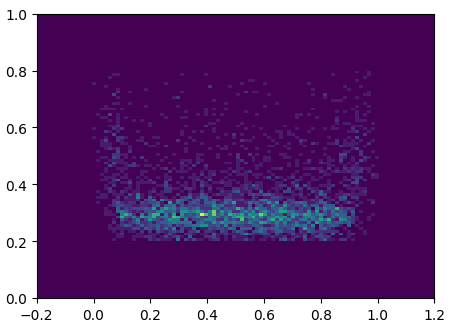}
                 \end{subfigure}
                 \hfill
                 \begin{subfigure}[b]{0.45\textwidth}
                     \centering
                     \includegraphics[width=\textwidth, height=7.5cm, keepaspectratio]{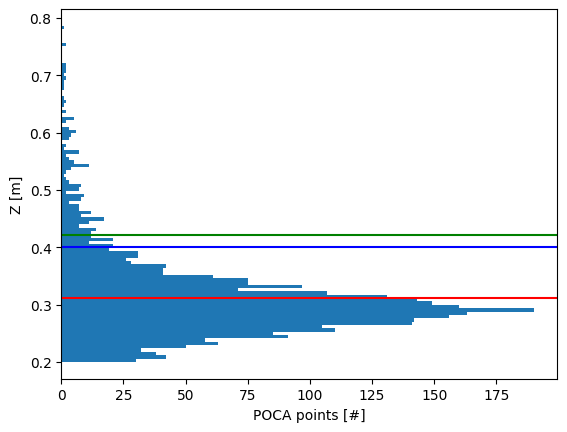}
                 \end{subfigure}
                    \caption{Muography of a simulation ladle furnace filled with steel up to \SI{0.4}{\metre}. The bottom wall of the furnace goes from \SIrange{0.2}{0.3}{\metre} in $z$ axis. On the left, the front view of \poca reconstruction. On the right, \poca distribution in vertical axis ($z$) and filling level inference example: true steel level (blue), $\overline{z_{p}}$ (red), and $h_{pred}$ (green) for the particular case of a bias correction on $z_{pred}$, with $a=2$ and $b=-0.2$ (see \autoref{:sec:debias} for details). }
                    \label{fig:MeanpocaInference}
            \end{figure}

            In the spirit of an end-to-end optimisation pipeline, the inference must output the fill-height prediction and receive optimisation loss gradients based on this. In developing a suitable inference method, we noticed that the mean value of the $z$-positions of \pocas ($\overline{z_{p}}$) was monotonic with the fill-height, due to hard scatterings occurring inside the liquid-metal part of the furnace volume, rather than the air and slag regions. This meant that it could serve as an indicator of the mass density distribution centre in the vertical Z axis (see \autoref{fig:MeanpocaInference}). As a benefit, this inference approach does not require computing \xo predictions for the voxels, and thus is more computationally efficient.
    
            This can be adapted to an inference method suitable for detector optimisation by computing a weighted average of the \poca $z$-positions, according to the muon-trajectory properties. The basis for this weight is the muon-trajectory efficiency ($\epsilon_p$), divided by the squared uncertainty of its \poca $z$-position ($(\Delta z_p)^2$).
            
            However, we also observed that \pocas located in the furnace walls could bias the prediction, and reduce precision. Additionally, we know we have uncertainties on the $xy$ positions of the \pocas. In a similar manner to the \xo inference described in \autoref{sec:tomopt_inference}, we define Gaussian distributions in $xy$ centred on the \pocas and scaling with their $xy$ uncertainties, and integrate them over each voxel in the transverse area of the furnace. We then use a pair of double-sigmoid-based functions in $xy$ to ascribe a weight to each voxel, that smoothly decreases to zero at the furnace walls, and maximises to one at the centre. This is used to multiply the integral in each voxel per \poca. The $xy$ weight of each \poca ($w_{xy,p}$) is defined as sum of its weighted voxel integrals, and this enters as a multiplicative factor of the overall \poca weight:
            \begin{equation}
                w_p = \frac{\epsilon_p}{\left(\Delta z_p\right)^2}{w_{xy,p}}.
            \end{equation}
            This weight has the effect of emphasising the $z$ positions of \pocas that are well measured in $z$, located near the centre of the furnace, and have a high trajectory efficiency.
        
            The predicted height is then taken as the weighted average of \poca $z$ positions:
            \begin{equation}
                z_{\mathrm{pred.}} = \frac{\sum_p w_p z_p}{\sum_p, w_p}.
            \end{equation}
            Whilst the mean of \poca $z$ positions is actually more likely to correspond to the centre of the fill-material, this redefinition to the fill-height can be accomplished by the linear transformation described in the next section.
    
        \subsubsection{Bias correction}\label{:sec:debias}
            The method described above, does not perfectly predict the fill-height: as mentioned, the mean of \poca $z$ positions is likely to correspond to the centre of the fill material, but with some finite precision; the \poca method is inherently biased through its ascription of the entire muon scattering to a single point; additionally, the bias on the prediction is detector-configuration-dependent.
    
            Provided that the biased prediction increases with the true fill-height, bias correction can be performed for a given detector configuration using many volumes of different fill-heights and adjusting the predictions (e.g via a continuous function, or bin-based corrections) such that on average, the predictions match the targets. Knowing this, what we really want the inference and detectors to produce is a prediction that has low in standard deviation when repeated multiple times for the same volume, whilst still increasing in mean value with the true fill-height.
    
            Eventually, though, we will need the detector to be used for actual measurements of fill-heights. At this point, the predictions from the inference must be de-biased as well as possible, in order to correspond to the true targets. The de-biasing can take the form of a parametric function of the predictions, and for simplicity we opt for a linear correction of the form:
            \begin{equation}
                h_{\mathrm{pred.}} = a\times z_{\mathrm{pred.}}+b,
            \end{equation}
            where parameters $a$ and $b$ are computed based on the predictions and targets from a large set of known example volumes at various fill-heights. The effect of this de-biassing can be seen in \autoref{fig:ladle:init_perf}.
    
            In order to improve monitoring capabilities during the optimisation, we refine the values of $a$ and $b$ after each update to the detector and display the resulting MSE of de-biased predictions. In order to avoid having to re-scan a large number of volumes, this refinement is performed using predictions and targets collected during the current batch of volumes.

        \subsubsection{Loss function}
            Even with the de-biasing, the detectors can still produce predicted heights that contain a residual bias, and optimisation under MSE risks being sensitive to these residual biases, resulting in unstable optimisation. As mentioned, the aim of the detector optimisation is to eventually allow the inference to produce precise predictions that increase with the true fill-height. A loss function that is suitable for this is one that captures the distribution of predictions for each target fill-height, i.e.:
            \begin{equation}
                \mathcal{L} = \frac{1}{N_l}\sum_l\sigma\!(p_l),
            \end{equation}
            where $\sigma\!(p_l)$ is the standard deviation of predictions for fill-height index $l$. If the spread of predictions becomes smaller, the loss will decrease, and vice versa, thus the loss encourages precise measurements of the fill-heights.

            However, there is nothing to ensure that the predictions increase in value with the fill-height: given sufficient flexibility in the detector (and potentially inference), it may be possible that the system outputs the same value for all predictions, thus proving zero spread, and so a perfect loss value. In order to penalise such a collapse, we need to include a factor that encourages a large distribution in predictions for different fill-heights, thus the final loss function becomes:
            \begin{equation}
                \mathcal{L} = \frac{1}{N_l}\frac{\sum_l\sigma\!(p_l)}{\sigma\!(\bar{p}_l)},
                \label{eqn:spread_range}
            \end{equation}
            where $\sigma\!(\bar{p}_l)$ is the standard deviation of the mean prediction for each fill-height.

            If the system were to output the same value for every volume (thus minimising the spread of predictions within each fill-height), the spread of mean predictions for each fill-height would also collapse to zero and act to increase the loss value. The loss function is therefore minimised when the distributions of predictions in each fill-height is small, and when the mean predictions between fill-heights is as large as possible. Thus the predictions become easier to de-bias: if the predicted height for a given volume is within a certain range, we can be more sure it corresponds to a particular true fill-height. We refer to \autoref{eqn:spread_range} as the ``spread over range" loss function.

            \begin{figure}[h]
            \centering
                \includegraphics[width=\textwidth, height=7.5cm, keepaspectratio]{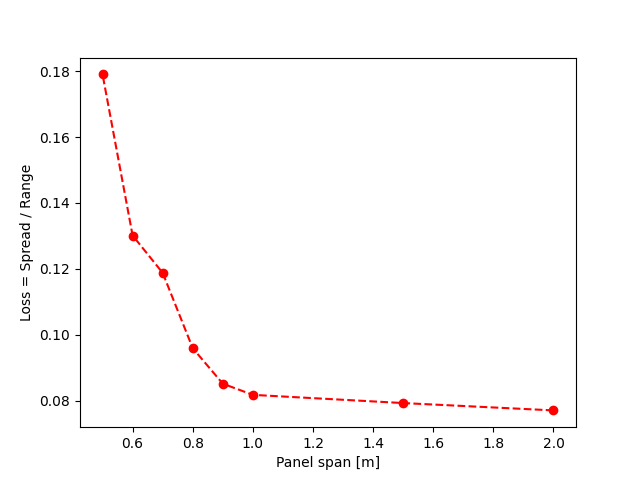}
                \caption{``Spread over range" loss function tested for different detection surfaces, and using the $h_{pred}$ inference method with unweighted $z_{pred}$ and a linear de-biasing of $a=2$ and $b=-0.2$.}
                \label{fig:LossFunction}
            \end{figure}

            In the \autoref{fig:LossFunction}, the evolution of the proposed loss function with relation to the span of detection panels is shown. The centre of the panels in the horizontal axes X and Y, is the same that the one of the furnace ladle. A sharp loss decrease is noticed when detection panel surface is increased within the limits of the furnace ladle interior, where the liquid steel is located, i.e., the muography target. However, if the panel span exceeds the width of the ladle interior, the loss reduction starts to be less noticeable. We found that this effect is due to the smaller proportion of muons that cross the target when detectors are larger than the furnace interior width, and also because of the contribution of furnace walls to muon deflection. Their presence modifies the density distribution in $z$ axis, distorting our predictions, although this effect is smoothed by the weighted \poca inference (see \autoref{sec:tomopt_inference}). Either way, the ``spread over range" loss is continuously and endlessly reduced when increasing the detector surface, the expected behaviour regarding the performance of a muography, since a larger surface allows to collect a higher number of meaningful muons, even if they traverse the target in a transversal direction. This generally improves the precision of inference algorithms. Note that in the ``spread over range" part of the loss function is not included the loss component of detection surface cost (see \autoref{sec:tomopt:optimisation}).
    
    \subsection{Optimisation process and results}

        \subsubsection{Passive-volume data}
            Using \SI{10}{\centi\metre} voxels, we generate volumes at eight different fill-heights to simulate from ladles that are almost empty to ones that are almost full. In every case, we add a \SI{10}{\centi\metre} layer of slag on top. When optimising the detector, each of these volumes will be evaluated five times (40 volumes per \textit{volume batch}) -- whilst the volumes will always be the same, stochasticity in prediction will still be present due to the muon generation. When evaluating the performance of the detectors, each volume is evaluated 12 times. 1000 muons will be used per evaluation.

        \subsubsection{Initial configuration}
            In order to test the optimisation capabilities of \tomopt, we initially set the detector in a knowingly sub-optimal configuration: one in which the detector panels are off-set from the passive volume in $x-y$, thus lowering the flux muons that are well detected, and interact with the passive volume; and closely space in $z$, thus increasing the uncertainty in \poca reconstruction.
            
\begin{figure}[ht]
        	\begin{center}
        		\begin{subfigure}[t]{\sfSmall\textwidth}
        			\begin{center}
        				\includegraphics[width=\textwidth]{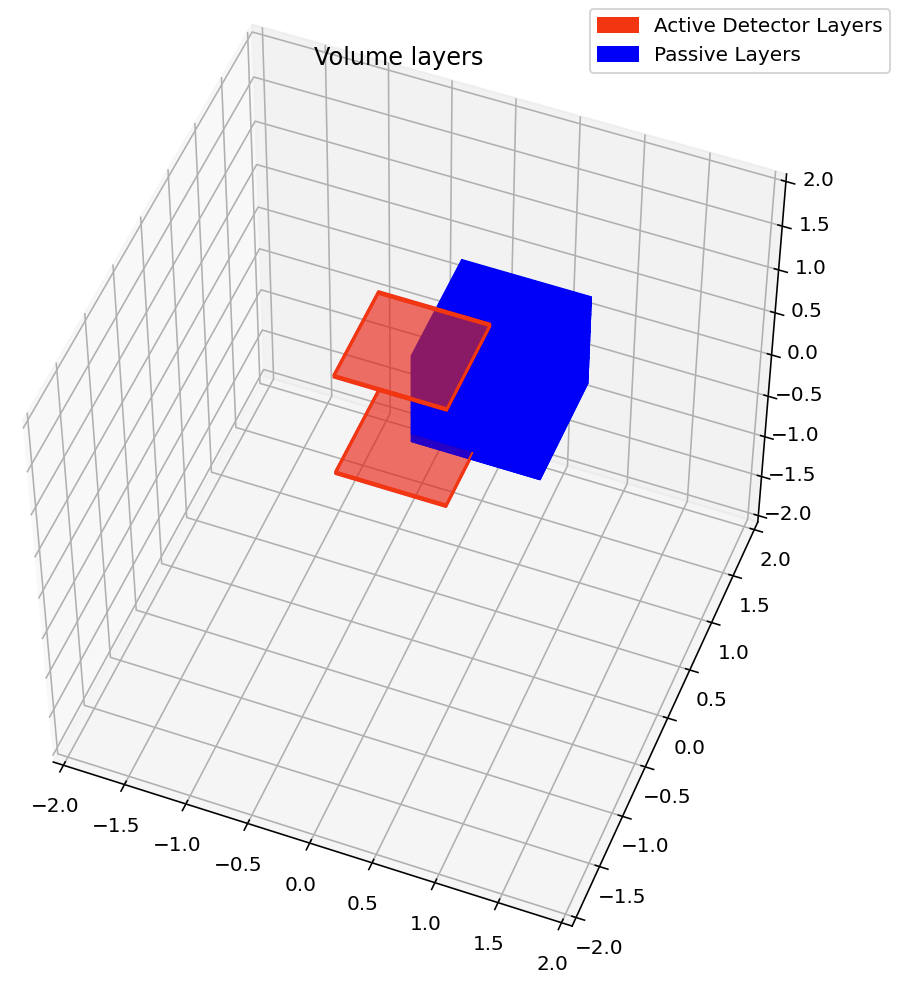}
        			\end{center}
        		\end{subfigure}
        		\begin{subfigure}[t]{\sfSmall\textwidth}
        			\begin{center}
        				\includegraphics[width=\textwidth]{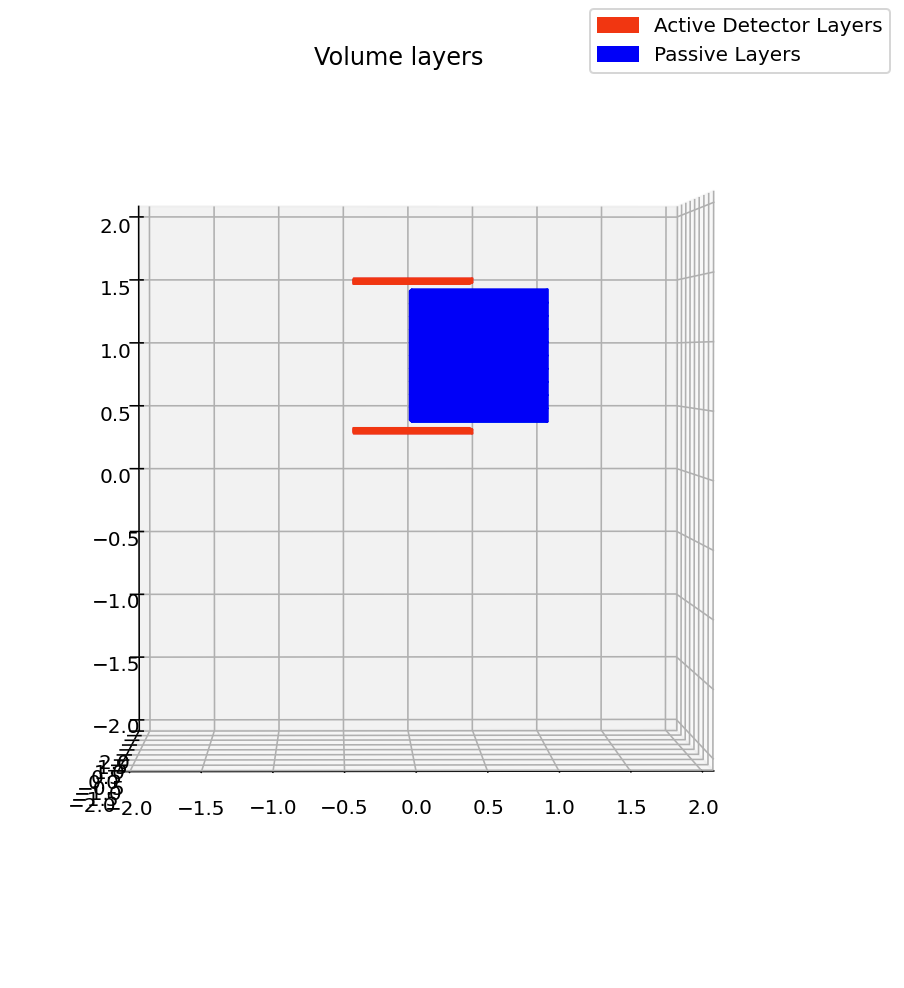}
        			\end{center}
        		\end{subfigure}
        		\caption{Initial layout of the detector panels above and below the passive volume.}
        		\label{fig:ladle:init_det}
        	\end{center}
            \end{figure}

            The detector itself (illustrated in \autoref{fig:ladle:init_det}) consists of four equally sized square panel above the passive volume, and another four panels below. These panels begin centred in $x-y$ over the corner of the passive volume, and with an (unrealistic) separation in $z$ of \SI{1}{\centi\metre} between each panel. There is a \SI{5}{\centi\metre} separation between the passive volume to the nearest panels. The total cost of the detector is fixed at six arbitrary units, and panels cost one arbitrary unit per \si{\metre\squared}. The available budget is initially distributed equally to all panels, thus each panel measures \SI{86.6}{\centi\metre} in both $x$ and $y$ length. Panels are defined to have nominal efficiencies of \SI{90}{\%}, and $x-y$ resolutions of \SI{0.1}{\milli\metre}.

\begin{figure}[ht]
        	\begin{center}
        		\begin{subfigure}[t]{\sfSmall\textwidth}
        			\begin{center}
        				\includegraphics[width=\textwidth]{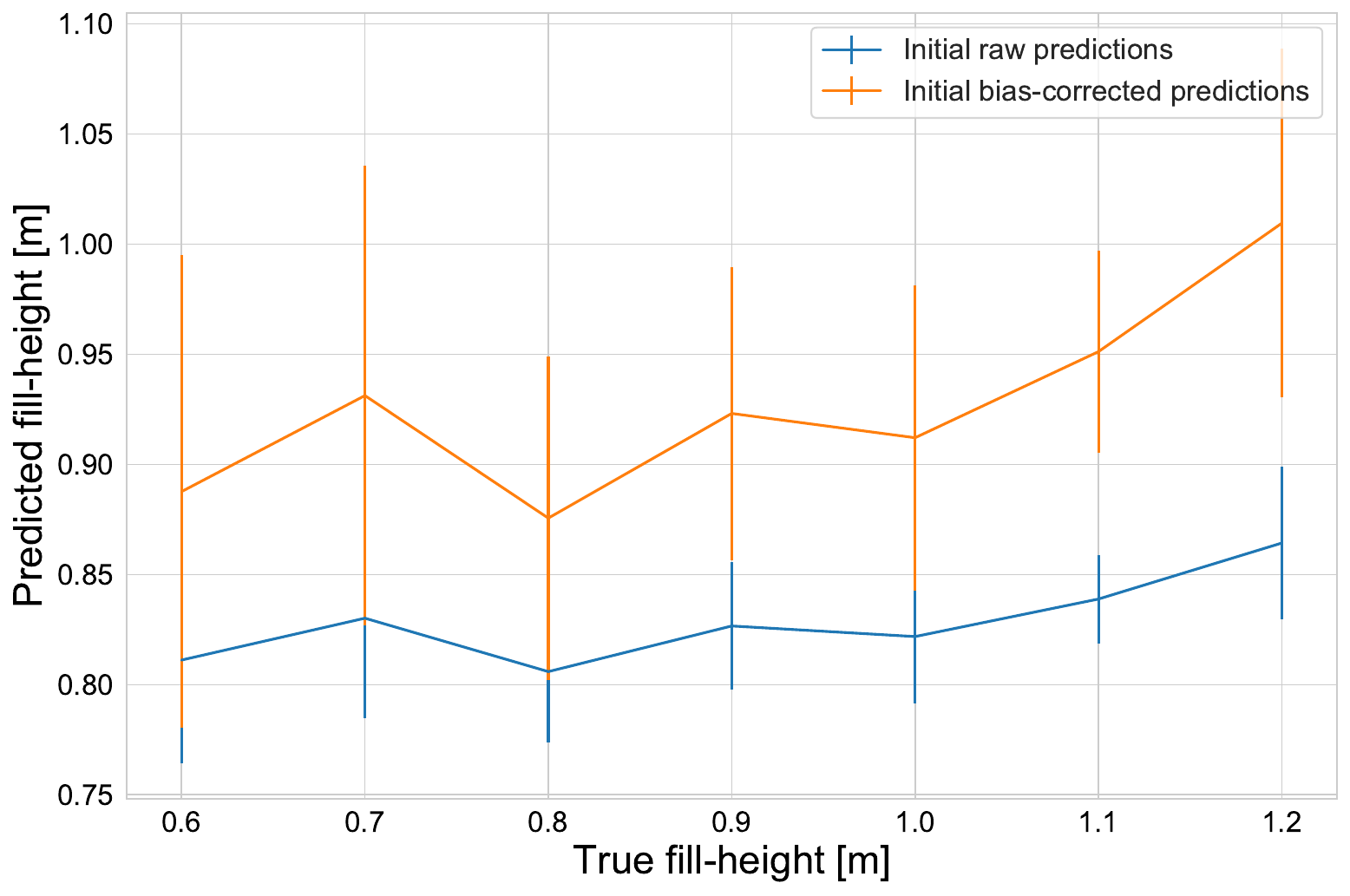}
        			\end{center}
        		\end{subfigure}
        		\begin{subfigure}[t]{\sfSmall\textwidth}
        			\begin{center}
        				\includegraphics[width=\textwidth]{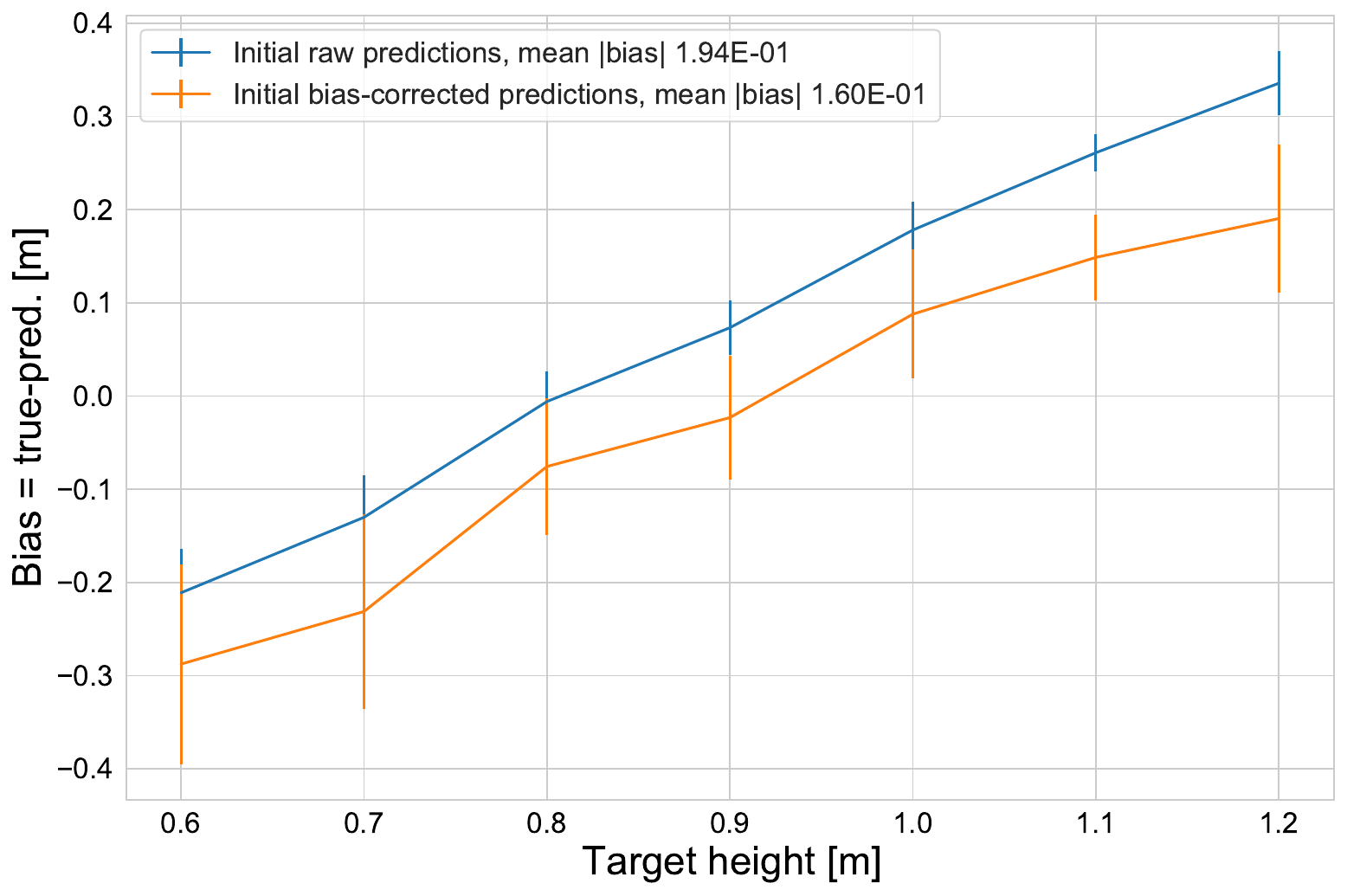}
        			\end{center}
        		\end{subfigure}
                    \begin{subfigure}[t]{\sfSmall\textwidth}
        			\begin{center}
        				\includegraphics[width=\textwidth]{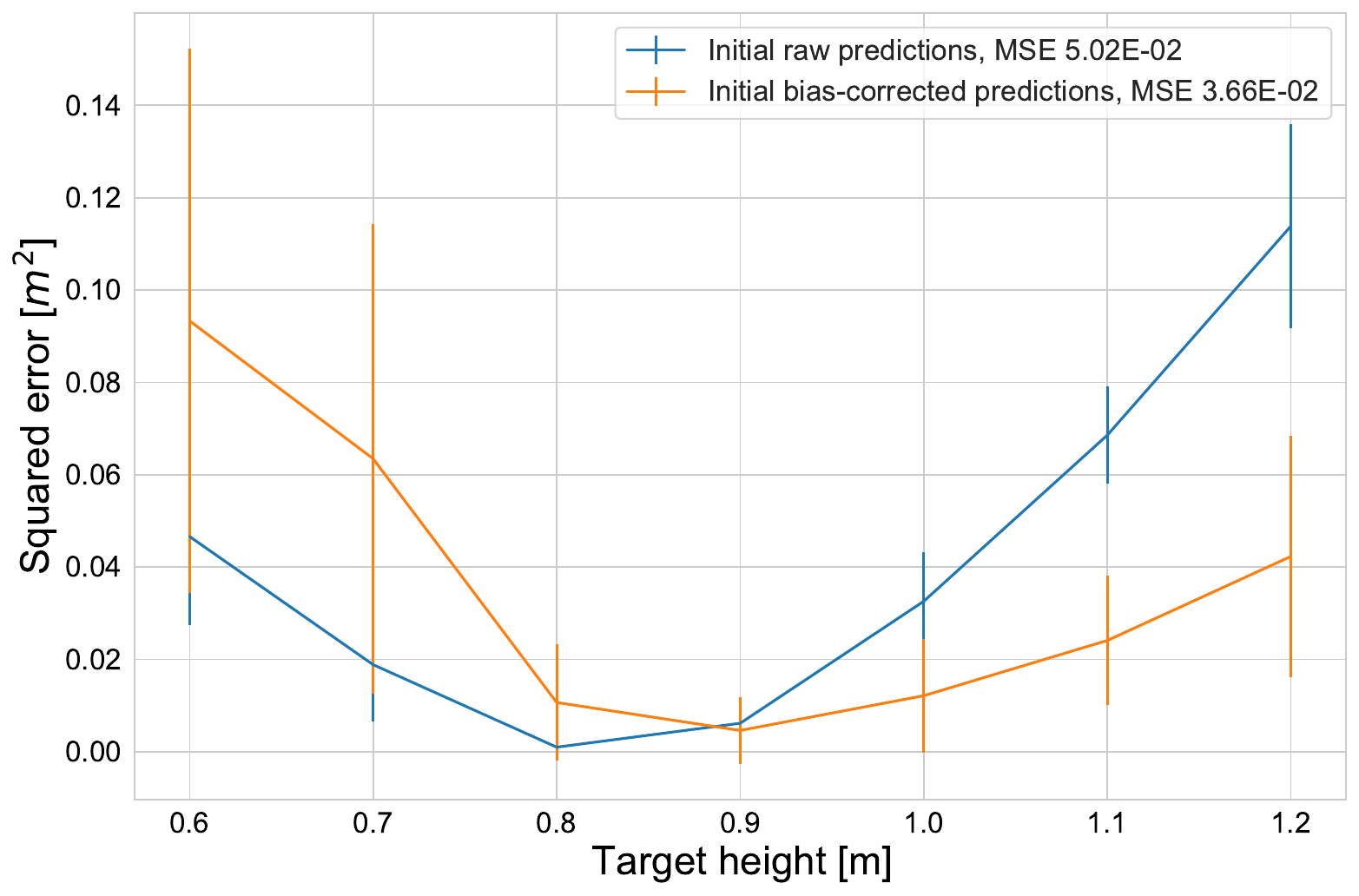}
        			\end{center}
        		\end{subfigure}
        		\caption{Performance of the initial detector before and after the de-biasing process.}
        		\label{fig:ladle:init_perf}
        	\end{center}
            \end{figure}

            The initial performance for the detector is shown in \autoref{fig:ladle:init_perf}. The ``raw" predictions are the result of the inference with no de-biasing. We can observe a slight slope to their values w.r.t. the fill-height, however the large spread in predictions for the same target, means that it is difficult to distinguish different fills. The de-biasing process via a linear correction is able slightly reduce the MSE, but cannot do much more that shift the predictions. Averaging across the range of fill-heights considered the initial detector has a mean MSE of \SI{0.037}{\metre\squared}, and an average absolute error on fill-level prediction of \SI{16}{\centi\metre}.

            Our intuitive expectation is that from this position, the panels should move to cover the centre of the passive volume, and increase in $z$-separation. There is, however, a trade-off in increasing the $z$-separation: whilst it generally increases the \poca resolution, it also decreases the hit-reconstruction for muons entering or leaving the passive volume at large angles. We therefore expect the optimisation to find a natural point at which the improved resolution is balanced by the decreased high-angle-muon flux. Additionally, we have no confident intuition over how the panels should be resized to best improve performance given a constant budget.

        \subsubsection{Optimisation callbacks}
            As mentioned in \autoref{sec:tomopt:optimisation}, \tomopt has a callback system that allows the user to augment the basic optimisation loop to include different functionality, and to help stabilise the optimisation process. For this task, the following callbacks are used:
            \begin{itemize}
                \item \textsc{MuonResampler} -- When muons are generated, they may not ever interact with the passive volume, and so form a source of \textit{background}, which would ideally be removed by some filtering process. In the interests of simplicity, we are not considering background processes here\footnote{Given the small number of muons used, relative to the incoming flux and acquisition time, a highly pure sample of muons is expected to be acquirable by even a low-efficiency trigger/filter.}. The \textsc{MuonResampler} callback acts to regenerate muons until all muons in the batch will interact with the passive volume at some point. This therefore improves the generation efficiency.
            \item \textsc{LinearCorrection} -- The de-biasing process mentioned in \autoref{:sec:debias} is implemented as a callback that can modify in-place the outputs of the inference module, collect raw predictions and targets during the volume batch, and then re-fit its correction after each detector update.
            \item \textsc{SpreadRangeLoss} -- By default, \tomopt computes the loss per volume and tracks a running sum of losses over the volume batch. However the ``spread over range" loss can only be computed for a population of volume predictions. This callback collects the predictions over the course of the volume batch and then sets the value of the loss prior to back-propagation.
            \item \textsc{NoMoreNaNs} -- Optimisations can take several hours, and whilst \tomopt is fairly resilient, given the number of muons being propagated and the complexity of the reconstruction and inference process, it is inevitable that invalid (\textit{NaN}) gradients are occasionally produced. These would otherwise kill the optimisation process, however after back-propagation but before the parameter updates, this callback goes through the parameters and sets any \textit{NaN} gradients to be zero.
            \item \textsc{OptConfig} -- The learning rate of optimisers in gradient descent is one of their most important parameters, however it can be difficult to set suitably with limited \textit{a priori} knowledge. Whilst techniques such as the ``learning rate range finder test" from Ref.~\cite{smith_2015} have been developed, they are designed for DNNs, where the learnable parameters lack a physical meaning. Here, our parameters do have physical meaning (e.g. the detector position is a learnable parameter measured in metres). We are thus able to do something better: we can specify the desired change per update of parameters in their physical units, and over the course of a \textit{warm-up phase} (in which all parameters are fixed) compute the average gradient received by each parameter. From these averages, we can then compute the learning rate that is likely to produce the desired update rate. As an example we can say that we want the panel $xy$ positions to update at a rate of \SI{1}{\centi\metre} per iteration. We can then monitor the median gradients for a few iterations and then set an appropriate learning rate that results in a change of about \SI{1}{\centi\metre} per iteration according to $\gamma = \sfrac{\SI{1}{\centi\metre}}{\bar{\nabla}_{xy-\mathrm{pos.}}\mathcal{L}}$.
            \item \textsc{PanelCentreing} --  In order to help stabilise the optimisation in \tomopt, we use this callback: after panel $xy$ positions have been updated, it goes through and for panels above and below the passive volume separately, it computes the mean position in $x-y$ of panels and then sets panels to that position. Panels below the passive volume can still have different positions to those above, but all panels above and below each have a common $x-y$ position. Whilst this does limit the originality of detector configuration that \tomopt can explore, it is currently necessary to avoid instabilities and divergences in the optimisation.
            \item \textsc{PanelMetricLogger} -- This provides real-time feedback to the user, and displays the current layout of the detector, the history of the loss values, and other diagnostic information.
            \item \textsc{EpochSave} -- This automatically saves the state of the detector after every update, allowing the user to manually pick the best performing detector in case of performance divergence, or to recover in case of errors
            \item \textsc{OneCycle} -- This is a learning rate scheduler that adapts the learning rate (and optionally momentum/$\beta_1$ coefficients) or the optimisers according to the 1-cycle schedule described in Refs.~\cite{smith_2017, smith_2018}. This is mainly used to allow the learning rate to gradually decrease over the course of the optimisation, allowing the detector configuration to better converge, whilst allowing larger updates near the beginning to training to move quickly from sub-optimal to near-optimal positions.
            \item \textsc{SigmoidPanelSmoothnessSchedule} -- as mentioned in \autoref{sec:tomopt:opt_mode}, the physical panels are approximated differentiably during the \textit{training epochs} using a sigmoid-based model. It was also suggested that different values of the \textit{smoothness} of the model allows different parameters of the detector panels mode or less of an effect on the loss. This callback allows the smoothness to be decreased across the optimisation process, meaning that close to the end of the training, the panels better approximate physical detectors, and the loss is more sensitive to the exact $z$ position of the panels.
            \end{itemize}

        \subsubsection{Optimisation process}
            Five parameter families are considered: $xy$ positions of the detector panels and the $z$ positions of the panels are considered separately for panels above and below the volume (four parameter families); and the budget assignment per panel (one parameter family). Over the course of five epochs, we monitor the gradients received by the trainable parameters in these families and compute nominal learning rates that correspond to position updates at a rate of \SI{1}{\centi\metre} per epoch, and budget weights at a rate of 0.1 (dimensionless).

            \paragraph{Stage one}
                Over the course of 10 epochs, the detector is updated, using 1-cycle learning rate annealing to scale the nominal learning rates over the course of the optimisation: initially they increase to allow large updates of the detector, and towards the end the decrease to much smaller values, allowing the panels to converge and stabilise. Additionally, the sigmoid detector model is annealed from an initially smooth configuration to a sharper, more physical distribution of efficiency and resolution. On a 2018 Macbook Pro with an Intel i7, this takes about 30 minutes to run (no GPU is used).
    
\begin{figure}[ht]
            	\begin{center}
            		\begin{subfigure}[t]{\sfSmall\textwidth}
            			\begin{center}
            				\includegraphics[width=\textwidth]{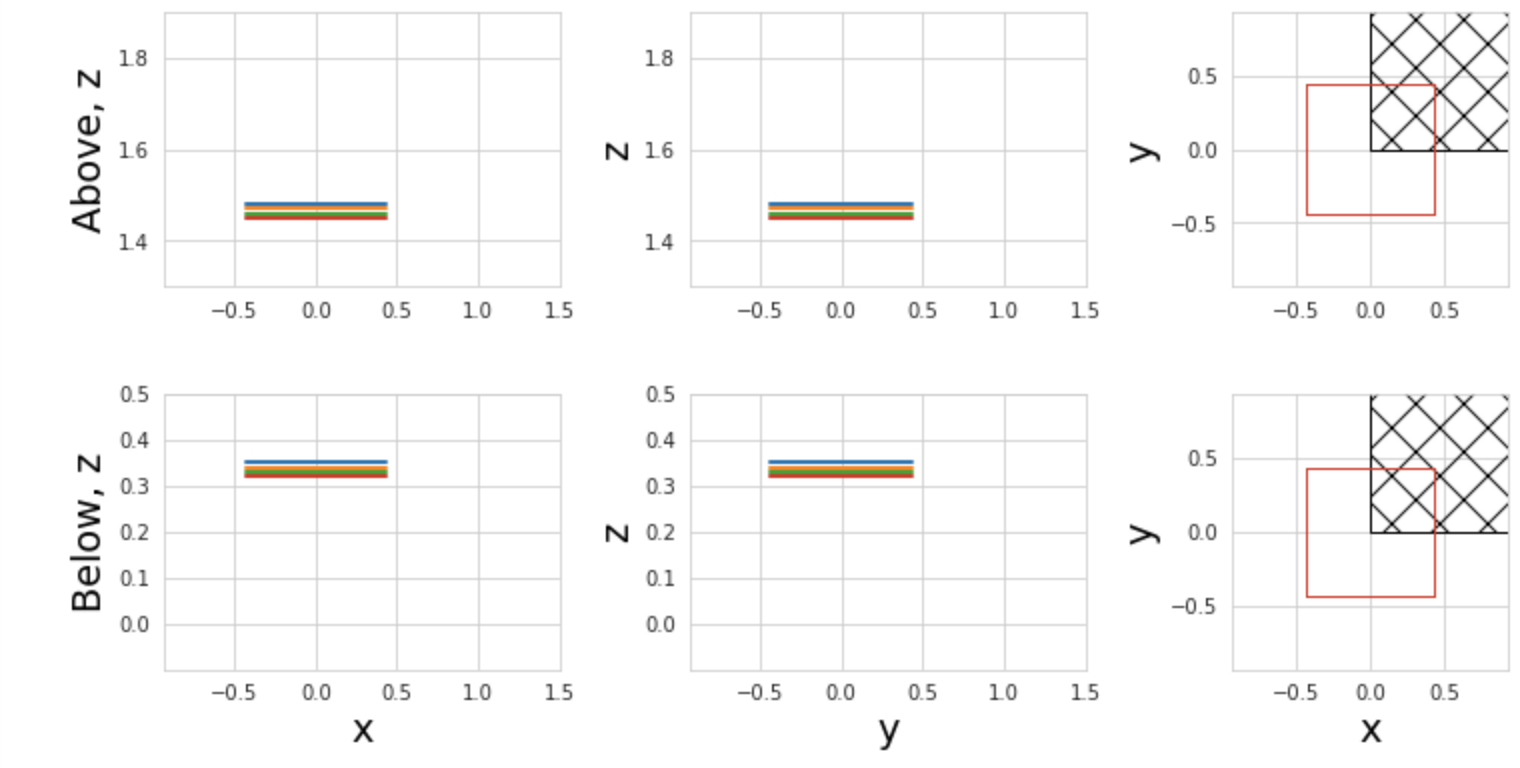}
                                \caption{Initial detector configuration.}
            			\end{center}
            		\end{subfigure}
            		\begin{subfigure}[t]{\sfSmall\textwidth}
            			\begin{center}
            				\includegraphics[width=\textwidth]{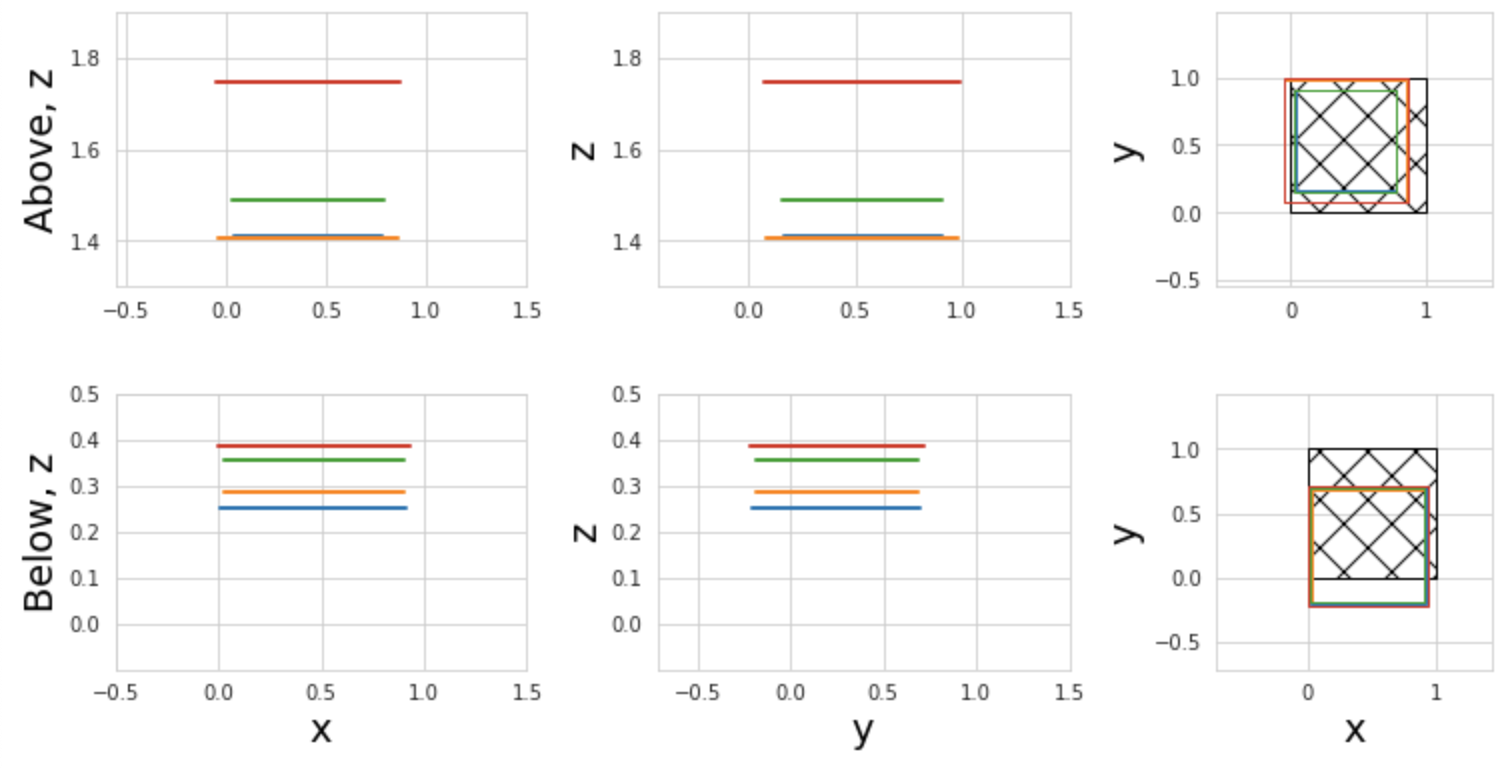}
                                \caption{Detector configuration after stage one optimisation process.}
            			\end{center}
            		\end{subfigure}
            		\caption{Comparison of detector configurations before and after stage one of the optimisation process. The coloured lines/squares indicate the positions and sizes of the panels, and the hatched area indicates the position and size of the passive volume. The top rows indicate panels above the passive volumes, and the bottom rows indicate panels below the passive volume.}
            		\label{fig:ladle:det_opt_s1}
            	\end{center}
                \end{figure}

                Figure~\ref{fig:ladle:det_opt_s1} illustrates the detector configuration before and after this stage of the optimisation process. From this we can see that the panels have indeed moved to be more central in $x-y$, and have expanded in $z$.
    
\begin{figure}[ht]
            	\begin{center}
            		\begin{subfigure}[t]{\sfSmall\textwidth}
            			\begin{center}
            				\includegraphics[width=\textwidth]{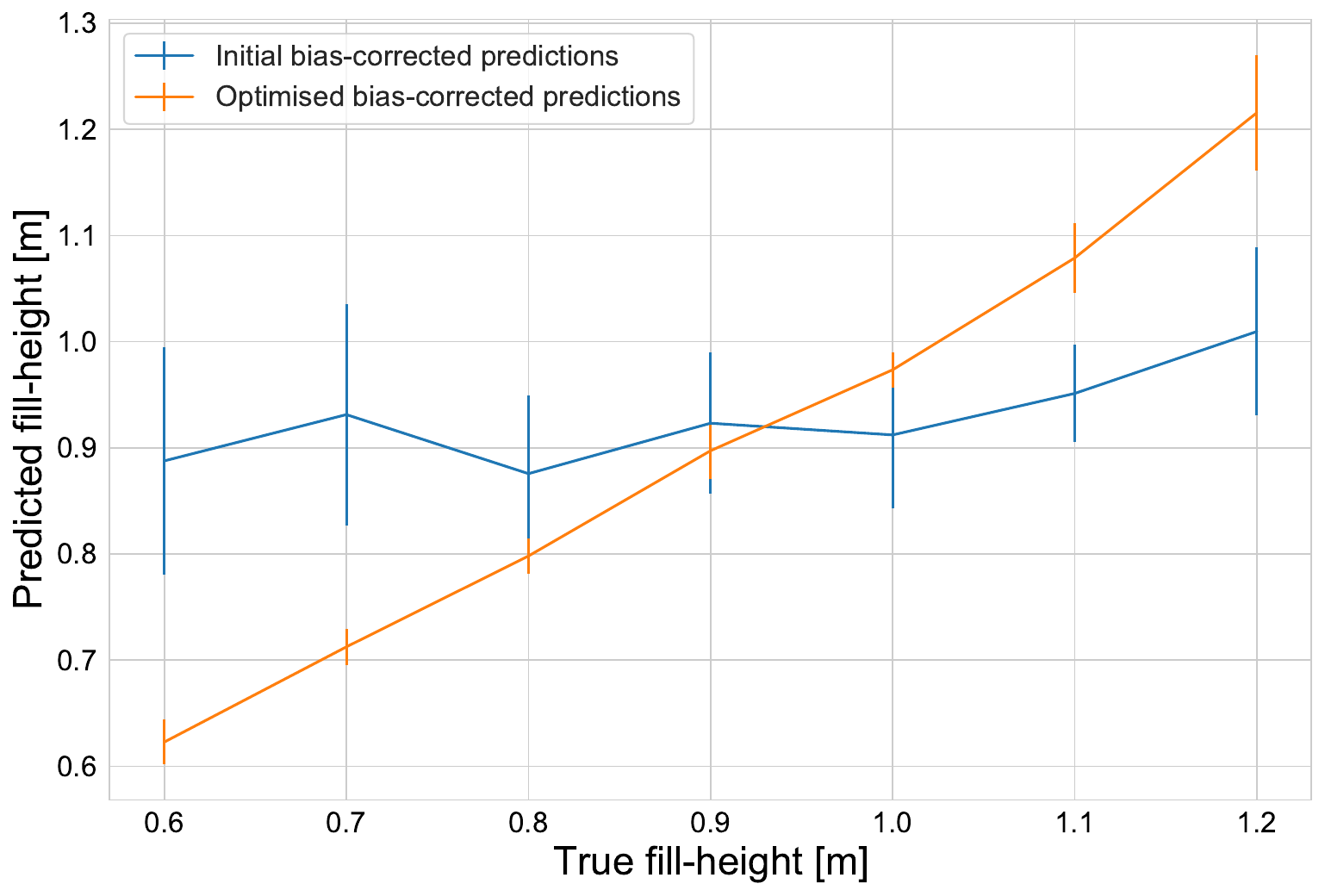}
            			\end{center}
            		\end{subfigure}
            		\begin{subfigure}[t]{\sfSmall\textwidth}
            			\begin{center}
            				\includegraphics[width=\textwidth]{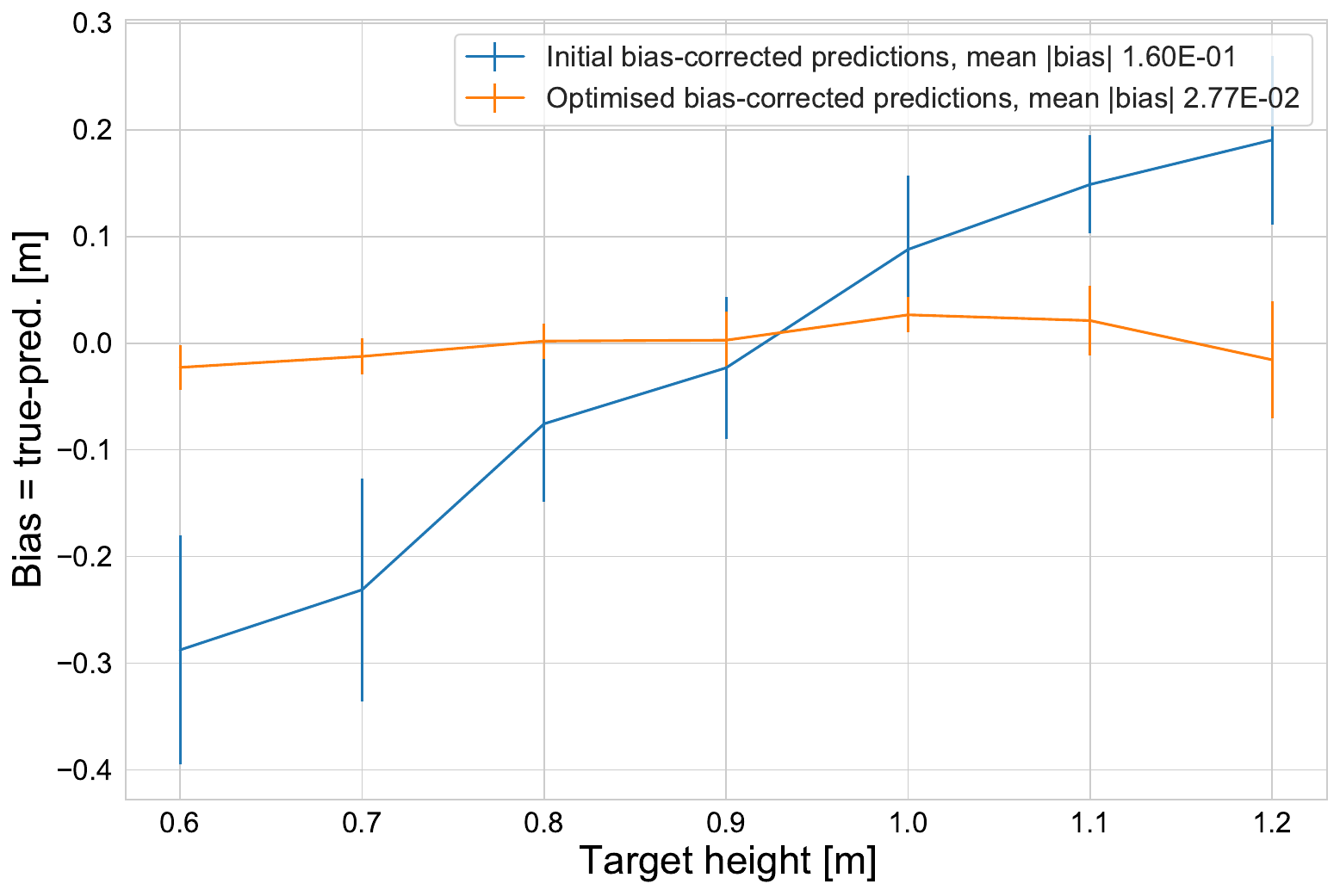}
            			\end{center}
            		\end{subfigure}
                        \begin{subfigure}[t]{\sfSmall\textwidth}
            			\begin{center}
            				\includegraphics[width=\textwidth]{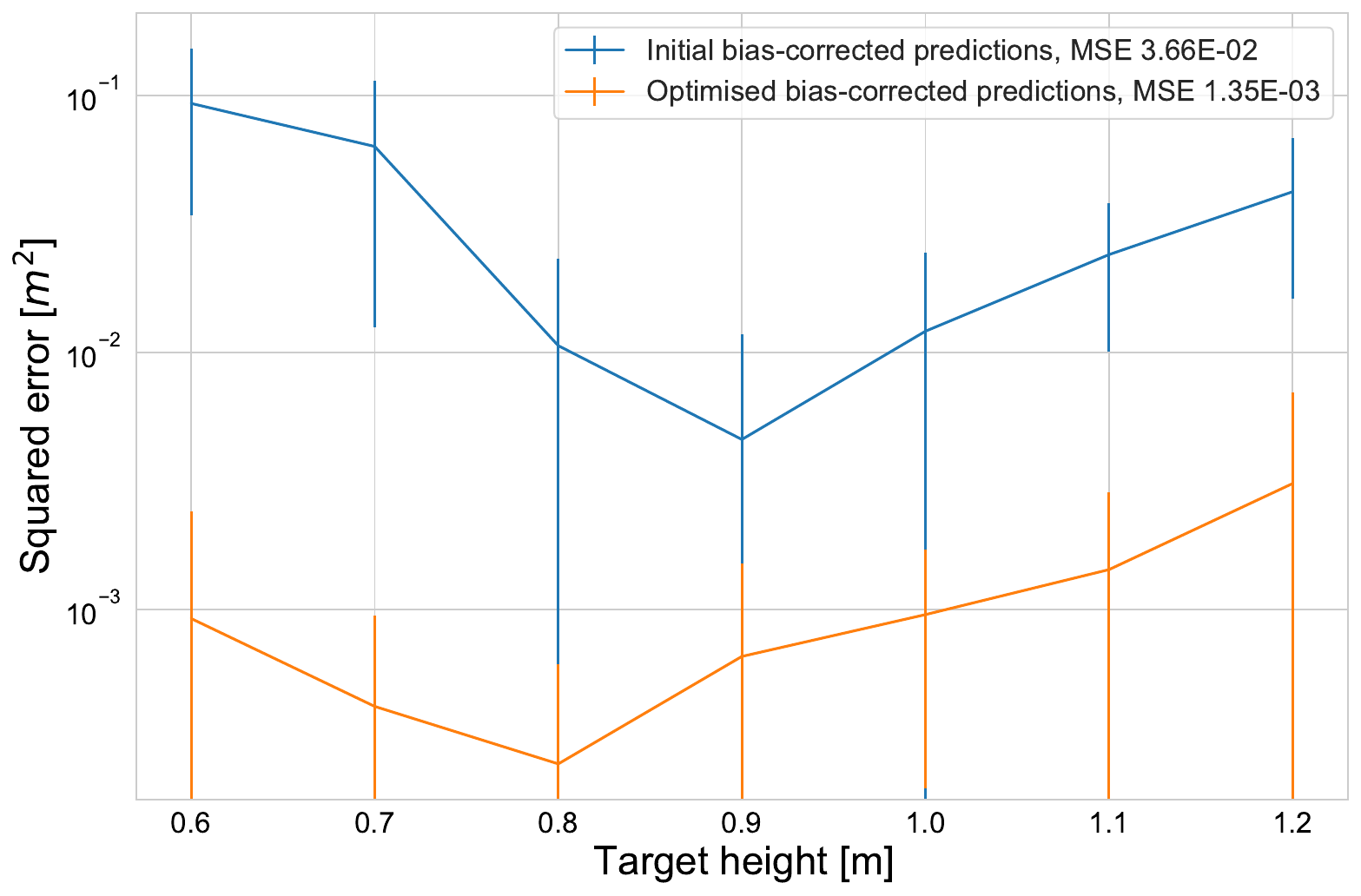}
            			\end{center}
            		\end{subfigure}
            		\caption{Performance of the optimised detector after stage 1 compared to the initial detector (both shown after the de-biasing process). Heights shown include the space below the passive volume.}
            		\label{fig:ladle:det_opt_s1_perf}
            	\end{center}
                \end{figure}

                Figure~\ref{fig:ladle:det_opt_s1_perf} compares the performance of the optimised detector to the initial detector, and indicates a significant improvement in both the precision and bias: now predictions are clearly distinguishable by true fill-height. The optimised detector has a mean MSE of \SI{0.0014}{\metre\squared}, and an average absolute error on fill-level prediction of \SI{2.8}{\centi\metre}.
    
            \paragraph{Stage two}
                The previous optimisation process showed that \tomopt can take a poorly designed detector and quickly shift it into a configuration that provides much better performance. In this second stage we will demonstrate that \tomopt can be used to refine detectors that are already thought to be well performing, by altering only subsets of the total available parameters. To do so, we will use our detector from stage one.

                During stage one, we saw that the panels moved to become central in $x-y$. We will take this experimental evidence, along with our domain knowledge of the problem, to assume that the ideal position for the detectors is directly above and below the passive volume. We will manually shift the panel to the centre and fix them there, allowing only the $z$ and budget assignments to slowly change.

\begin{figure}[ht]
            	\begin{center}
            		\begin{subfigure}[t]{\sfSmall\textwidth}
            			\begin{center}
            				\includegraphics[width=\textwidth]{content/figures/optimised.png}
                                \caption{Detector configuration after stage one optimisation process.}
            			\end{center}
            		\end{subfigure}
            		\begin{subfigure}[t]{\sfSmall\textwidth}
            			\begin{center}
            				\includegraphics[width=\textwidth]{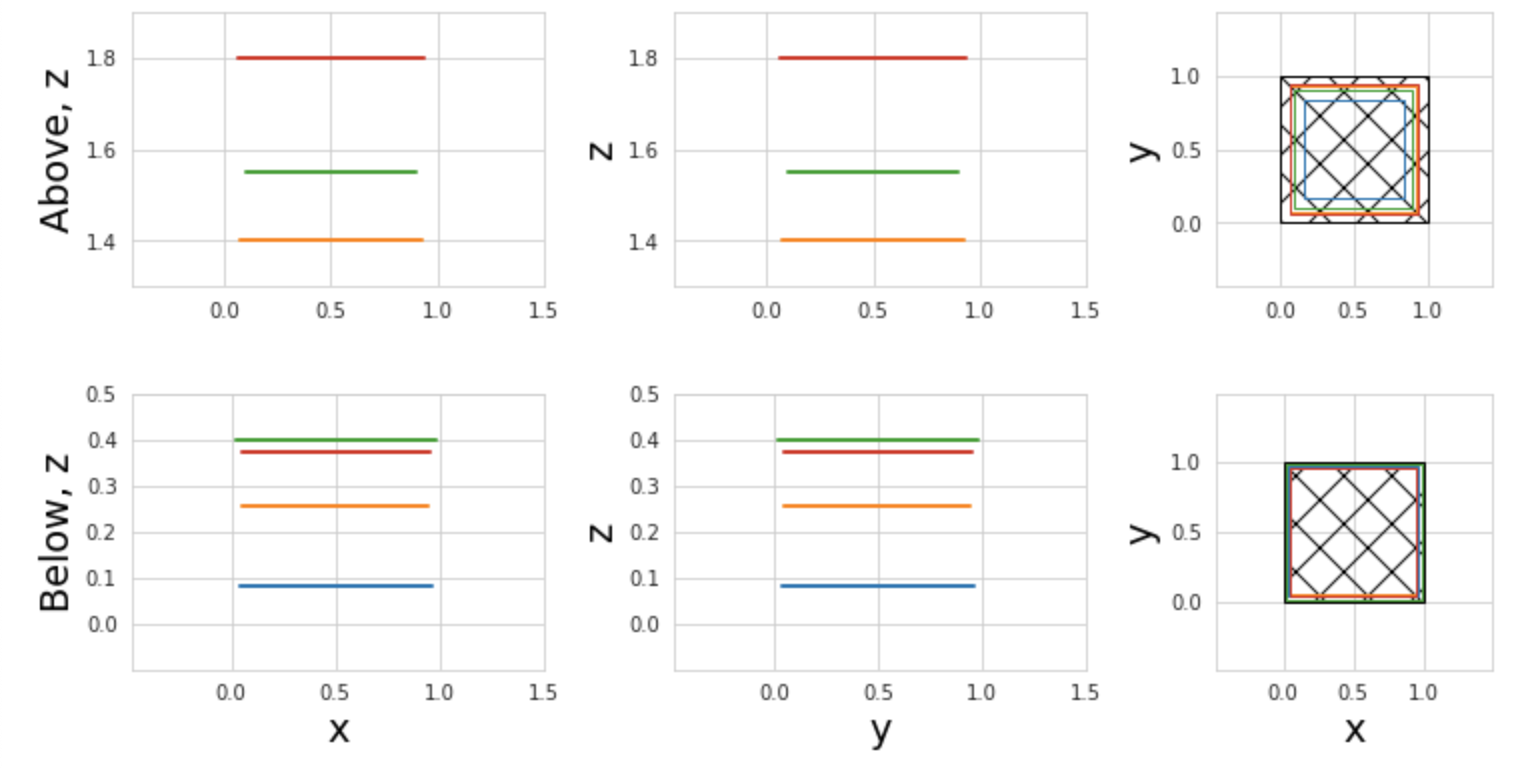}
                                \caption{Detector configuration after stage two optimisation process.}
            			\end{center}
            		\end{subfigure}
            		\caption{Comparison of detector configurations after stage one (left) and stage two (right) of the optimisation process.}
            		\label{fig:ladle:det_opt_s2}
            	\end{center}
                \end{figure}

\begin{figure}[ht]
            	\begin{center}
            		\begin{subfigure}[t]{\sfSmall\textwidth}
            			\begin{center}
            				\includegraphics[width=\textwidth]{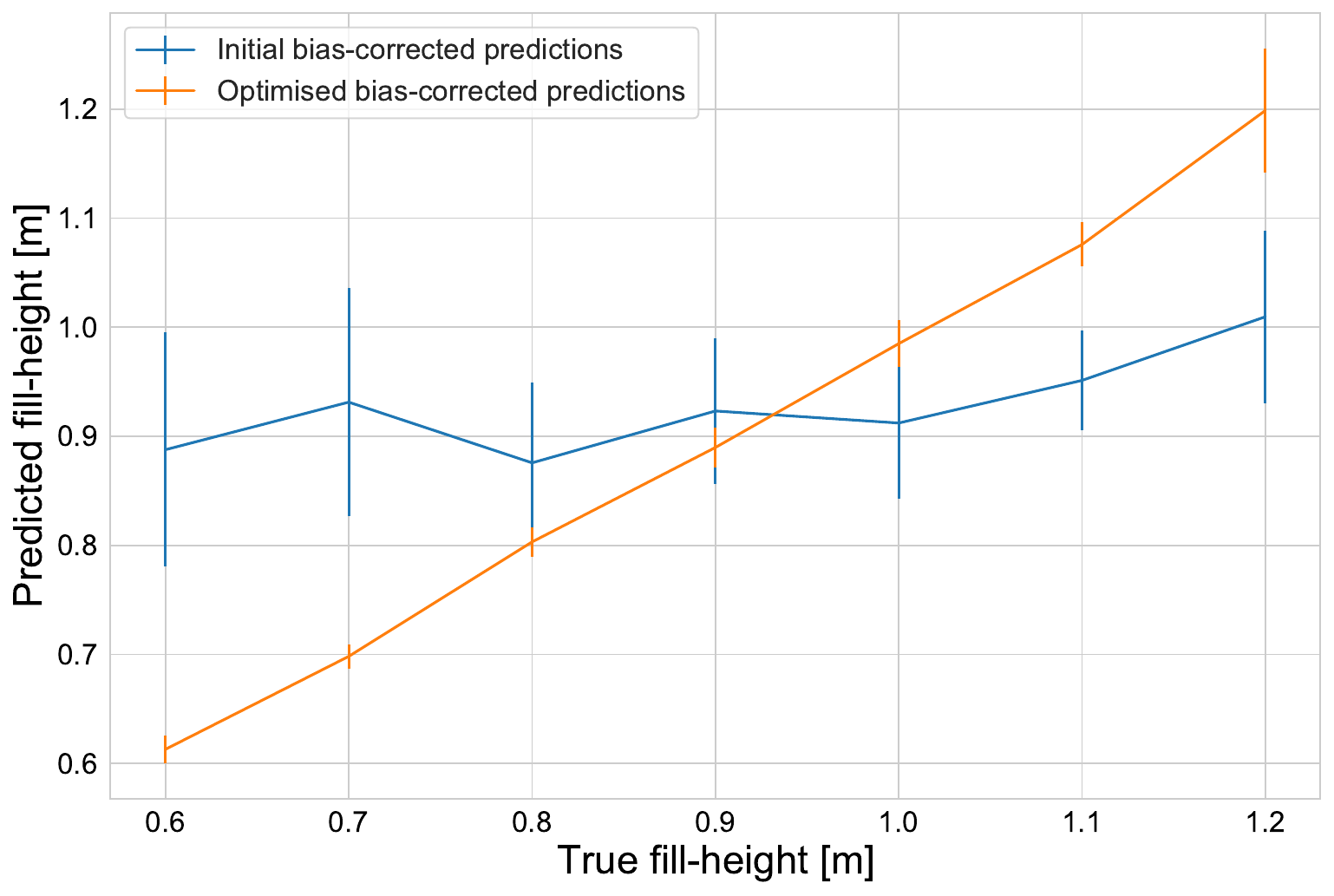}
            			\end{center}
            		\end{subfigure}
            		\begin{subfigure}[t]{\sfSmall\textwidth}
            			\begin{center}
            				\includegraphics[width=\textwidth]{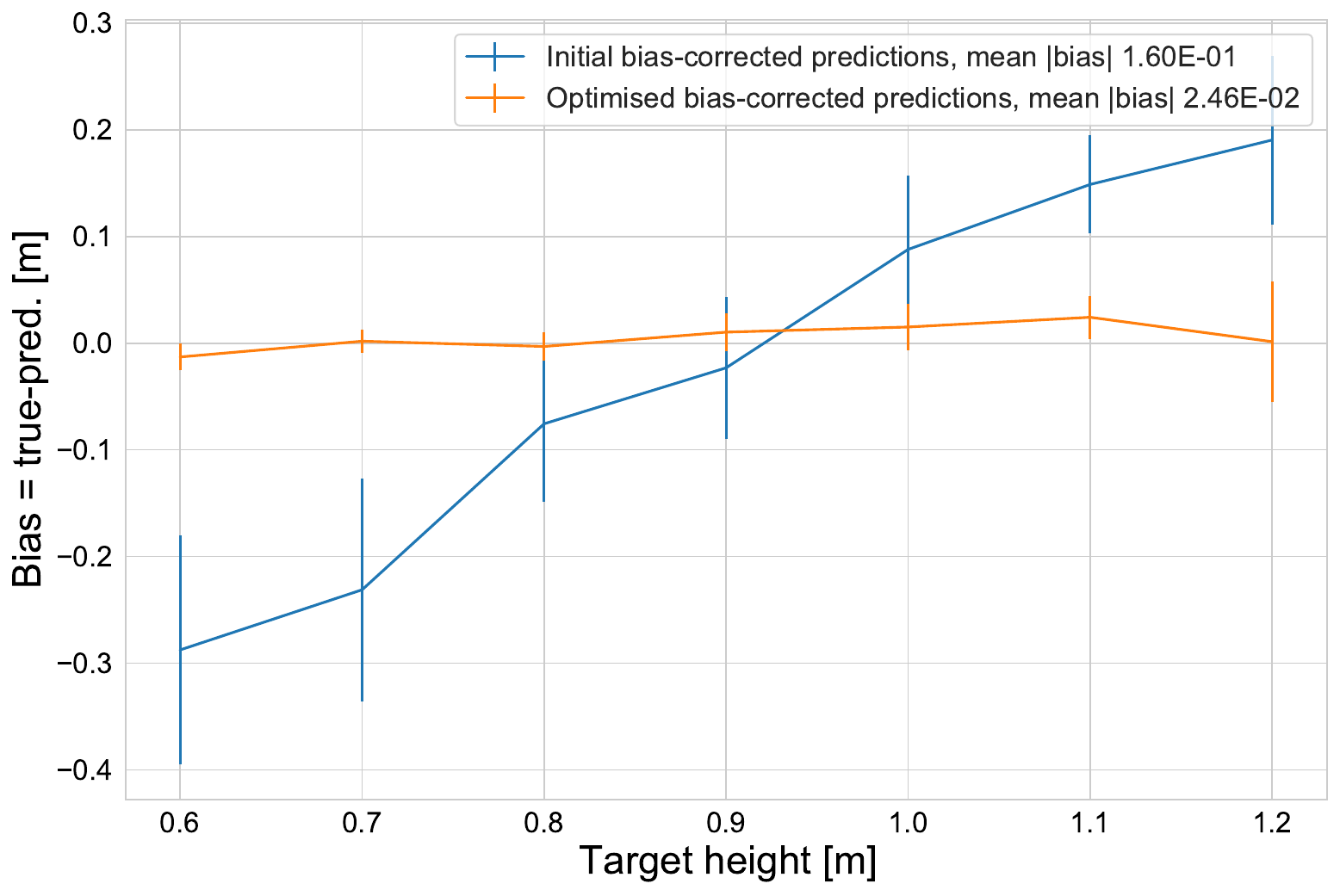}
            			\end{center}
            		\end{subfigure}
                        \begin{subfigure}[t]{\sfSmall\textwidth}
            			\begin{center}
            				\includegraphics[width=\textwidth]{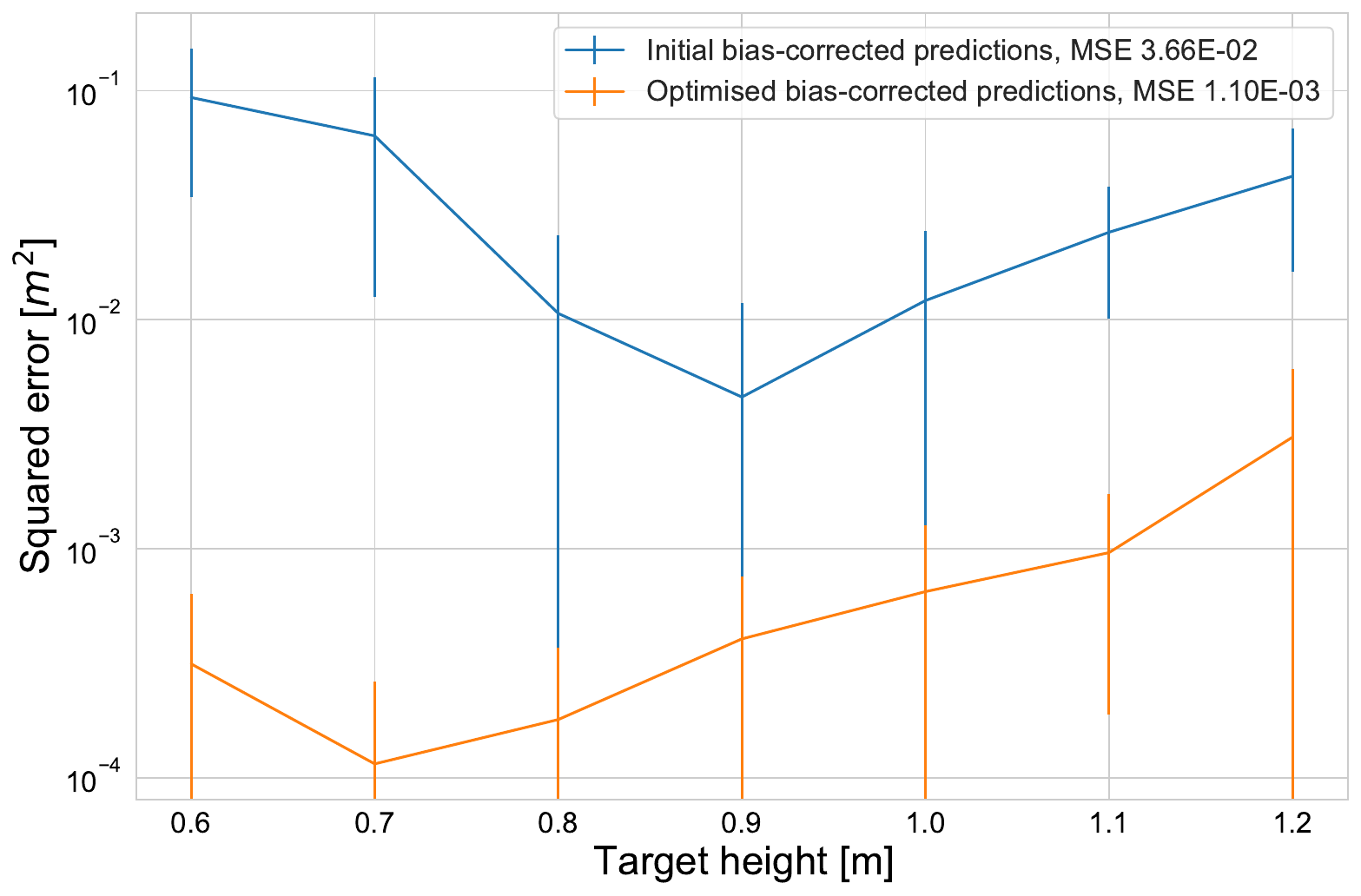}
            			\end{center}
            		\end{subfigure}
            		\caption{Performance of the optimised detector after stage 2 compared to the initial detector (both shown after the de-biasing process). Heights shown include the space below the passive volume.}
            		\label{fig:ladle:det_opt_s2_perf}
            	\end{center}
                \end{figure}

                Once again, we compute suitable nominal learning rates for our three remaining parameter families by monitoring the gradients they receive for five epochs. Over another 10 epochs we allow the detector to further be refined, this time with smaller update steps. Figure~\ref{fig:ladle:det_opt_s2} illustrates the detector after this refinement stage, and \autoref{fig:ladle:det_opt_s2_perf} its performance. Mainly we can see that the panels below the passive volume have moved to cover a larger $z$ range. While this could come at the expense of missing some high-angle muons leaving the passive volume, the panels have slightly increased in size by decreasing the intermediate panels above the passive volume. The refined detector has a mean MSE of \SI{0.0011}{\metre\squared}, and an average absolute error on fill-level prediction of \SI{2.5}{\centi\metre}.

    \subsection{Comparison to human baselines}
\begin{figure}[ht]
            \begin{center}
                \begin{subfigure}[t]{\sfSmall\textwidth}
                    \begin{center}
                        \includegraphics[width=\textwidth]{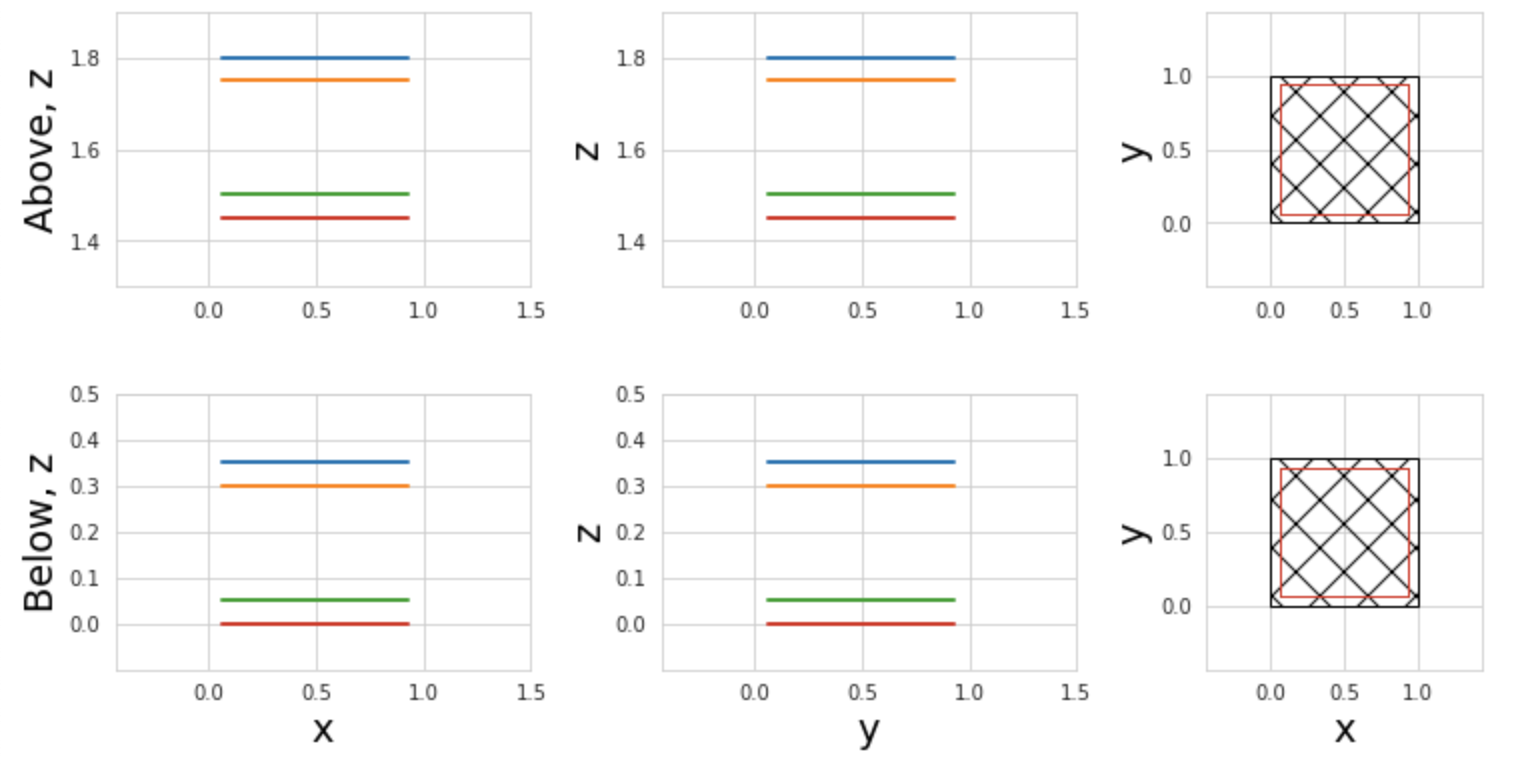}
                            \caption{Baseline detector 1.}
                    \end{center}
                \end{subfigure}
                \begin{subfigure}[t]{\sfSmall\textwidth}
                    \begin{center}
                        \includegraphics[width=\textwidth]{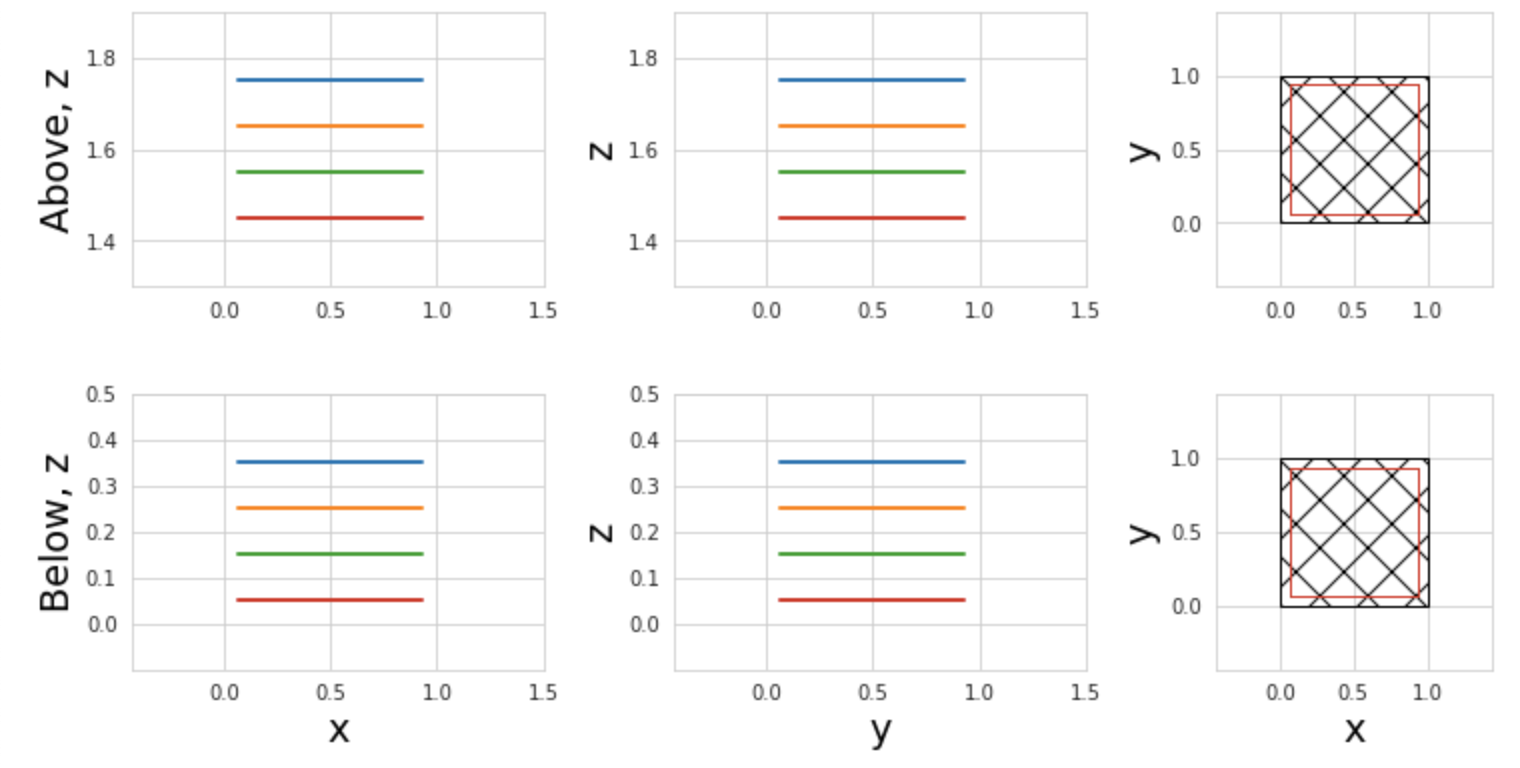}
                            \caption{Baseline detector 2.}
                    \end{center}
                \end{subfigure}
                \caption{Configurations for the human-designed baseline detectors. The coloured lines/squares indicate the positions and sizes of the panels, and the hatched area indicates the position and size of the passive volume. The top rows indicate panels above the passive volumes, and the bottom rows indicate panels below the passive volume} 
                \label{fig:ladle:baseline}
            \end{center}
        \end{figure}

        Having  demonstrated that \tomopt can successfully optimise detectors, we now want to see whether the resulting detector is good compared to a human-designed detector. For this we will take two baseline configurations:
        \begin{itemize}
            \item Baseline 1: Detector panels are placed in pairs with a \SI{5}{\centi\metre} separation between panels, and a \SI{25}{\centi\metre} separate between pairs above and below the passive volume. This configuration aims to maximise the precision of the muon trajectory reconstruction by allowing precise measurements of the muon positions over a long baseline.
            \item Baseline 2: Detector panels are placed with an even \SI{10}{\centi\metre} spacing. Baseline 1 potentially misses some high-angle muons due to the large distance to the second pair panels. This baseline attempt to strike a balance between precision and exposure.
        \end{itemize}
        These are illustrated in \autoref{fig:ladle:baseline}.

\begin{figure}[ht]
            \begin{center}
                \begin{subfigure}[t]{\sfSmall\textwidth}
                    \begin{center}
                        \includegraphics[width=\textwidth]{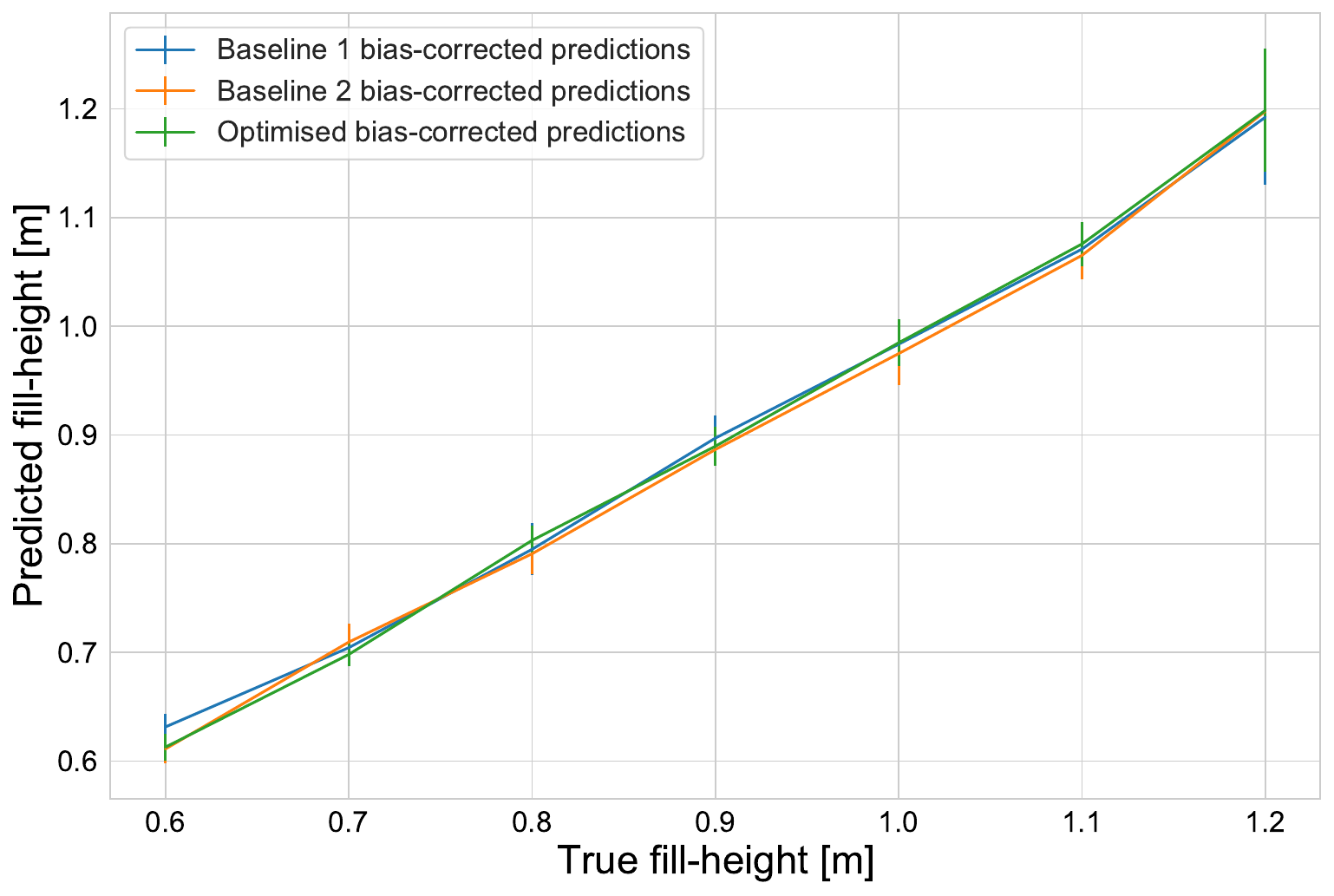}
                    \end{center}
                \end{subfigure}
                \begin{subfigure}[t]{\sfSmall\textwidth}
                    \begin{center}
                        \includegraphics[width=\textwidth]{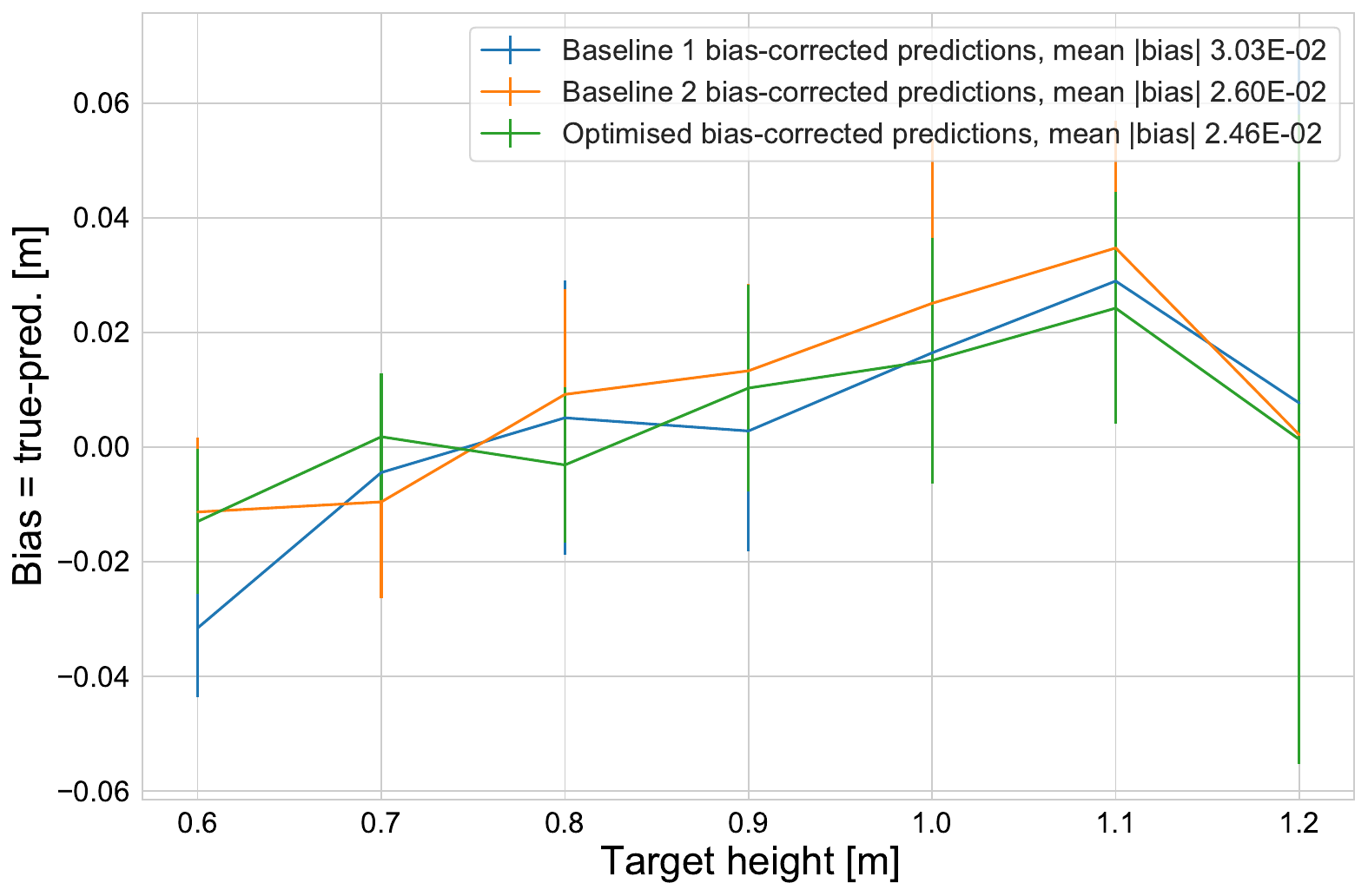}
                    \end{center}
                \end{subfigure}
                    \begin{subfigure}[t]{\sfSmall\textwidth}
                    \begin{center}
                        \includegraphics[width=\textwidth]{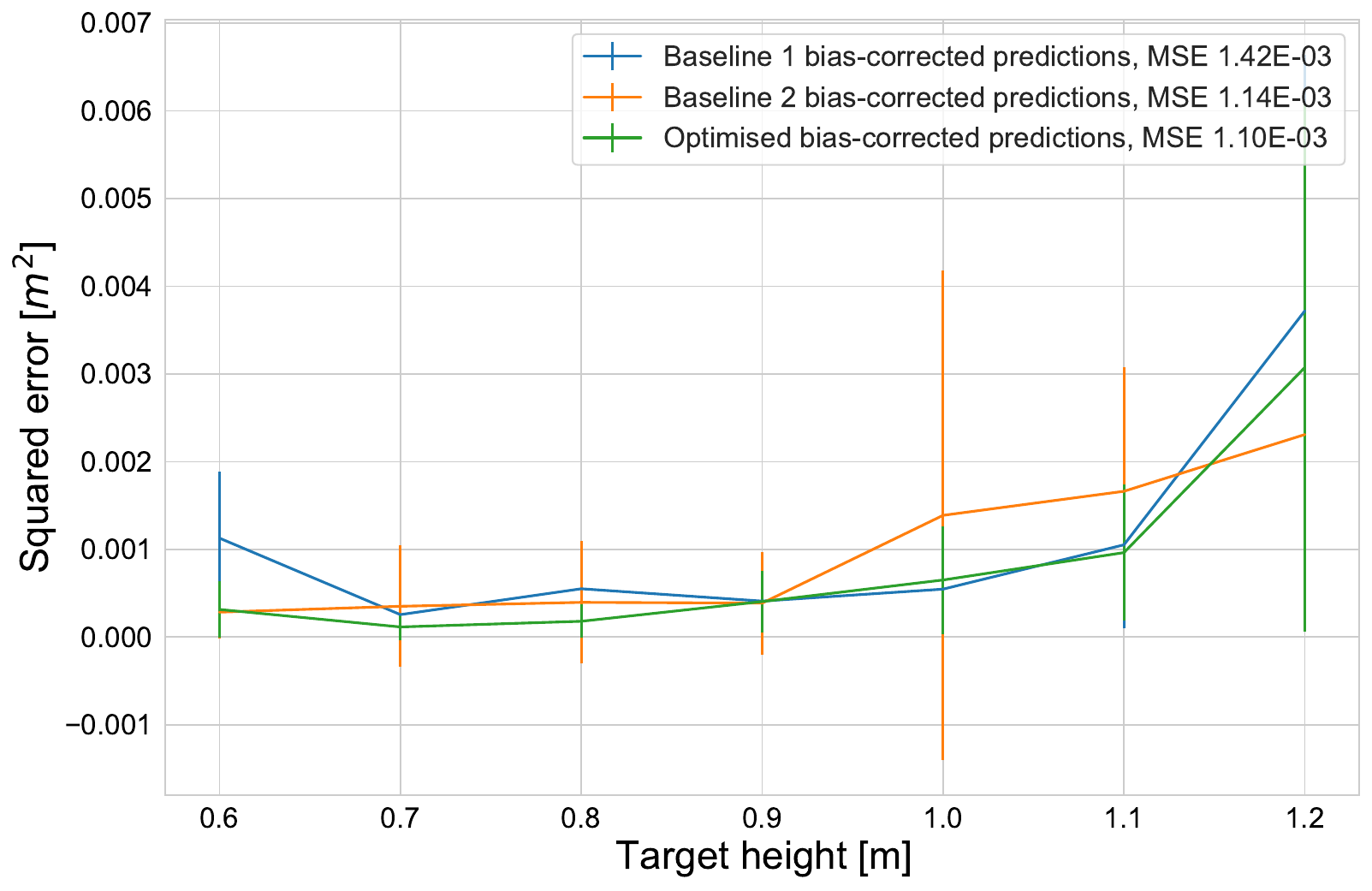}
                    \end{center}
                \end{subfigure}
                \caption{Performance of the optimised detector after stage 2 compared to human-designed baseline detectors (both shown after the de-biasing process). Heights shown include the space below the passive volume.}
                \label{fig:ladle:baseline_perf}
            \end{center}
            \end{figure}

        Figure~\ref{fig:ladle:baseline_perf} compares the performances of the stage-two optimised detector against the two baseline detectors. Whilst all three detectors offer similar levels of performance, our optimised detector is able to not only match human performance, but slightly and consistently outperforms it.

\FloatBarrier
\section{Conclusions}
\label{sec:conclusions}

In this document, we have described \tomopt; a system that allows muon-tomography detector designers to specify their end goal numerically as a task-specific loss function, and through the use of automatic differentiation optimise their designs to provide the best possible performance. Since the updates to the detector are made in the presence of both the inference algorithm and end-goal, the designers can be sure that changes made have a genuine impact on the final performance, rather than focussing on optimising task-agnostic proxy metrics (such as track resolution), and hoping that these eventually translate to better actual performance for the task at hand.

We demonstrated the capabilities of \tomopt in a simulation of one of the typical industrial use-cases -- that of estimating the fill-level of furnace ladles at a metal refinery. Using task-specific inference and losses, we were able to optimise an initial, poorly designed detector into one that was able to not-only match, but slightly outperform two human-designed baseline detectors in terms of fill-height prediction precision. In this example we performed the optimisation such that the detector always conformed to a fixed cost, however alternatively the optimisation could be performed to minimise the detector cost and not exceed a specified budget.

We believe that \tomopt provides the first demonstration of end-to-end-differentiable and inference-aware optimisation of particle physics instruments, and represents an important step in changing the way we design experimental apparatus. Given the number of free parameters considered here, it is worth mentioning that non-differentiable, black-box optimisers such as Bayesian optimisation (BO) could be well applicable. However given the parameter spaces of larger detectors, such as those at the CERN LHC, where perhaps more than 1000 parameters might be available for control, we believe that differentiable optimisation will be required. With that in mind, our primary of this paper was to demonstrate the viability of measurement-aware detector design in a differentiable pipeline.

In practical application on larger detectors, BO could potentially be used on a reduced parameter set to act as an initial global search to find a decent initialisation, after which full differentiable optimisation can be used on the larger parameter space to refine the detector. Given BO's capability to more easily optimise categorical parameters, this initial search may be additionally beneficial in fixing such parameters prior to differentiable optimisation.

Looking to the immediate future of this work, there are many possible extensions. So far, we have relied on ``classical" inference algorithms, but inference via deep-neural networks are naturally compatible with \tomopt's requirement on differentiable processes. We will be exploring their use in future publications, as these can help to ease the burden of designing task-specific differentiable inference algorithms. Additionally, as a quick simulator of the physics, \tomopt can be exploited for developing and testing advanced (non-differentiable) inference algorithms for muon-tomography in general. The available degrees of freedom of detectors, in terms of placement, technology, and modelling needs to be expanded, to allow for e.g. detectors to be placed on the side of the passive volume, or otherwise rotated. Finally, we will be looking to expand the repertoire of benchmark scenarios used for demonstration and testing. 
\clearpage
\section{Appendix}
\label{sec:appendix}

\begin{figure}[ht]
    \centering
    \includegraphics[width=.55\linewidth]{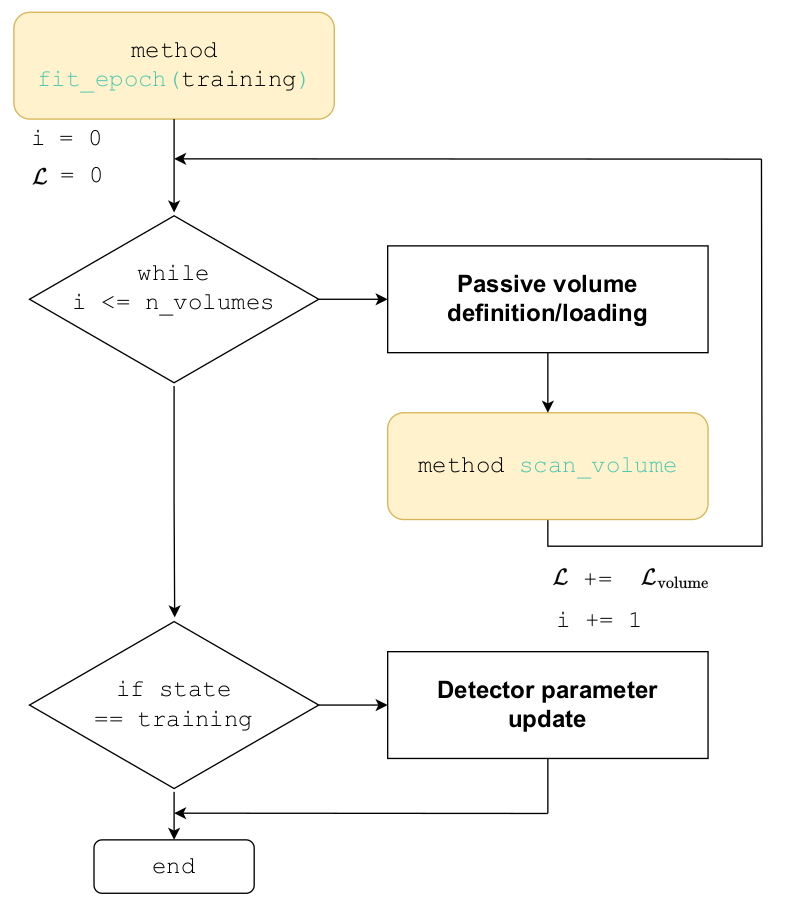}
    \includegraphics[width=.25\linewidth]{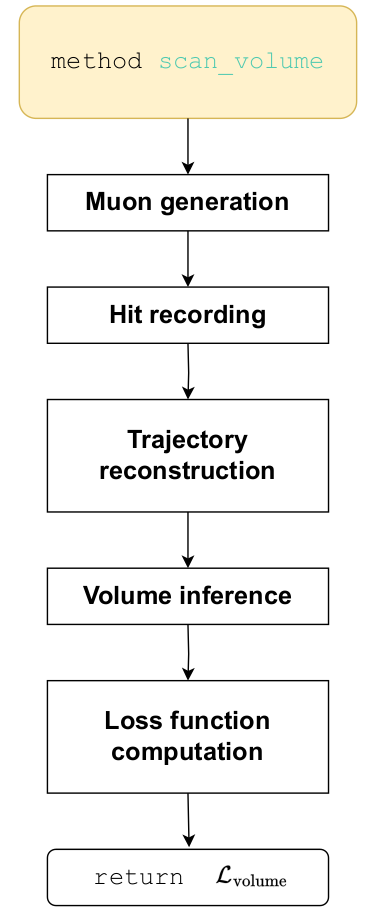}
    \includegraphics[width=.6\linewidth]{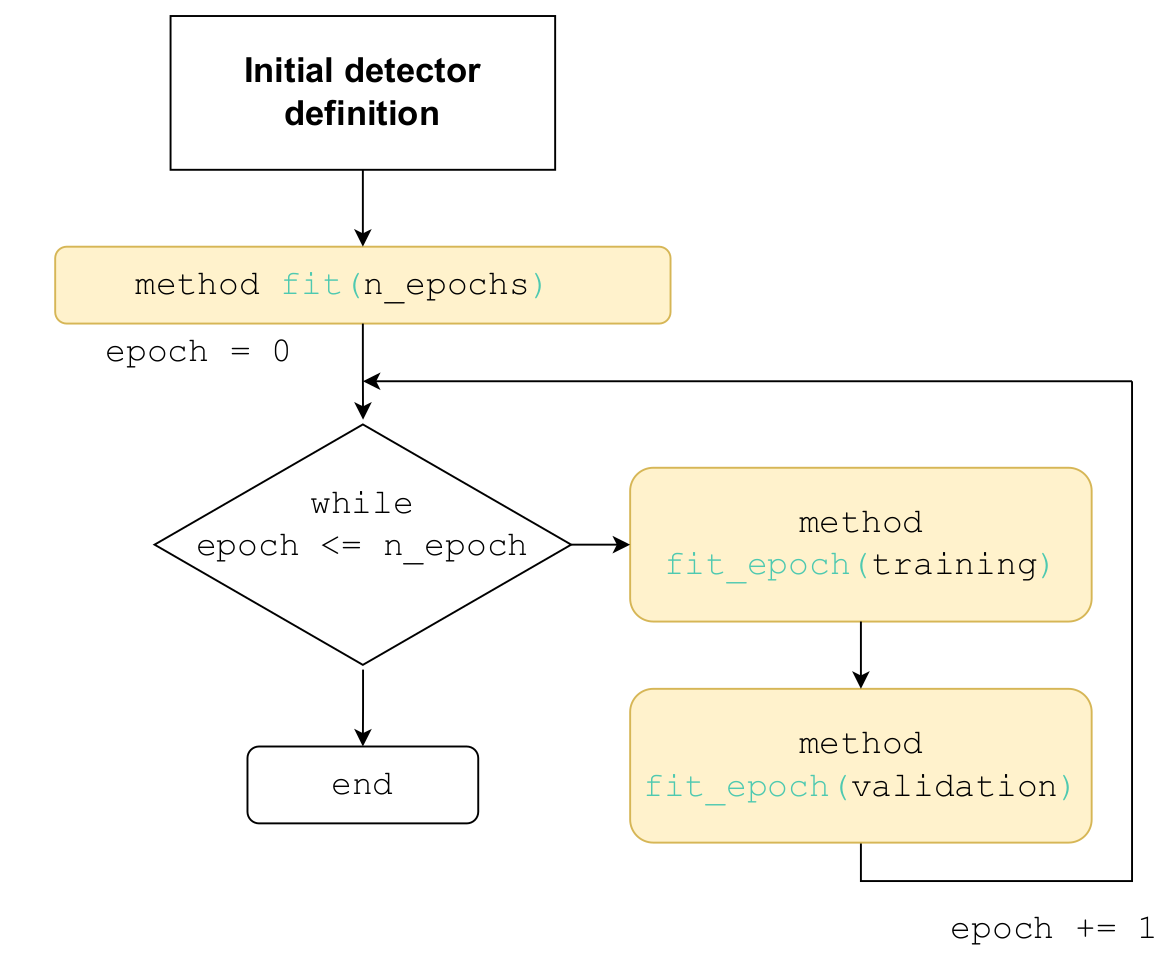}
    \caption{Scan loop for a batch of passive volumes (left), scan loop for muons over a single passive volume (right) and breakdown of the fitting procedure of detectors in \tomopt (below).}
    \label{fig:fit}
\end{figure}

\begin{table}[H]
\caption{Description of variables and learnable parameters from \autoref{fig:fit}, \autoref{fig:all_diag} and \autoref{fig:all_diag_1}.}
\begin{center}
\begin{tabular}{c c c } 
 \hline
 \hline
 Symbol & Definition & Unit  \\ 
\hline
$\theta$ & Muon zenith angle & $\text{rad},\:^\circ$ \\
$\phi$ & Muon azimuthal angle & $\text{rad},\:^\circ$ \\
$X_0$ & Radiation length & m \\

\hline
 $S_{gen}$ & Muon generation surface &   \\ 
 $dx$, $dy$ & Generation surface span in $x$, $y$ & \text{m} \\
 $\text{E}_{\text{range}}$ & Muon generation energy range & \text{GeV} \\
 $\theta_{\text{range}}$ & Muon generation zenith angle range & \text{GeV} \\
 $N_{\text{muons}}$ & Number of generated muons &  \\
 \hline
 $\epsilon_{\text{max}}$ & Panel maximum efficiency  &  \\
 $\sigma_{\text{max}}$ & Panel maximum spatial resolution in $x$ and $y$  & m \\
 $x_{\text{span}, j}, y_{\text{span}, j}$ & Panel $j$ span in $x$ and $y$ as a learnable parameter & m\\
 $x_j$, $y_j$, $z_j$ & Panel $j$ position in $x$, $y$, $z$ as a learnable parameter & m\\

 \hline
 $\epsilon_{\text{eff}}$ & Reconstructed hit effective uncertainty & \\
 $x_{\text{rec}}$, $y_{\text{rec}}$ & Reconstructed muon hit & m\\
 
 $\epsilon_{\mu}$ & Per-muon efficiency &  \\
 $\epsilon_{\text{batch}}$ & Muon batch efficiency, sum of the per-muon efficiencies & \\
 \hline
 
 $\mathcal{L}$ & Total loss function & \\
 $\mathcal{L}_{\text{cost}}$ & Detector cost loss function & \\
 $\mathcal{L}_{\text{error}}$ & Error loss function & \\
$\beta$ & Loss function cost weight & \\
 $\mathcal{L}_{\text{volume}}$ & Volume loss function & \\
 
 \hline
 
  $X^k$ & \makecell{Set of all learnable parameters \\$\{..., (x_j,y_j,z_j,x_{\text{span},j}, y_{\text{span},j}), ...\}$ at iteration $k$} & \\
 $\nabla_{X}$ & Loss function gradients w.r.t learnable parameters $X$& \\
 \hline
 \hline
\end{tabular}
\end{center}
\end{table}

\begin{figure}[ht]
    \centering
    \includegraphics[trim={0 2cm 0 5cm}, width=1\linewidth]{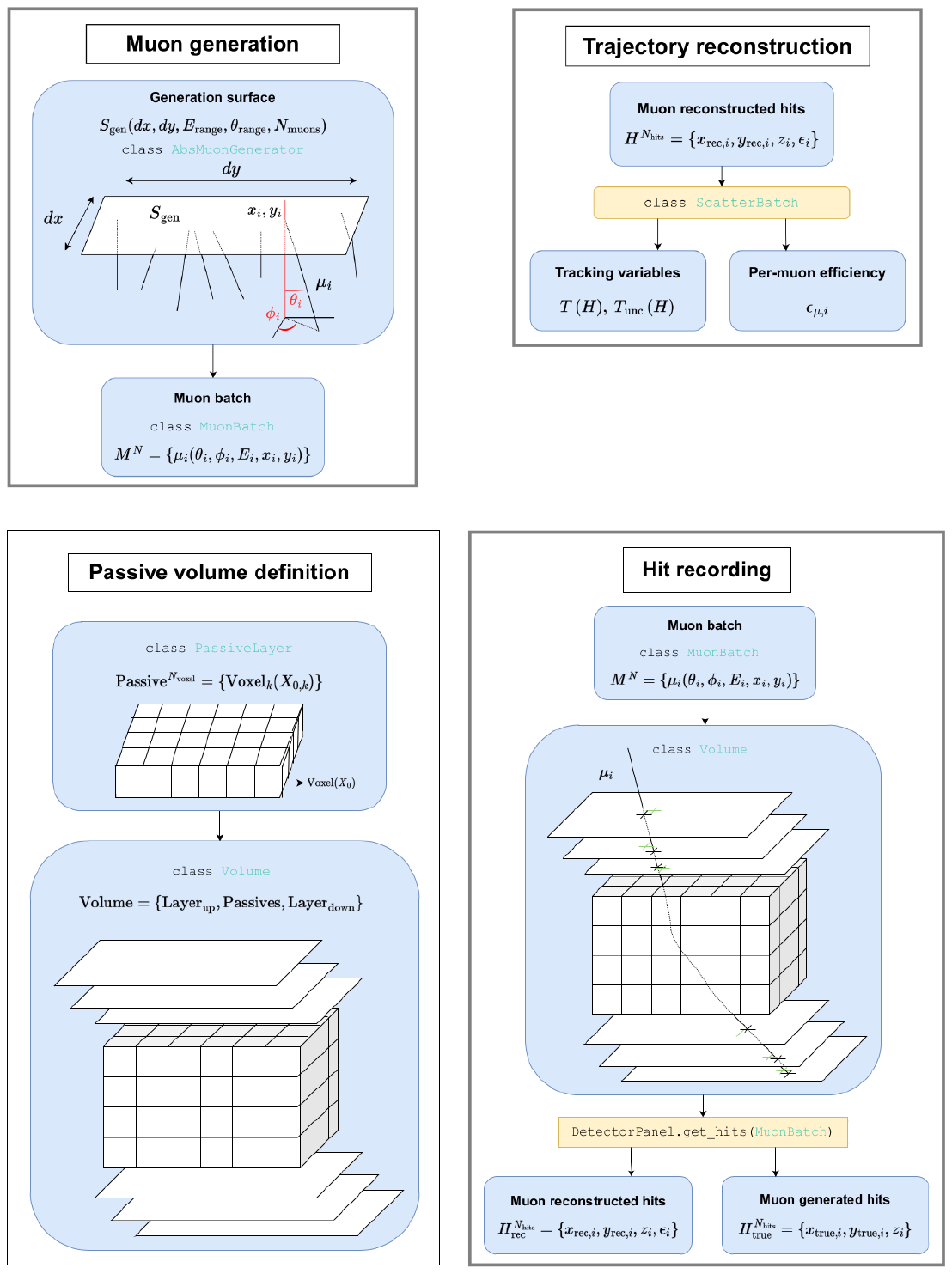}
    \caption{Simplified representation of the various computation steps from the \textit{fit epoch} and \textit{scan volume} methods, from \autoref{fig:fit}.}
    \label{fig:all_diag}
\end{figure}

\begin{figure}[ht]
    \centering
    \includegraphics[trim={0 9.5cm 0 5cm}, width=1\linewidth]{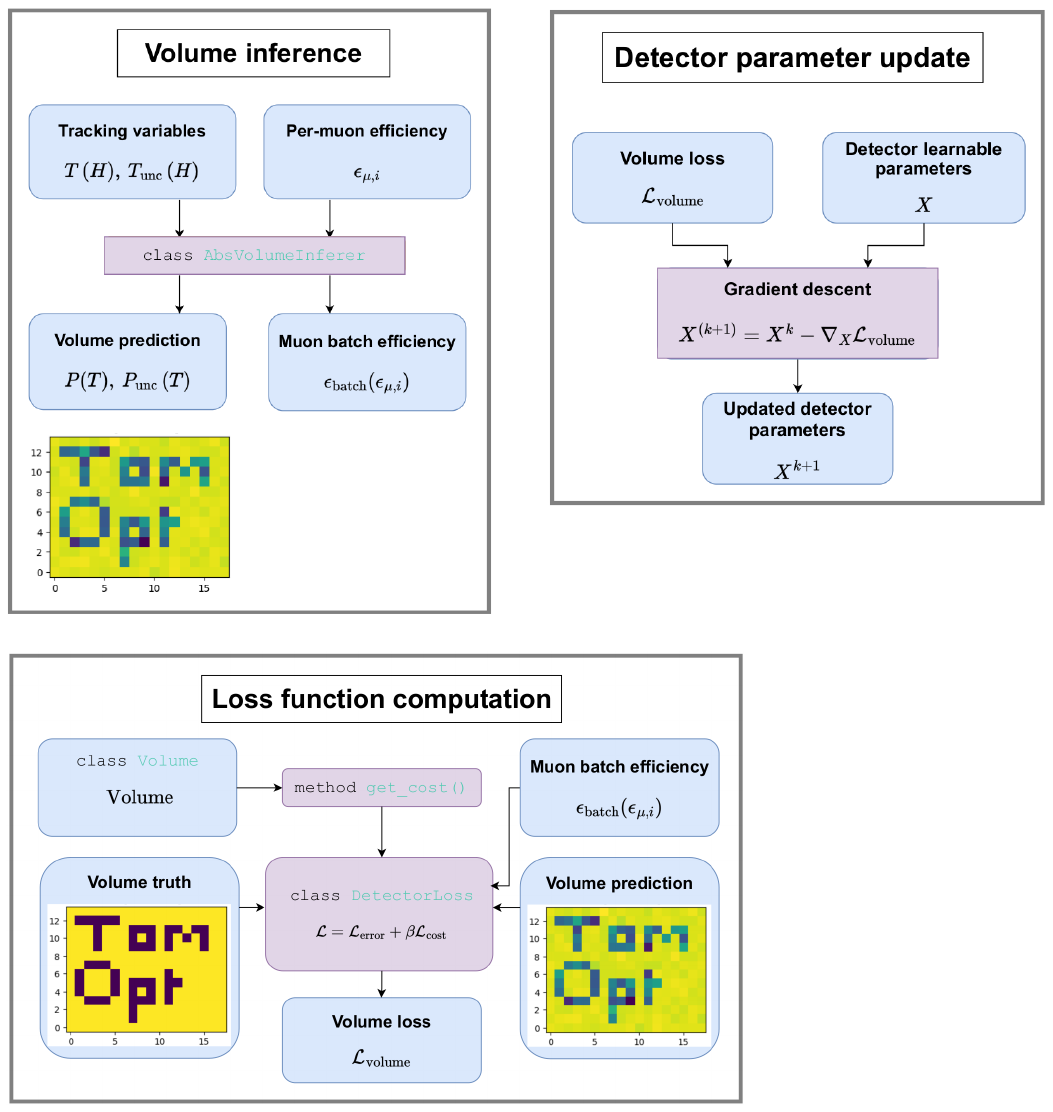}
    \caption{Simplified representation of the various computation steps from the \textit{fit epoch} and \textit{scan volume} methods, from \autoref{fig:fit}.}    
    \label{fig:all_diag_1}
\end{figure}


\FloatBarrier
\clearpage
\section*{Acknowledgements}
This work was partially supported by the EU Horizon 2020 Research and Innovation Programme under grant agreement No. 101021812 (``SilentBorder''), and by the Fonds de la Recherche Scientifique - FNRS under Grants No. T.0099.19 and J.0070.21. The collaboration of Aitor Orio was co-funded by the Spanish Ministry of Science and Innovation through the program “Ayudas para contratos para la formación de doctores en empresas (Doctorados Industriales) 2018” (Grant reference: DIN2018-009886). Pietro Vischia’s work was partially supported by the FNRS under Grant No. 40000963 and by the Ramón y Cajal program under the Project No. RYC2021-033305-I funded by MCIN/AEI/10.13039/501100011033 and by the European Union NextGenerationEU/PRTR.
Max Lamparth was supported by the Deutsche Forschungsgemeinschaft (DFG, German Research Foundation) under Germany's Excellence Strategy -- EXC-2094 -- 390783311. We are grateful to Pablo Canteli Morales for the useful comments he provided us.
\small{\bibliography{content/main}}

\providecommand{\href}[2]{#2}\begingroup\raggedright\begin{thebibliography}{10}

\bibitem{tomopt_git}
G.~C. Strong {\em et~al.},  {\em TomOpt: Differential Muon Tomography
  Optimisation}, Feb., 2024.
\newblock \url{https://doi.org/10.5281/zenodo.10673885}.
  \url{https://github.com/GilesStrong/tomopt}.

\bibitem{George1955}
E.~P. George,  {\em Cosmic Rays Measure Overburden of Tunnel}, Commonwealth
  Engineer {\bfseries July 1} (1955) 455.

\bibitem{Alvarez1970}
L.~W. Alvarez {\em et~al.},  {\em Search for Hidden Chambers in the Pyramids},
  Science {\bfseries 167} (1970) 832.

\bibitem{Morishima2017}
K.~Morishima {\em et~al.},  {\em {Discovery of a big void in Khufu's Pyramid by
  observation of cosmic-ray muons}},
  \href{http://dx.doi.org/10.1038/nature24647}{Nature {\bfseries 552} no.~7685,
  (2017) 386},
\href{http://arxiv.org/abs/1711.01576}{{\ttfamily arXiv:1711.01576
  [physics.ins-det]}}.

\bibitem{procureur2023precise}
S.~Procureur {\em et~al.},  {\em {Precise characterization of a corridor-shaped
  structure in Khufu’s Pyramid by observation of cosmic-ray muons}}, Nature
  Communications {\bfseries 14} no.~1, (2023) 1144.

\bibitem{MuographyBook}
L.~Ol{\'a}h, H.~K. Tanaka, and D.~Varga, eds., {\em Muography: Exploring
  Earth's Subsurface with Elementary Particles}.
\newblock John Wiley \& Sons, 2022.

\bibitem{Rutherford1911}
E.~Rutherford,  {\em The scattering of $\alpha$ and $\beta$ particles by matter
  and the structure of the atom},
  \href{http://dx.doi.org/10.1080/14786440508637080}{Lond. Edinb. Dubl. Phil.
  Mag. {\bfseries 21} no.~125, (1911) 669}.

\bibitem{LynchDahl1991}
G.~R. Lynch and O.~I. Dahl,  {\em {Approximations to multiple Coulomb
  scattering}}, \href{http://dx.doi.org/10.1016/0168-583X(91)95671-Y}{Nucl.
  Inst. Meth. B {\bfseries 58} (1991) 6}.

\bibitem{Borozdin2003}
K.~N. Borozdin {\em et~al.},  {\em Radiographic imaging with cosmic-ray muons},
  \href{http://dx.doi.org/10.1038/422277a}{Nature {\bfseries 422} (2003) 277}.

\bibitem{barnes2023cosmic}
S.~Barnes, A.~Georgadze, A.~Giammanco, M.~Kiisk, V.~A. Kudryavtsev,
  M.~Lagrange, and O.~L. Pinto,  {\em Cosmic-Ray Tomography for Border
  Security}, Instruments {\bfseries 7} no.~1, (2023) 13.

\bibitem{weekes2021material}
M.~Weekes, A.~Alrheli, D.~Barker, D.~Kiko{\l}a, A.~K. Kopp, M.~Mhaidra,
  J.~Stowell, L.~Thompson, and J.~J. Velthuis,  {\em Material identification in
  nuclear waste drums using muon scattering tomography and multivariate
  analysis}, Journal of Instrumentation {\bfseries 16} no.~05, (2021) P05007.

\bibitem{morris2014horizontal}
C.~L. Morris, J.~Bacon, K.~Borozdin, J.~Fabritius, H.~Miyadera, J.~Perry, and
  T.~Sugita,  {\em Horizontal cosmic ray muon radiography for imaging nuclear
  threats}, Nucl. Instr. Meth B {\bfseries 330} (2014)42--46.

\bibitem{IAEA2022}
{International Atomic Energy Agency},  {\em Muon imaging: Present Status and
  Emerging Applications}, IAEA TECDOC 2012, IAEA, Vienna, 2022.
\newblock \url{https://www.iaea.org/publications/15182/muon-imaging}.

\bibitem{Mrdja_2016}
D.~Mrdja, I.~Bikit, K.~Bikit, J.~Slivka, J.~Hansman, L.~Ol{\'{a}}h, and
  D.~Varga,  {\em First cosmic-ray images of bone and soft tissue},
  \href{http://dx.doi.org/10.1209/0295-5075/116/48003}{{EPL} (Europhysics
  Letters) {\bfseries 116} no.~4, (nov, 2016) 48003}.
  \url{https://doi.org/10.1209/0295-5075/116/48003}.

\bibitem{Bikit_2016}
I.~Bikit, D.~Mrdja, K.~Bikit, J.~Slivka, N.~Jovancevic, L.~Ol{\'{a}}h,
  G.~Hamar, and D.~Varga,  {\em Novel approach to imaging by cosmic-ray muons},
  \href{http://dx.doi.org/10.1209/0295-5075/113/58001}{{EPL} (Europhysics
  Letters) {\bfseries 113} no.~5, (mar, 2016) 58001}.
  \url{10.1209/0295-5075/113/58001}.

\bibitem{TUMUTY_yifan_2018}
Z.~Yifan, Z.~Zhi, Z.~Ming, W.~Xuewu, and Z.~Ziran,  {\em Discrimination of
  drugs and explosives in cargo inspections by applying machine learning in
  muon tomography}, \href{http://dx.doi.org/10.11884/HPLPB201830.180062}{High
  Power Laser and Particle Beams {\bfseries 30} no.~30086002, (2018) 086002}.

\bibitem{xuan_tao_ji_2022}
J.~Xuan-Tao, L.~Si-Yuan, H.~Yu-He, Z.~Kun, Z.~Jin, P.~Xiao-Yu, X.~Min, and
  W.~Xiao-Dong,  {\em {A novel 4D resolution imaging method for low and medium
  atomic number objects at the centimeter scale by coincidence detection
  technique of cosmic-ray muon and its secondary particles}},
  \href{http://dx.doi.org/10.1007/s41365-022-00989-0}{Nuclear Science and
  Techniques {\bfseries 33} no.~2, (2022)2210--3147}.

\bibitem{Holma_Joutsenvaara_Kuusiniemi_2022}
M.~Holma, J.~Joutsenvaara, and P.~Kuusiniemi,  {\em {Trends in Publishing
  Muography Related Research: The Situation at the End of 2020}},
  \href{http://dx.doi.org/https://doi.org/10.31526/jais.2022.292}{Journal of
  Advanced Instrumentation in Science {\bfseries 2022} (2022) 292}.
  \url{http://journals.andromedapublisher.com/index.php/JAIS/article/view/292}.

\bibitem{Bonechi:2019ckl}
L.~Bonechi, R.~D'Alessandro, and A.~Giammanco,  {\em {Atmospheric muons as an
  imaging tool}}, \href{http://dx.doi.org/10.1016/j.revip.2020.100038}{Rev.
  Phys. {\bfseries 5} (2020) 100038},
  \href{http://arxiv.org/abs/1906.03934}{{\ttfamily arXiv:1906.03934
  [physics.ins-det]}}.

\bibitem{gaisser}
T.~K. {Gaisser}, {\em {Cosmic rays and particle physics.}}
\newblock {Cambridge University Press}, 1990.

\bibitem{guan}
M.~Guan, M.-C. Chu, J.~Cao, K.-B. Luk, and C.~Yang,  {\em A parametrization of
  the cosmic-ray muon flux at sea-level}, 2015.

\bibitem{shukla}
P.~Shukla and S.~Sankrith,  {\em {Energy and angular distributions of
  atmospheric muons at the Earth}},
  \href{http://dx.doi.org/10.1142/S0217751X18501750}{Int. J. Mod. Phys. A
  {\bfseries 33} no.~30, (2018) 1850175},
  \href{http://arxiv.org/abs/1606.06907}{{\ttfamily arXiv:1606.06907
  [hep-ph]}}.

\bibitem{pdg_model}
{\bfseries Particle Data Group} Collaboration, R.~L. Workman and Others,  {\em
  {Review of Particle Physics}},
  \href{http://dx.doi.org/10.1093/ptep/ptac097}{PTEP {\bfseries 2022} (2022)
  083C01}.

\bibitem{moliere1948theorie}
G.~Moliere,  {\em Theorie der streuung schneller geladener teilchen ii
  mehrfach-und vielfachstreuung}, Zeitschrift f{\"u}r Naturforschung A
  {\bfseries 3} no.~2, (1948)78--97.

\bibitem{bethe1953moliere}
H.~A. Bethe,  {\em Moliere's theory of multiple scattering}, Physical review
  {\bfseries 89} no.~6, (1953) 1256.

\bibitem{pytorch}
A.~Paszke {\em et~al.},  {\em PyTorch: An Imperative Style, High-Performance
  Deep Learning Library}, in {\em Advances in Neural Information Processing
  Systems 32}, H.~Wallach, H.~Larochelle, A.~Beygelzimer, F.~d\textquotesingle
  Alch\'{e}-Buc, E.~Fox, and R.~Garnett, eds., pp.~8024--8035.
\newblock Curran Associates, Inc., 2019.
\newblock
  \url{http://papers.neurips.cc/paper/9015-pytorch-an-imperative-style-high-performance-deep-learning-library.pdf}.

\bibitem{mode_npni}
A.~G. Baydin, K.~Cranmer, P.~de~Castro~Manzano, C.~Delaere, D.~Derkach,
  J.~Donini, T.~Dorigo, A.~Giammanco, J.~Kieseler, L.~Layer, G.~Louppe,
  F.~Ratnikov, G.~C. Strong, M.~Tosi, A.~Ustyuzhanin, P.~Vischia, and H.~Yarar,
   {\em Toward Machine Learning Optimization of Experimental Design},
  \href{http://dx.doi.org/10.1080/10619127.2021.1881364}{Nuclear Physics News
  {\bfseries 31} no.~1, (2021)25--28},
  \href{http://arxiv.org/abs/https://doi.org/10.1080/10619127.2021.1881364}{{\ttfamily
  https://doi.org/10.1080/10619127.2021.1881364}}.
  \url{https://doi.org/10.1080/10619127.2021.1881364}.

\bibitem{mode_whitepaper}
T.~Dorigo, A.~Giammanco, P.~Vischia, M.~Aehle, M.~Bawaj, A.~Boldyrev, P.~{de
  Castro Manzano}, D.~Derkach, J.~Donini, A.~Edelen, F.~Fanzago, N.~R. Gauger,
  C.~Glaser, A.~G. Baydin, L.~Heinrich, R.~Keidel, J.~Kieseler, C.~Krause,
  M.~Lagrange, M.~Lamparth, L.~Layer, G.~Maier, F.~Nardi, H.~E. Pettersen,
  A.~Ramos, F.~Ratnikov, D.~Röhrich, R.~R. {de Austri}, P.~M.~R. {del Árbol},
  O.~Savchenko, N.~Simpson, G.~C. Strong, A.~Taliercio, M.~Tosi,
  A.~Ustyuzhanin, and H.~Zaraket,  {\em Toward the end-to-end optimization of
  particle physics instruments with differentiable programming},
  \href{http://dx.doi.org/https://doi.org/10.1016/j.revip.2023.100085}{Reviews
  in Physics {\bfseries 10} (2023) 100085}.
  \url{https://www.sciencedirect.com/science/article/pii/S2405428323000047}.

\bibitem{giles_1st_mode_ws}
G.~Strong and T.~Dorigo,  {\em {TOMOPT: Differential Muon Tomography
  Optimisation}}, 1st MODE Workshop on Differentiable Programming (2021).
  \url{https://indico.cern.ch/event/1022938/timetable/?view=standard#7-tomopt-pytorch-based-differe}.
  Accessed 2023/09/23.

\bibitem{2nd_mode_challenge}
G.~Strong {\em et~al.},  {\em {Data Challenge for the 2nd MODE Workshop on
  Differentiable Programming}},
  \href{http://dx.doi.org/10.5281/zenodo.6947863}{2nd MODE Workshop on
  Differentiable Programming (2022)}.
  \url{https://indico.cern.ch/event/1145124/contributions/4806648/}. Repository
  \url{https://github.com/GilesStrong/mode_diffprog_22_challenge}.

\bibitem{giles_2nd_mode_ws}
G.~Strong, T.~Dorigo, A.~Giammanco, M.~L. Pietro~Vischia, Jan~Kieseler,
  F.~Nardi, H.~Zaraket, M.~Lamparth, F.~Fanzago, O.~Savchenko, and
  A.~Bordignon,  {\em {TomOpt: Differentiable Optimisation of Muon-Tomography
  Detectors}}, 2nd MODE Workshop on Differentiable Programming (2022).
  \url{https://indico.cern.ch/event/1145124/contributions/4983513/}. Accessed
  2023/09/23.

\bibitem{AGOSTINELLI2003250}
S.~Agostinelli {\em et~al.},  {\em Geant4—a simulation toolkit},
  \href{http://dx.doi.org/10.1016/S0168-9002(03)01368-8}{Nuclear Instruments
  and Methods in Physics Research Section A: Accelerators, Spectrometers,
  Detectors and Associated Equipment {\bfseries 506} no.~3, (2003)250--303}.

\bibitem{1610988}
J.~Allison {\em et~al.},  {\em Geant4 developments and applications},
  \href{http://dx.doi.org/10.1109/TNS.2006.869826}{IEEE Transactions on Nuclear
  Science {\bfseries 53} no.~1, (2006)270--278}.

\bibitem{ALLISON2016186}
J.~Allison {\em et~al.},  {\em Recent developments in Geant4},
  \href{http://dx.doi.org/10.1016/j.nima.2016.06.125}{Nuclear Instruments and
  Methods in Physics Research Section A: Accelerators, Spectrometers, Detectors
  and Associated Equipment {\bfseries 835} (2016)186--225}.

\bibitem{4271541}
L.~J. Schultz, G.~S. Blanpied, K.~N. Borozdin, A.~M. Fraser, N.~W. Hengartner,
  A.~V. Klimenko, C.~L. Morris, C.~Orum, and M.~J. Sossong,  {\em Statistical
  Reconstruction for Cosmic Ray Muon Tomography},
  \href{http://dx.doi.org/10.1109/TIP.2007.901239}{IEEE Transactions on Image
  Processing {\bfseries 16} no.~8, (2007)1985--1993}.

\bibitem{Qasim:2019otl}
S.~R. Qasim, J.~Kieseler, Y.~Iiyama, and M.~Pierini,  {\em {Learning
  representations of irregular particle-detector geometry with
  distance-weighted graph networks}},
  \href{http://dx.doi.org/10.1140/epjc/s10052-019-7113-9}{Eur. Phys. J. C
  {\bfseries 79} no.~7, (2019) 608},
  \href{http://arxiv.org/abs/1902.07987}{{\ttfamily arXiv:1902.07987
  [physics.data-an]}}.

\bibitem{giles_5th_iml}
G.~Strong,  {\em {2-Level Graphs for Muon-Tomography}}, 5th CERN
  Inter-experiment Machine Learning Workshop (2022).
  \url{https://indico.cern.ch/event/1078970/timetable/?view=standard#42-two-level-graphs-for-muon-t}.
  Accessed 2023/09/23.

\bibitem{Linnainmaa_70}
S.~Linnainmaa,  {\em The representation of the cumulative rounding error of an
  algorithm as a {Taylor} expansion of the local rounding errors}, Master's
  thesis, Univ. Helsinki, 1970.

\bibitem{Werbos_81}
P.~J. Werbos,  {\em Applications of Advances in Nonlinear Sensitivity
  Analysis}, in {\em Proceedings of the 10th IFIP Conference, 31.8 - 4.9, NYC},
  pp.~762--770.
\newblock 1981.

\bibitem{backprop}
D.~E. {Rumelhart}, G.~E. {Hinton}, and R.~J. {Williams},  {\em {Learning
  representations by back-propagating errors}},
  \href{http://dx.doi.org/10.1038/323533a0}{Nature {\bfseries 323} (Oct.,
  1986)533--536}.

\bibitem{gradient_descent}
J.~Hadamard, {\em M{\'e}moire sur le probl{\`e}me d'analyse relatif {\`a}
  l'{\'e}quilibre des plaques {\'e}lastiques encastr{\'e}es}.
\newblock M{\'e}moires pr{\'e}sent{\'e}s par divers savants {\`a}
  l'Acad{\'e}mie des sciences de l'Institut de France: {\'E}xtrait. Imprimerie
  nationale, 1908.
\newblock \url{http://books.google.com.au/books?id=BTEPAAAAIAAJ}.

\bibitem{adam}
D.~P. Kingma and J.~Ba,  {\em Adam: {A} Method for Stochastic Optimization}, in
  {\em 3rd International Conference on Learning Representations, {ICLR} 2015,
  San Diego, CA, USA, May 7-9, 2015, Conference Track Proceedings}, Y.~Bengio
  and Y.~LeCun, eds.
\newblock 2015.
\newblock \url{http://arxiv.org/abs/1412.6980}.

\bibitem{Martinez2012}
P.~Martínez, C.~Díez, P.~Gómez, and A.~Orio,  {\em Process monitoring:
  measurement of the metal–slag interface in furnace ladles}, IAEA TECDOC
  SERIES. Muon Imaging. Present Status and Emerging Applications {\bfseries
  2012} (2022)34--35.
  \url{https://www.iaea.org/publications/15182/muon-imaging}.

\bibitem{smith_2015}
L.~N. {Smith},  {\em Cyclical Learning Rates for Training Neural Networks}, in
  {\em 2017 IEEE Winter Conference on Applications of Computer Vision (WACV)},
  pp.~464--472.
\newblock 2017.
\newblock \href{http://arxiv.org/abs/1506.01186}{{\ttfamily arXiv:1506.01186}}.
\newblock \url{http://arxiv.org/abs/1506.01186}.

\bibitem{smith_2017}
L.~N. Smith and N.~Topin, \href{http://dx.doi.org/10.1117/12.2520589}{ {\em
  {Super-convergence: very fast training of neural networks using large
  learning rates}},} in {\em Artificial Intelligence and Machine Learning for
  Multi-Domain Operations Applications}, T.~Pham, ed., vol.~11006, pp.~369 --
  386, International Society for Optics and Photonics.
\newblock SPIE, 2019.
\newblock \href{http://arxiv.org/abs/1708.07120}{{\ttfamily arXiv:1708.07120}}.
\newblock \url{https://doi.org/10.1117/12.2520589}.

\bibitem{smith_2018}
L.~N. Smith,  {\em A disciplined approach to neural network hyper-parameters:
  Part 1 - learning rate, batch size, momentum, and weight decay}, CoRR
  {\bfseries abs/1803.09820} (2018),
  \href{http://arxiv.org/abs/1803.09820}{{\ttfamily arXiv:1803.09820}}.
  \url{http://arxiv.org/abs/1803.09820}.

\end{thebibliography}\endgroup

\end{document}